\documentclass[11pt,aps,prd,superscriptaddress,showpacs,byrevtex,floatfix]{revtex4}

\usepackage{epsfig}
\usepackage{amssymb}

\newcommand{\BF}{\ensuremath{{\cal{B}}}}
\newcommand{\Am}{\ensuremath{{\cal{A}}}}
\newcommand{\UFS}{\ensuremath{\Upsilon(4S)}}
\newcommand{\bbbar}{\ensuremath{B\bar{B}}}
\newcommand{\qqbar}{\ensuremath{q\bar{q}}}

\newcommand{\de}{\ensuremath{\Delta E}}
\newcommand{\mb}{\ensuremath{M_{\rm bc}}}

\newcommand{\mom}{GeV/$c$}
\newcommand{\mass}{MeV/$c^2$}
\newcommand{\Mass}{GeV/$c^2$}
\newcommand{\Masssq}{(GeV/$c^2$)$^2$}

\newcommand{\bckpp}{\ensuremath{B^+\to K^+\pi^+\pi^-}}
\newcommand{\bckkk}{\ensuremath{B^+\to K^+K^+K^-}}

\newcommand{\Kpp}{\ensuremath{K\pi\pi}}
\newcommand{\KKK}{\ensuremath{KKK}}
\newcommand{\kpp}{\ensuremath{K^+\pi^+\pi^-}}
\newcommand{\kkk}{\ensuremath{K^+K^+K^-}}

\newcommand{\pipi}{\ensuremath{\pi^+\pi^-}}
\newcommand{\kpkm}{\ensuremath{K^+K^-}}
\newcommand{\kcpi}{\ensuremath{K^+\pi^-}}

\newcommand{\chic}{\ensuremath{\chi_{c0}}}
\newcommand{\mkkmin}{\ensuremath{M(K^+K^-)_{\rm min}}}
\newcommand{\mkkmax}{\ensuremath{M(K^+K^-)_{\rm max}}}

\newcommand{\mkkmins}{\ensuremath{M^2(K^+K^-)_{\rm min}}}
\newcommand{\mkkmaxs}{\ensuremath{M^2(K^+K^-)_{\rm max}}}
\newcommand{\mkkssks}{\ensuremath{M^2(K^+K^+)}}

\newcommand{\sfs}{\ensuremath{s_{12}}}
\newcommand{\sft}{\ensuremath{s_{13}}}
\newcommand{\sst}{\ensuremath{s_{23}}}

\def\nima#1#2#3{{Nucl.\ Instr.\ and Meth.} {\bf A#1}, #3 (#2)}

\def\npb#1#2#3{{ Nucl.\ Phys.}  {\bf B#1}, #3 (#2)}
\def\plb#1#2#3{{ Phys.\ Lett.}  {\bf B#1}, #3 (#2)}

\def\prd#1#2#3{{ Phys.\ Rev.}   {\bf D#1}, #3 (#2)}

\def\prl#1#2#3{{ Phys.\ Rev.\ Lett.} {\bf #1}, #3 (#2)}

\def\zpc#1#2#3{{ Zeit.\ Phys.} {\bf C#1}, #3 (#2)}
\begin{document}

\vbox{ \vbox{    \hbox{   }
                 \hbox{\hspace*{120mm} \bf Belle Preprint 2004-40}
                 \hbox{\hspace*{120mm} \bf KEK Preprint 2004-81}
}}

\title{\quad\\[0.5cm]
\LARGE \bf \boldmath Dalitz analysis of the three-body charmless decays \\
$\bckpp$ and $\bckkk$}

\affiliation{Budker Institute of Nuclear Physics, Novosibirsk}
\affiliation{Chiba University, Chiba}
\affiliation{Chonnam National University, Kwangju}
\affiliation{University of Cincinnati, Cincinnati, Ohio 45221}
\affiliation{Gyeongsang National University, Chinju}
\affiliation{University of Hawaii, Honolulu, Hawaii 96822}
\affiliation{High Energy Accelerator Research Organization (KEK), Tsukuba}
\affiliation{Hiroshima Institute of Technology, Hiroshima}
\affiliation{Institute of High Energy Physics, Chinese Academy of Sciences, Beijing}
\affiliation{Institute of High Energy Physics, Vienna}
\affiliation{Institute for Theoretical and Experimental Physics, Moscow}
\affiliation{J. Stefan Institute, Ljubljana}
\affiliation{Kanagawa University, Yokohama}
\affiliation{Korea University, Seoul}
\affiliation{Kyungpook National University, Taegu}
\affiliation{Swiss Federal Institute of Technology of Lausanne, EPFL, Lausanne}
\affiliation{University of Ljubljana, Ljubljana}
\affiliation{University of Maribor, Maribor}
\affiliation{University of Melbourne, Victoria}
\affiliation{Nagoya University, Nagoya}
\affiliation{Nara Women's University, Nara}
\affiliation{National Central University, Chung-li}
\affiliation{National United University, Miao Li}
\affiliation{Department of Physics, National Taiwan University, Taipei}
\affiliation{H. Niewodniczanski Institute of Nuclear Physics, Krakow}
\affiliation{Nihon Dental College, Niigata}
\affiliation{Niigata University, Niigata}
\affiliation{Osaka City University, Osaka}
\affiliation{Osaka University, Osaka}
\affiliation{Panjab University, Chandigarh}
\affiliation{Peking University, Beijing}
\affiliation{Princeton University, Princeton, New Jersey 08544}
\affiliation{Saga University, Saga}
\affiliation{University of Science and Technology of China, Hefei}
\affiliation{Seoul National University, Seoul}
\affiliation{Sungkyunkwan University, Suwon}
\affiliation{University of Sydney, Sydney NSW}
\affiliation{Tata Institute of Fundamental Research, Bombay}
\affiliation{Toho University, Funabashi}
\affiliation{Tohoku Gakuin University, Tagajo}
\affiliation{Tohoku University, Sendai}
\affiliation{Department of Physics, University of Tokyo, Tokyo}
\affiliation{Tokyo Institute of Technology, Tokyo}
\affiliation{Tokyo Metropolitan University, Tokyo}
\affiliation{Tokyo University of Agriculture and Technology, Tokyo}
\affiliation{University of Tsukuba, Tsukuba}
\affiliation{Virginia Polytechnic Institute and State University, Blacksburg, Virginia 24061}
\affiliation{Yonsei University, Seoul}
  \author{A.~Garmash}\affiliation{Princeton University, Princeton, New Jersey 08544} 
  \author{K.~Abe}\affiliation{High Energy Accelerator Research Organization (KEK), Tsukuba} 
  \author{K.~Abe}\affiliation{Tohoku Gakuin University, Tagajo} 
  \author{H.~Aihara}\affiliation{Department of Physics, University of Tokyo, Tokyo} 
  \author{M.~Akatsu}\affiliation{Nagoya University, Nagoya} 
  \author{Y.~Asano}\affiliation{University of Tsukuba, Tsukuba} 
  \author{V.~Aulchenko}\affiliation{Budker Institute of Nuclear Physics, Novosibirsk} 
  \author{T.~Aushev}\affiliation{Institute for Theoretical and Experimental Physics, Moscow} 
  \author{T.~Aziz}\affiliation{Tata Institute of Fundamental Research, Bombay} 
  \author{S.~Bahinipati}\affiliation{University of Cincinnati, Cincinnati, Ohio 45221} 
% ...
\collaboration{The Belle Collaboration}

\begin{abstract}
We report results on the Dalitz analysis of three-body charmless $\bckpp$ and
$\bckkk$ decays based on a $140$~fb$^{-1}$ data sample collected with the
Belle detector. Measurements of branching fractions for quasi-two-body
decays to scalar-pseudoscalar states: $B^+\to f_0(980)K^+$,
$B^+\to K^*_0(1430)^0\pi^+$, and to vector-pseudoscalar states:
$B^+\to K^*(892)^0\pi^+$, $B^+\to\rho(770)^0K^+$, $B^+\to\phi K^+$ are
presented. Upper limits on decays to some pseudoscalar-tensor final
states are reported. We also report the measurement of the $B^+\to\chic K^+$
branching fraction in two $\chic$ decays channels: $\chic\to \pi^+\pi^-$ and
$\chic\to K^+K^-$.

\end{abstract}

\maketitle


\section{Introduction}

Studies of $B$ meson decays to three-body charmless hadronic final states are
a natural extension of studies of decays to two-body charmless final states.
Some of the final states considered so far as two-body (for example $\rho \pi$,
$K^*\pi$, etc.) proceed via quasi-two-body processes involving a wide
resonance state that immediately decays in the simplest case to two particles,
thereby producing a three-body final state. Multiple resonances occurring
nearby in phase space will interfere and a full amplitude analysis is required
to extract correct branching fractions for the intermediate quasi-two-body
states. $B$ meson decays to three-body charmless hadronic final states may
provide new possibilities for $CP$ violation
searches~\cite{b2hhhcp,garmash2,hazumi}.

Observations of $B$ meson decays to various three-body charmless hadronic final
states have already been reported by the Belle~\cite{garmash,chang,garmash2},
CLEO~\cite{eckhart} and BaBar~\cite{aubert} collaborations. First results on
the distribution of signal events over the Dalitz plot in the three-body
$\bckpp$ and $\bckkk$ decays are described in Ref.~\cite{garmash}. With a data
sample of $29.1$~fb$^{-1}$ a simplified analysis technique was used due to
lack of statistics. Using a similar technique, the BaBar collaboration has
reported results of their analysis of the Dalitz plot for the decay $\bckpp$
with a $56.4$~fb$^{-1}$ data sample~\cite{babar-dalitz}. With the large data
sample that is now available, we can perform a full amplitude analysis. The
analysis described in this paper is based on a 140\,fb$^{-1}$ data sample
containing 152 million $B\bar{B}$ pairs, collected  with the Belle detector
operating at the KEKB asymmetric-energy $e^+e^-$ collider~\cite{KEKB} with a
center-of-mass (c.m.) energy at the $\Upsilon(4S)$ resonance (on-resonance
data). The beam energies are 3.5 GeV for positrons and 8.0 GeV for electrons.
For the study of the $e^+e^-\to q\bar{q}$ continuum background, we use
8.3\,fb$^{-1}$ of data taken 60~MeV below the $\Upsilon(4S)$ resonance
(off-resonance data).

The paper is organized as follows: Section~\ref{sec:detector} gives a brief
description of the Belle detector; the event reconstruction procedure and
background suppression techniques are described in Sections~\ref{sec:ev_rec}
and~\ref{sec:background}, respectively; Section~\ref{sec:khh}
describes results on the three-body signal yields measurement and qualitative
analysis of the two-particle mass spectra, while Section~\ref{sec:aa} is
devoted to the amplitude analysis of the observed three-body signals; final
results of the analysis are given in Section~\ref{sec:results} and 
discussed in Section~\ref{sec:discussion}.


\section{The Belle detector}
\label{sec:detector}

  The Belle detector~\cite{Belle} is a large-solid-angle magnetic spectrometer
based on a 1.5~T superconducting solenoid magnet. Charged particle tracking is
provided by a three-layer silicon vertex detector and a 50-layer central
drift chamber (CDC) that surround the interaction point. The charged particle
acceptance covers laboratory polar angles between $\theta=17^{\circ}$ and
$150^{\circ}$, corresponding to about 92\% of the total solid angle in the
c.m.\ frame. The momentum resolution is determined from cosmic rays and
$e^+ e^-\to\mu^+\mu^-$ events to be $\sigma_{p_t}/p_t=(0.30\oplus0.19p_t)\%$,
where $p_t$ is the transverse momentum in GeV/$c$.

  Charged hadron identification is provided by $dE/dx$ measurements in the CDC,
an array of 1188 aerogel \v{C}erenkov counters (ACC), and a barrel-like array
of 128 time-of-flight scintillation counters (TOF); information from the three
subdetectors is combined to form a single likelihood ratio, which is then used
for pion, kaon and proton discrimination. At large momenta ($>2.5$~GeV/$c$)
only the ACC and CDC are used to separate charged pions and kaons since here
the TOF provides no additional discrimination. Electromagnetic showering
particles are detected in an array of 8736 CsI(Tl) crystals (ECL) that covers
the same solid angle as the charged particle tracking system. The energy
resolution for electromagnetic showers is
$\sigma_E/E = (1.3 \oplus 0.07/E \oplus 0.8/E^{1/4})\%$, where $E$ is in GeV.
Electron identification in Belle is based on a combination of $dE/dx$
measurements in the CDC, the response of the ACC, and the position, shape and
total energy deposition (i.e., $E/p$) of the shower detected in the ECL.
The electron identification efficiency is greater than 92\% for tracks with
$p_{\rm lab}>1.0$~GeV/$c$ and the hadron misidentification probability is below
0.3\%. The magnetic field is returned via an iron yoke that is instrumented to
detect muons and $K^0_L$ mesons. We use a GEANT-based Monte Carlo (MC)
simulation to model the response of the detector and determine its
acceptance~\cite{GEANT}.


\section{Event Reconstruction}
\label{sec:ev_rec}

Charged tracks are selected with a set of track quality requirements based on
the number of CDC hits and on the distances of closest approach to the
interaction point (IP). We also require that the track momenta transverse to
the beam be
greater than 0.1~GeV/$c$ to reduce the low momentum combinatorial background.
For charged kaon identification, we impose a requirement on the particle
identification variable that has 86\% efficiency and a 7\% fake rate from
misidentified pions. Charged tracks that are positively identified as electrons
or protons are excluded. Since the muon identification efficiency and fake rate
vary significantly with the track momentum, we do not veto muons to avoid
additional systematic errors.

\begin{figure}[t]
  \includegraphics[width=0.48\textwidth]{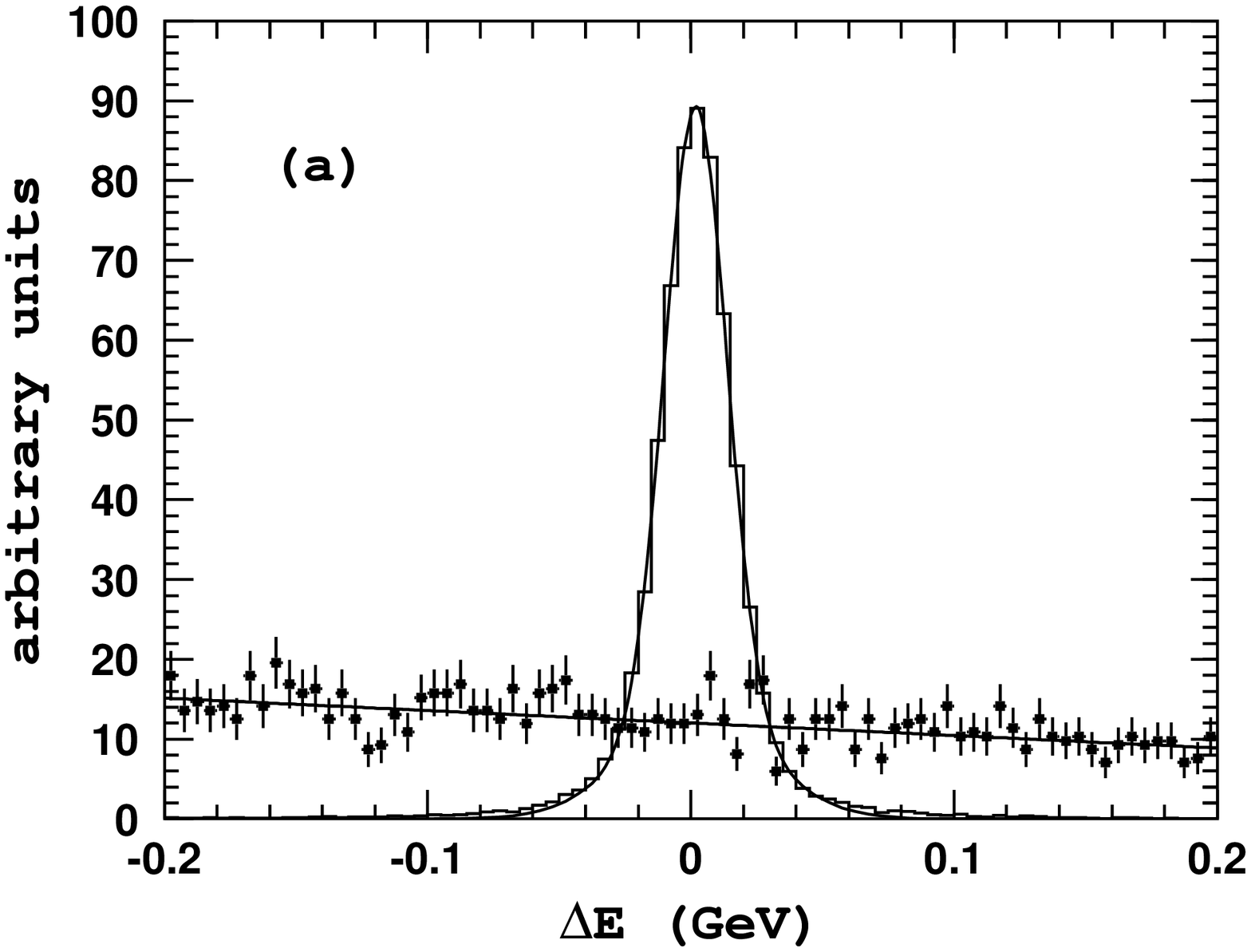} \hfill
  \includegraphics[width=0.48\textwidth]{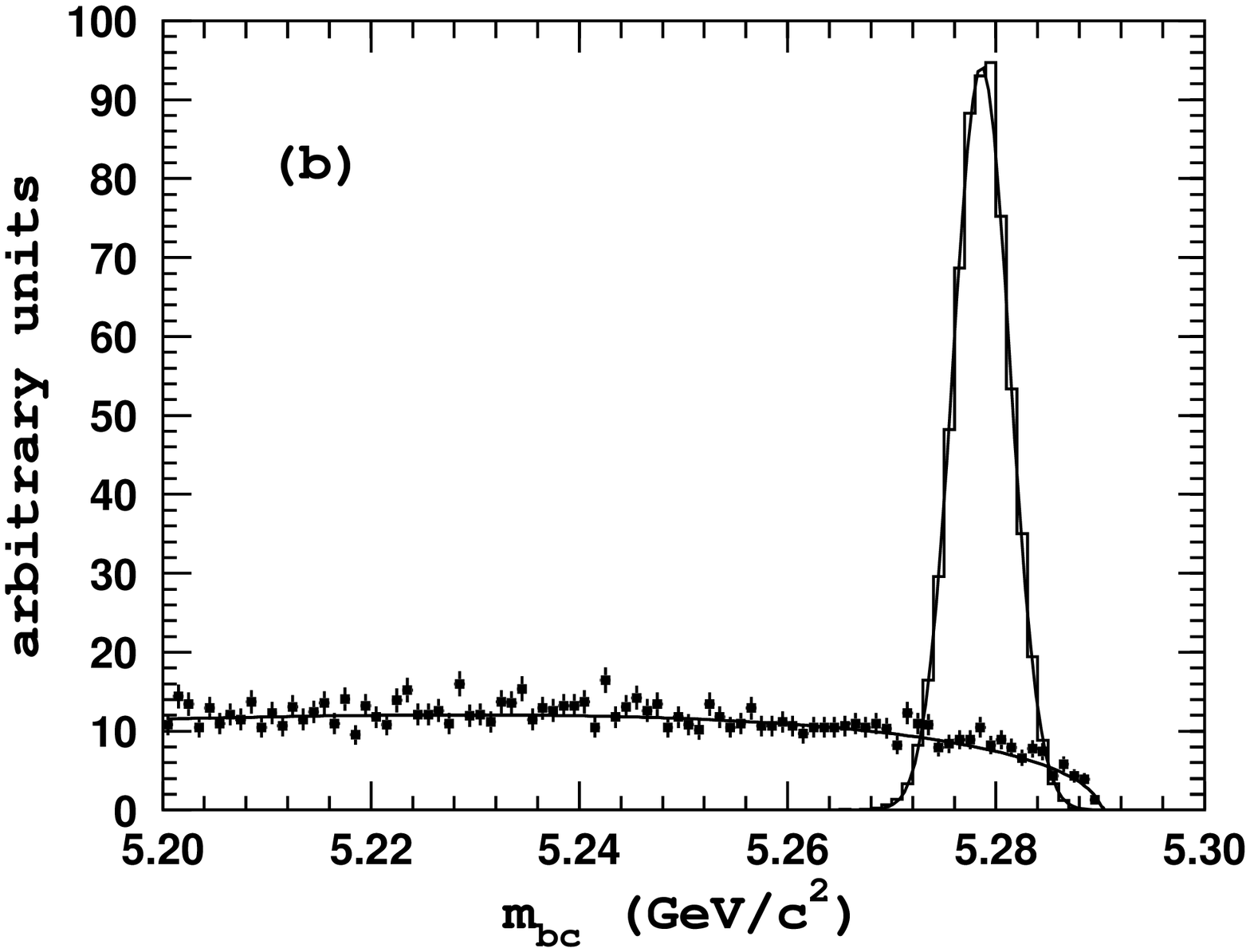} 
  \caption{(a) Energy difference $\de$ and (b) beam constrained mass $\mb$ 
           distributions for $\bckpp$ signal MC events 
           (histograms) and $\qqbar$ background in off-resonance data 
           (points). Curves represent the results of the fits.}
  \label{fig:demb_examp}
\end{figure}

We identify $B$ candidates using two variables: the energy difference $\de$
and the beam-energy constrained mass $\mb$. $\de$ is calculated as
$ \de = E_B - E^*_{\rm beam} = 
\left (\sum_i\sqrt{c^2{\bf p}_i^2 + c^4m_i^2} \right ) - E^*_{\rm beam},$
where the summation is over all particles from a $B$ candidate; and ${\bf p}_i$
and $m_i$ are their c.m.\ three-momenta and masses, respectively. Since the
$\UFS$ decays to a pair of $B$ mesons with no additional particles, each $B$
meson carries half of the c.m.\ energy
$\sqrt{s}/2 = E^*_{\rm beam} = 5.29$~GeV, where $E^*_{\rm beam}$ is the beam
energy in the c.m.\ frame. The $\de$ distribution for the $\bckpp$ signal MC
events is shown in Fig.~\ref{fig:demb_examp}(a). Since there are only charged
particles in final states considered in this analysis, the $\de$ width is
governed by the track momentum resolution. A typical value of the $\de$
resolution is 15~MeV. The beam energy spread is about 3~MeV and gives
a negligible contribution to the total $\de$ width. The signal $\de$ shape is
fitted by a sum of two Gaussian functions with a common mean. In fits to the
experimental data, we fix the width and fraction of the second Gaussian
function from MC simulation. The width of the main Gaussian is floated. For
comparison, the $\de$ distribution for the off-resonance data is also shown
in Fig.~\ref{fig:demb_examp}(a), where the background is parametrized by a
linear function.


The beam-energy constrained mass variable $\mb$ is equivalent to the $B$
invariant mass with the measured $B$ candidate energy $E_B$ replaced by the
beam energy $E^*_{\rm beam}$:
$ \mb = \frac{1}{c^2}\sqrt{E^{*2}_{\rm beam}-c^2{\bf P}_B^2} = 
        \frac{1}{c^2}\sqrt{E^{*2}_{\rm beam}-c^2(\sum_i {\bf p}_i)^2},$
where ${\bf P}_B$ is the $B$ candidate momentum in the c.m.\ frame. The average
$B$ meson momentum in the c.m.\ frame is about 0.34~\mom~which is much smaller
than its total energy. Thus, the uncertainty in the measured ${\bf P}_B$ gives
a small contribution to the $\mb$ width, which is dominated by the beam energy
spread. The $\mb$ width is about 3~MeV/$c^2$ and is nearly independent of the
final state unless photons are included. The
$\mb$ distribution for the signal $\bckpp$ MC events and for the off-resonance
data are shown in Fig.~\ref{fig:demb_examp}(b). The signal $\mb$ shape is well
described by a Gaussian function. The background shape is parametrized with an
empirical function $f(\mb)\propto x\sqrt{1-x^2}\exp[-\lambda(1-x^2)]$, where
$x=\mb/E^*_{\rm beam}$ and $\lambda$ is a parameter~\cite{ArgusF}.


\section{Background Suppression}
\label{sec:background}

  There are two sources of the background: the dominant one is
due to $e^+e^-\to~\qqbar$ ($q = u, d, s$ and $c$ quarks) continuum events
that have a cross-section about three times larger than that for the
$e^+e^-\to\UFS\to\bbbar$; the other one originates from other $B$ meson decays.
The background from continuum events is suppressed using variables that
characterize the event topology. Since the two $B$ mesons produced from an
$\UFS$ decay are nearly at rest in the c.m.\ frame, their decay
products are uncorrelated and the event tends to be spherical. In contrast,
hadrons from continuum $\qqbar$ events tend to exhibit a two-jet structure.
We use $\theta_{\rm thr}$, which is the angle between the thrust axis of
the $B$ candidate and that of the rest of the event, to discriminate between
the two cases. The distribution of $|\cos\theta_{\rm thr}|$ is strongly
peaked near $|\cos\theta_{\rm thr}|=1.0$ for $\qqbar$ events and is nearly
flat for $\bbbar$ events. A Fisher discriminant is utilized for the further
suppression of the continuum background. When combined, these two variables
reject about 98\% (92\%) of the continuum background in the $\bckpp$
($\bckkk$) decay while retaining 36\% (70\%) of the signal. (As the continuum
background in the three-kaon final state is much smaller a looser requirement
on the Fisher discriminant is imposed to retain the efficiency.) A detailed
description of the continuum suppression technique can be found
in~\cite{garmash2} and references therein.

  The understanding of the background that originates from other $B$ meson
decays is of great importance in the study of charmless $B$ decays. We study
the $\bbbar$ related background using a large sample (about 2.5 times the
experimental dataset) of MC generated $\bbbar$ generic events. We use the
CLEO generator~\cite{CLEO:QQ98} to simulate $B$ decays. Note that charmless
hadronic $B$ decays are not included in the QQ98 generator and are generated
separately. We find that the dominant $\bbbar$ background in the $\kpp$
final state that peaks in the signal region is due to $B^+\to\bar{D}^0\pi^+$,
$\bar{D}^0\to K^+\pi^-$ and also $B^+\to J/\psi(\psi(2S))K^+$,
$J/\psi(\psi(2S))\to \mu^+\mu^-$ decays. We veto $B^+\to\bar{D}^0\pi^+$
events by requiring $|M(K\pi)-M_D|>0.10$~GeV/$c^2$. The $B^+\to \bar{D}^0K^+$,
$\bar{D}^0\to\pi^+\pi^-$ signal is removed by requirement
$|M(\pi^+\pi^-)-M_D|>15$~MeV/$c^2$ ($\sim 2.5\sigma$). To suppress the
background due to $\pi/K$ misidentification, we also exclude candidates if
the invariant mass of any pair of oppositely charged tracks from the $B$
candidate is consistent with the $\bar{D}^0\to K^+\pi^-$ hypothesis within
15~MeV/$c^2$, regardless of the particle identification information.
Modes with $J/\psi(\psi(2S))$ in the final state contribute due to muon-pion
misidentification; the contribution from the $J/\psi(\psi(2S))\to e^+e^-$
submode is found to be negligible after the electron veto requirement.
We exclude $J/\psi(\psi(2S))$ background by requiring
$|M(\pi^+\pi^-)_{\mu^+\mu^-}-M_{J/\psi}|>0.07$~GeV/$c^2$ and
$|M(\pi^+\pi^-)_{\mu^+\mu^-}-M_{\psi(2S)}|>0.05$~GeV/$c^2$, where subscript
$\mu^+\mu^-$ indicates that the muon mass assignment was used for charged
tracks to calculate the two-particle invariant mass. Yet another small but
clearly visible background associated with $B^+\to J/\psi K^+$,
$J/\psi\to \mu^+\mu^-$ decays is due to a somewhat complicated particle
misidentification pattern: the charged kaon is misidentified as a pion, the
$\mu^+$ is misidentified as a kaon and the $\mu^-$ as another pion.
This background is excluded by applying a veto:
$|M(K^+\pi^-)_{\mu^+\mu^-}-M_{J/\psi}|>0.020$~GeV/$c^2$.
The most significant background from charmless
$B$ decays is found to originate from $B^+\to \eta'K^+$ followed by
$\eta'\to\pi^+\pi^-\gamma$. Another contribution comes from $B^+\to\rho^0\pi^+$
decay, where one of the final state pions is misidentified as a kaon. We take
these contributions into account when determining the signal yield.

The dominant background to the $\kkk$ final state from other $B$ decays is
found to come from the process $B\to Dh$, where $h$ stands for a charged pion
or kaon. To suppress this background, we reject events where any two-particle
invariant mass is consistent with $D^0\to \kpkm$ or $D^0\to K^-\pi^+$ within
15~MeV$c^2$ regardless of the particle identification information. We find no
charmless $B$ decay modes that produce a significant background to the $\kkk$
final state.


\section{Three-body Signal Yields}
\label{sec:khh}

\begin{figure}[t]
 \includegraphics[width=0.48\textwidth]{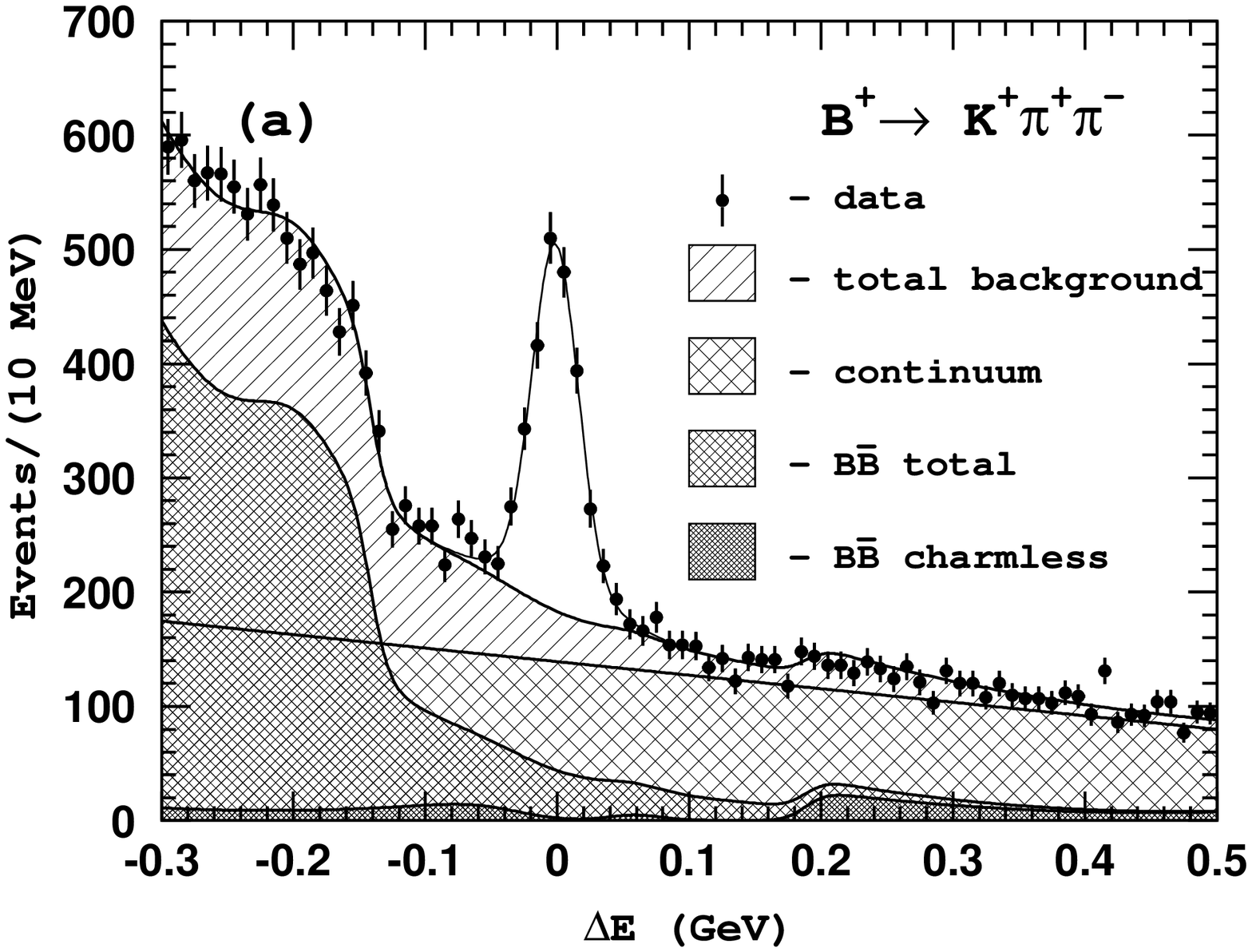} \hfill
 \includegraphics[width=0.48\textwidth]{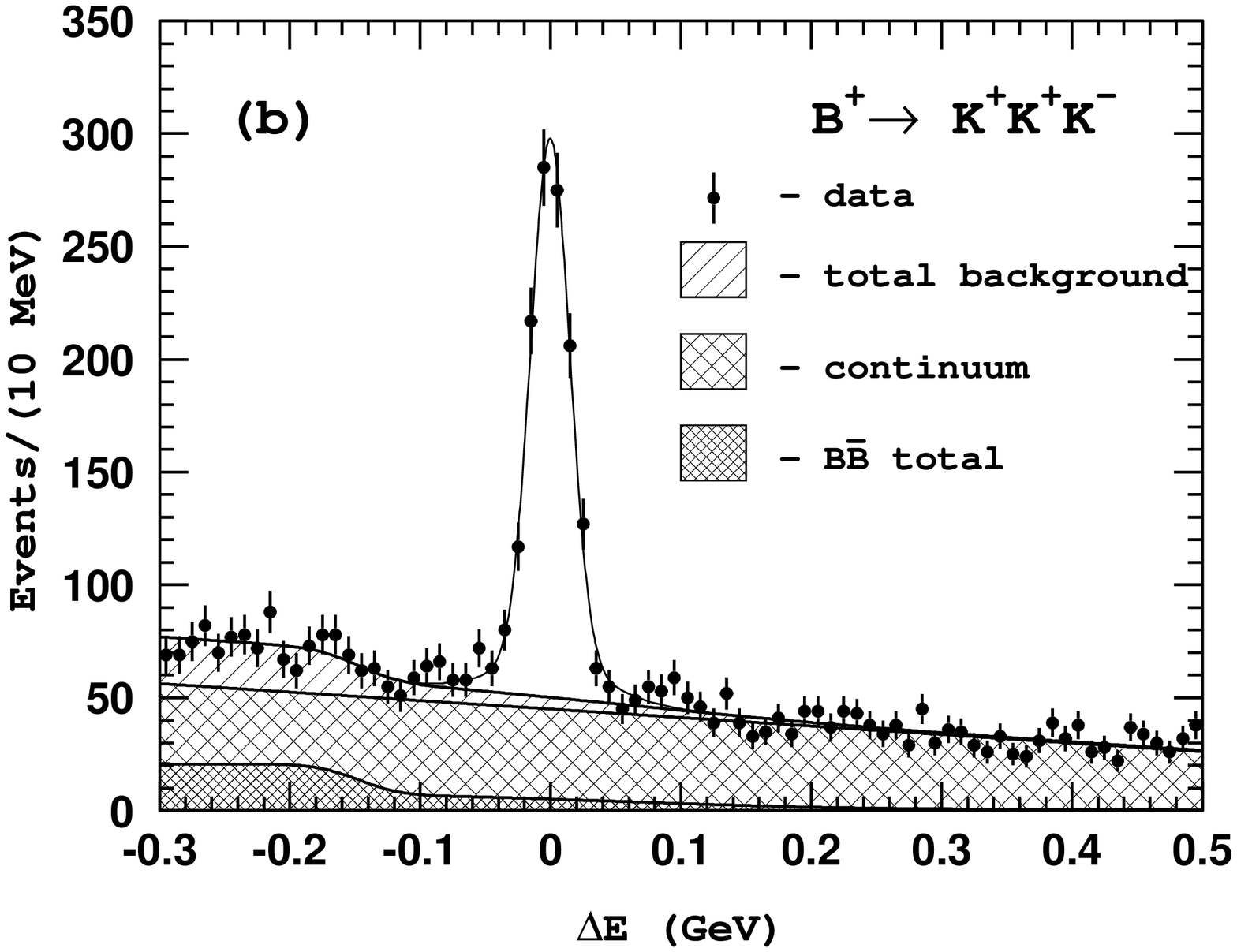}
 \caption{$\de$ distributions for (a) $\bckpp$ and
          (b) $\bckkk$ candidate events.
          Points with error bars are data; the curve is the fit 
          result; the hatched area is the background.}
  \label{fig:khh-DE}
\end{figure}

The $\de$ distributions for $\bckpp$ and $\bckkk$ candidates that pass all
the selection requirements and with $|\mb-M_B|<9$~MeV/$c^2$ are shown
in Fig.~\ref{fig:khh-DE}, where clear peaks in
the signal regions are observed. In the fit to the $\de$ distribution for the
$\kpp$ final state, we fix the shape and normalization of the charmless
$\bbbar$ background components from the measured branching fractions~\cite{PDG}
and known number of produced $\bbbar$ events. For the $\bbbar$ generic
component, we fix only the shape and let the normalization float. The slope
and normalization of the $\qqbar$ background component are free parameters.
Results of the fit are shown in Fig.~\ref{fig:khh-DE}, where different
components of the background are shown separately for comparison. There is a
large increase in the level of the $\bbbar$ related background in the
$\de<-0.15$~GeV region for the $\kpp$ final state. This is mainly due to the
decay $B\to D\pi$, $D\to K\pi\pi$. This decay mode produces the same final
state as the studied process plus one extra pion that is not included in the
energy difference calculation. The decay $B\to D\pi$, $D\to K\mu\nu_\mu$
also contributes due to muon-pion misidentification. The shape of this
background is described well by the MC simulation. In the fit to the $\de$
distribution for the $\kkk$ final state, we fix not only the shape but also
the normalization of the $\bbbar$ background. This is done because
the $\bbbar$ background in this final state is found to be much smaller than
the dominant $\qqbar$ background, thus the relative fraction of these two
contributions is poorly determined from the fit.
The signal yields obtained from fits are given in
Table~\ref{tab:defitall}.


\begin{table}[b]
  \caption{Results of fits to the $\de$ distributions with a double
           Gaussian for the signal (see Section~III).}
  \medskip
  \label{tab:defitall}
\centering
  \begin{tabular}{lccccc} \hline \hline
\hspace*{0mm}  Final state    \hspace*{1mm} & 
\hspace*{2mm}  $\sigma_1$     \hspace*{2mm} &
\hspace*{2mm}  $\sigma_2$     \hspace*{2mm} &
\hspace*{2mm}  Fraction of the     \hspace*{2mm} &
\hspace*{2mm}  Signal Yield   \hspace*{1mm}  \\
     &  MeV & MeV & main Gaussian & (events) \\
\hline
$~\kpp$ & $17.5\pm0.9$ & $ 35.0$ (fixed) & $0.84$ (fixed) & $1533\pm69$ \\
$~\kkk$ & $14.0\pm1.0$ & $ 40.0$ (fixed) & $0.85$ (fixed) & $1089\pm41$ \\
\hline \hline
  \end{tabular}
\end{table}

\begin{figure}[t]
 \includegraphics[width=0.48\textwidth]{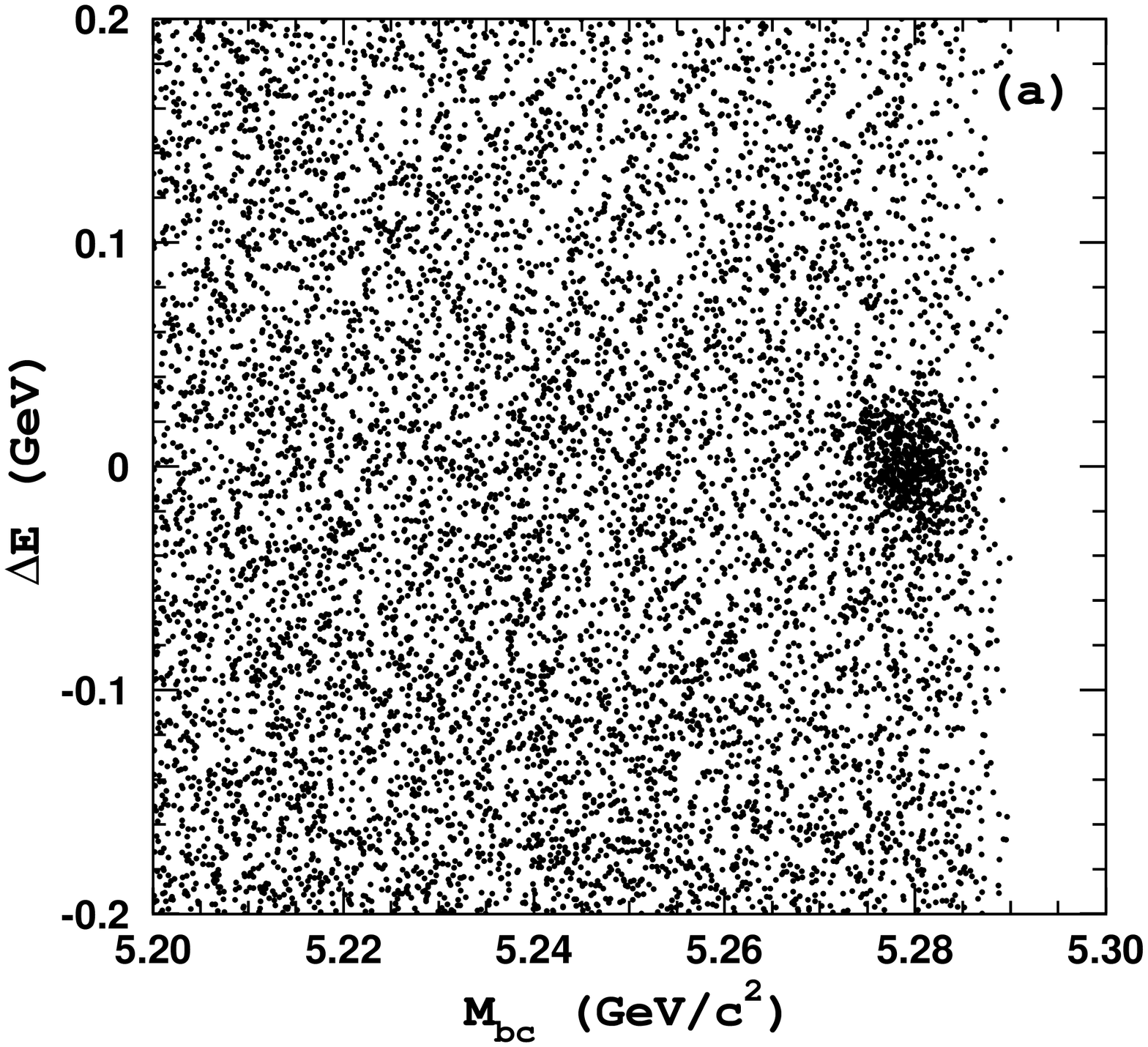} \hfill
 \includegraphics[width=0.48\textwidth]{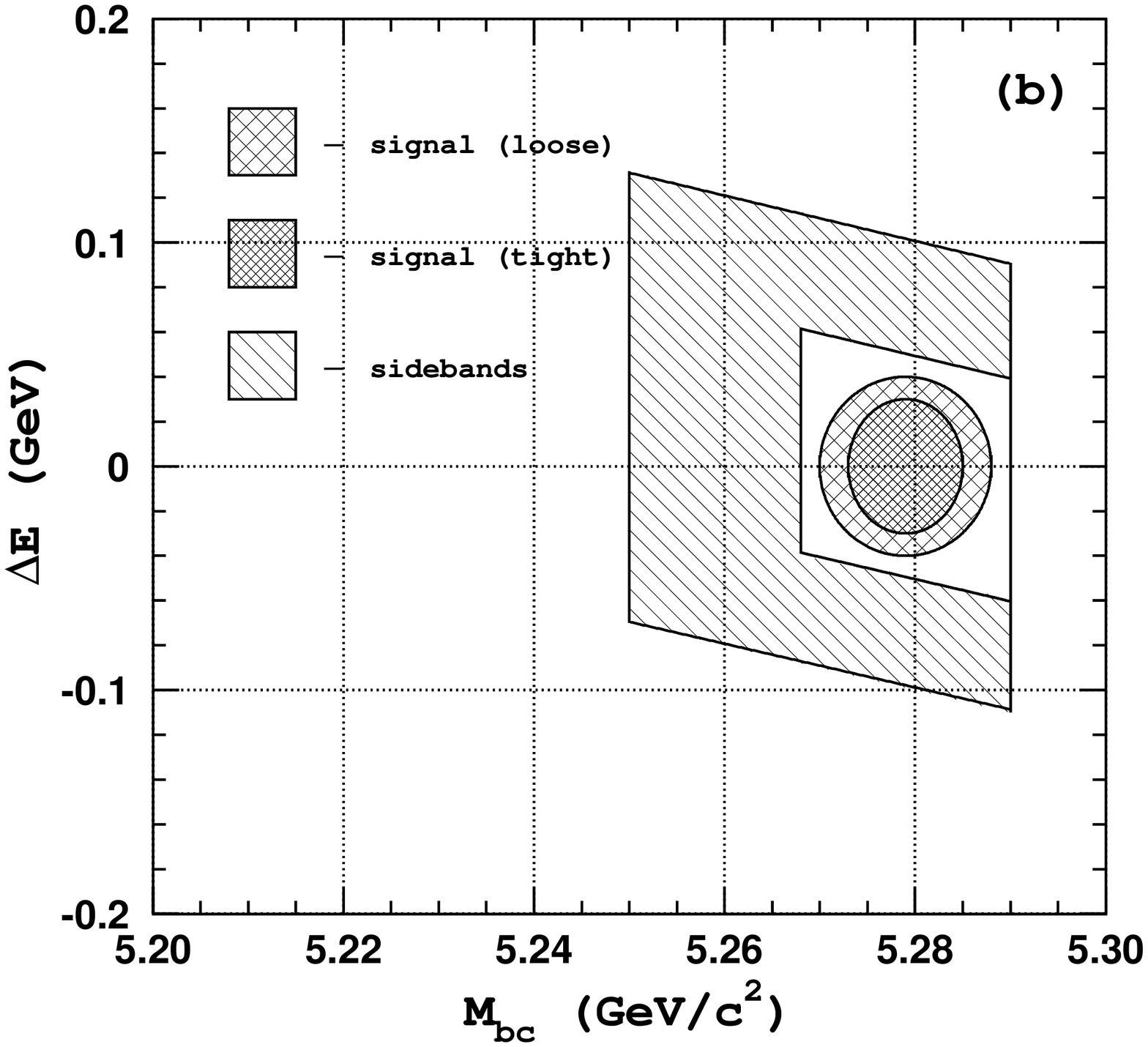}
 \caption{(a) Distribution of $\de$ versus $\mb$ for the $\bckkk$
          candidates in data. \mbox{(b) Definitions} of the signal and
          sideband regions in the $\mb-\de$ plane.}
 \label{fig:dE-Mbc}
\end{figure}

To examine possible quasi-two-body intermediate states in the observed $\bckpp$
and $\bckkk$ signals, we analyze two-particle invariant mass spectra. To
determine the distribution of the background we use events in the $\mb$
and $\de$ sidebands. The definition of the signal and sideband regions is
illustrated in Fig.~\ref{fig:dE-Mbc}. Defined in this way, the $\mb-\de$
sidebands are equivalent to the following sidebands in terms of the
three-particle invariant mass $M(Khh)$ and three-particle momentum $P(Khh)$
in the c.m.\ frame:
\[ 0.05 {\rm ~GeV}/c^2 < |M(Khh)-M_B| < 0.10 {\rm ~GeV}/c^2;
~~~P(Khh) < 0.48 {\rm ~GeV}/c \nonumber \]
and
\[|M(Khh)-M_B| < 0.10 {\rm ~GeV}/c^2;
~~~0.48 {\rm ~GeV}/c < P(Khh) < 0.65 {\rm ~GeV}/c. \nonumber\]
  The signal region is defined as an ellipse around the $\mb$ and $\de$ mean
values:
\[
\frac{(\mb-M_B)^2}{(n\sigma_{\mb})^2} + \frac{\de^2}{(n\sigma_{\de})^2} < 1,
\]
where $\sigma_{\mb}=3$~\mass ~and $\sigma_{\de}$ is equivalent to $\sigma_1$
in Table~\ref{tab:defitall}. We define two  signal regions: with loose
$(n = 3)$ and tight $(n = 2)$ requirements. Tight requirements reduce (compared
to the loose requirements) the background fraction in the data sample by about
65\% while retaining about 85\% of the signal. The efficiency of the loose
(tight) requirements that define the signal region is 0.923 (0.767) for the
$\kpp$ final state and 0.948 (0.804) for the $\kkk$ final state. The
total number of events in the signal region is 2584 (1809) for the $\kpp$ and
1400 (1078) for the $\kkk$ final state. To determine the relative fraction of
signal and background events in these samples, we use the results of the fits
to the $\de$ distributions (see Table~\ref{tab:defitall}). We find $1533\pm69$
signal $\bckpp$ events and $1089\pm41$ signal $\bckkk$ events. The relative
fraction of signal events in the signal region with loose (tight) requirements
is then determined to be $0.548\pm0.025$ ($0.650\pm0.032$) for the $\kpp$ and
$0.738\pm0.028$ ($0.828\pm0.033$) for the $\kkk$ final state. All final results
are obtained from fits to events in the signal region with loose $\de$ and
$\mb$ requirements. The subsample with tight requirements is used for a
cross-check only.


\begin{figure}[t]
  \centering
  \includegraphics[width=0.32\textwidth]{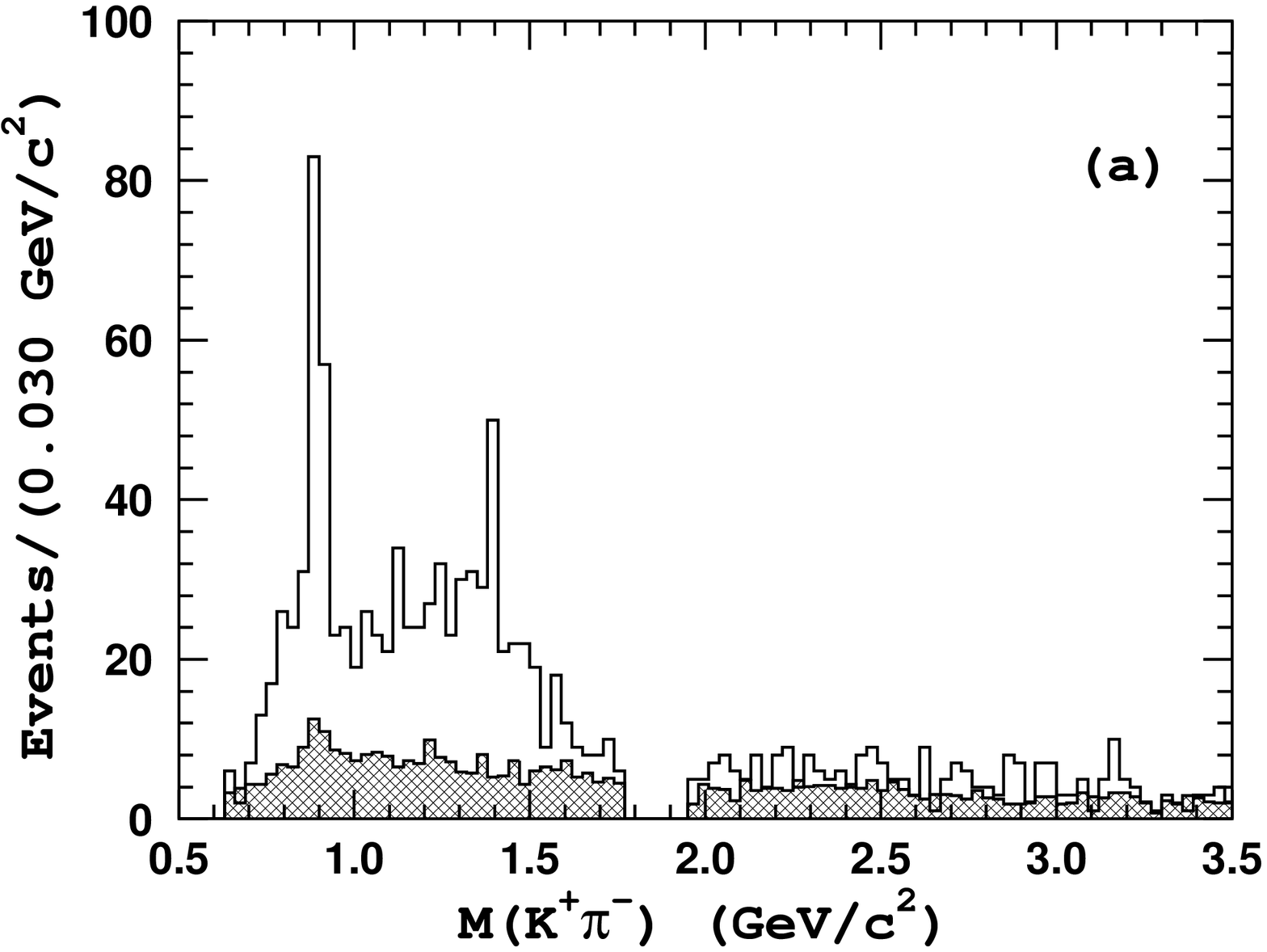} \hfill
  \includegraphics[width=0.32\textwidth]{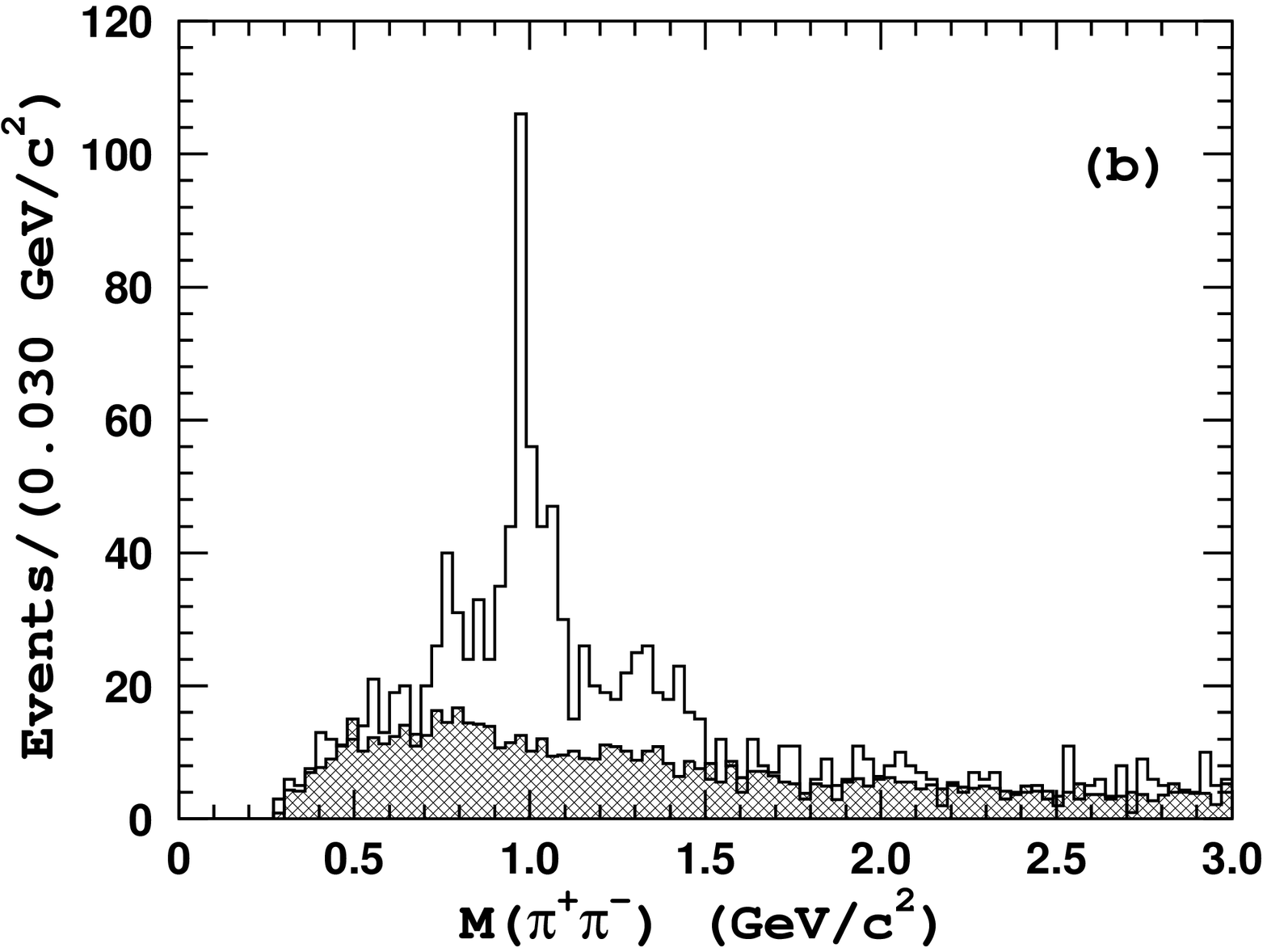} \hfill
  \includegraphics[width=0.32\textwidth]{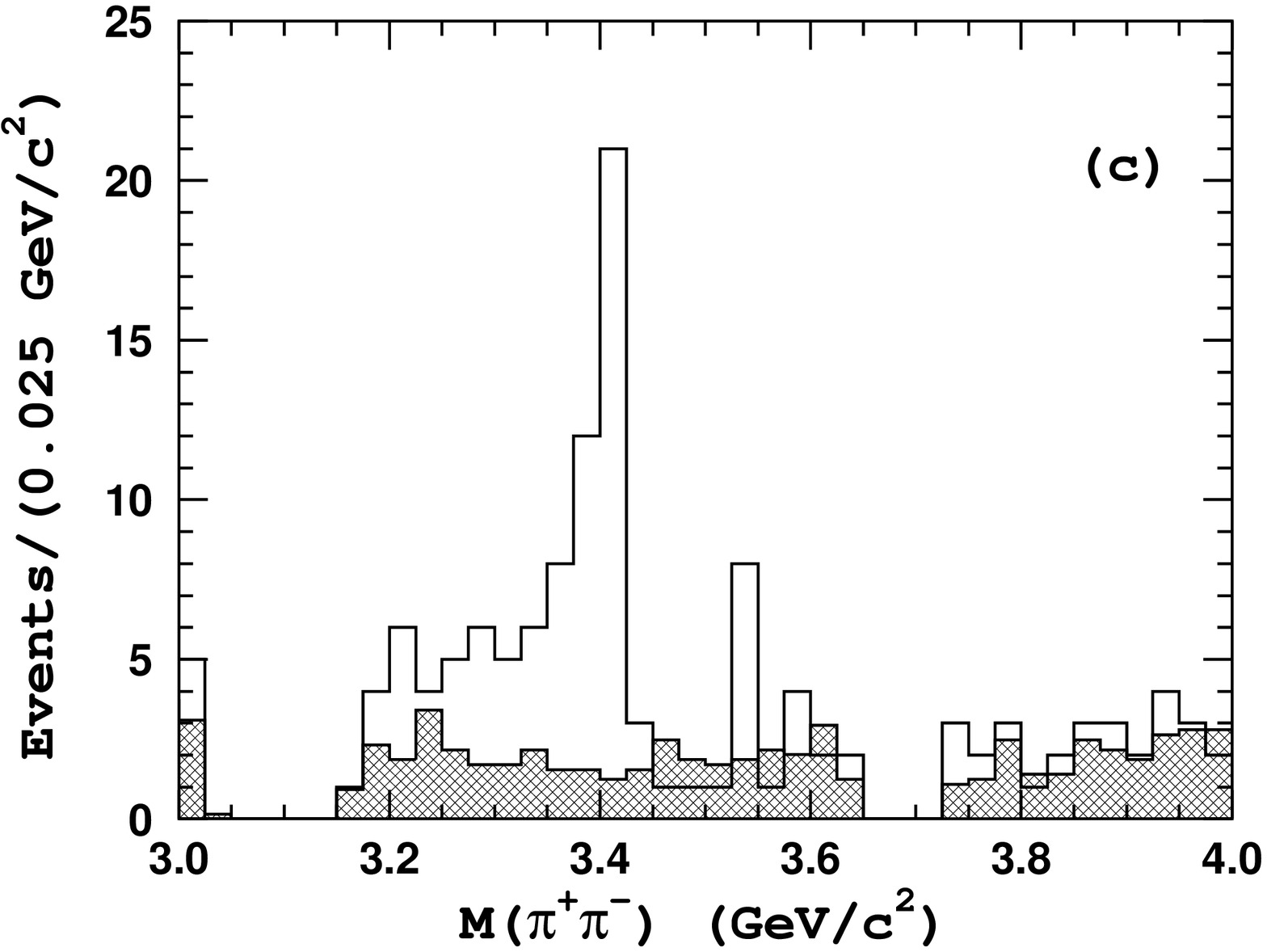}
  \caption{Two-particle invariant mass spectra for $\bckpp$ 
           candidates in the $B$ signal region (open histograms) and for
           background events in the $\de-\mb$ sidebands (hatched histograms). 
           (a) $M(K^+\pi^-)$ spectrum with $M(\pi^+\pi^-)>1.5$~\Mass;
           (b) $M(\pi^+\pi^-)$ with $M(K^+\pi^-)>2.0$~\Mass~ and
           (c) $M(\pi^+\pi^-)$ in the $\chic$ mass region with 
               $M(K^+\pi^-)>2.0$~\Mass.}
  \label{fig:kpp_hh}
  \centering
  \includegraphics[width=0.32\textwidth]{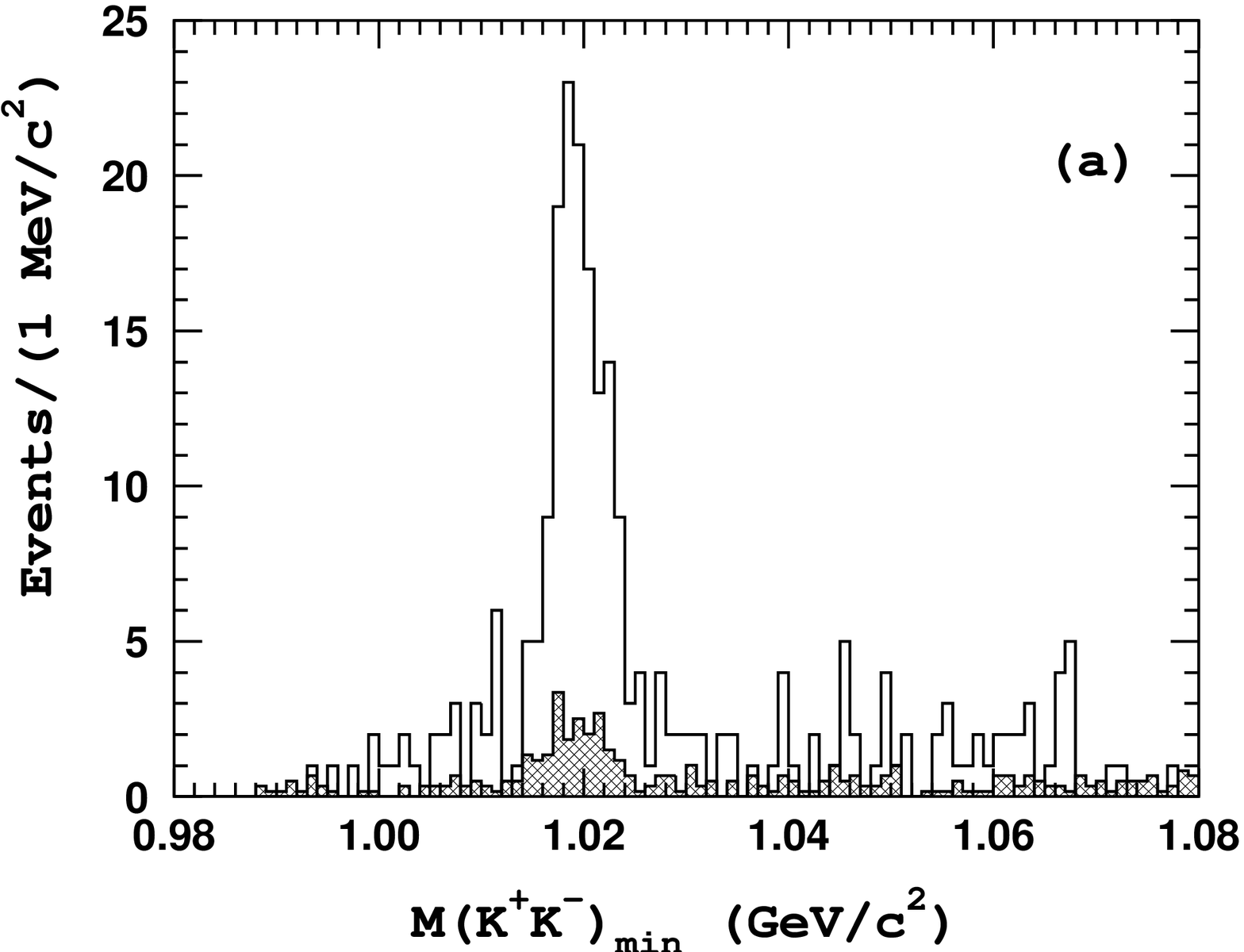} \hfill
  \includegraphics[width=0.32\textwidth]{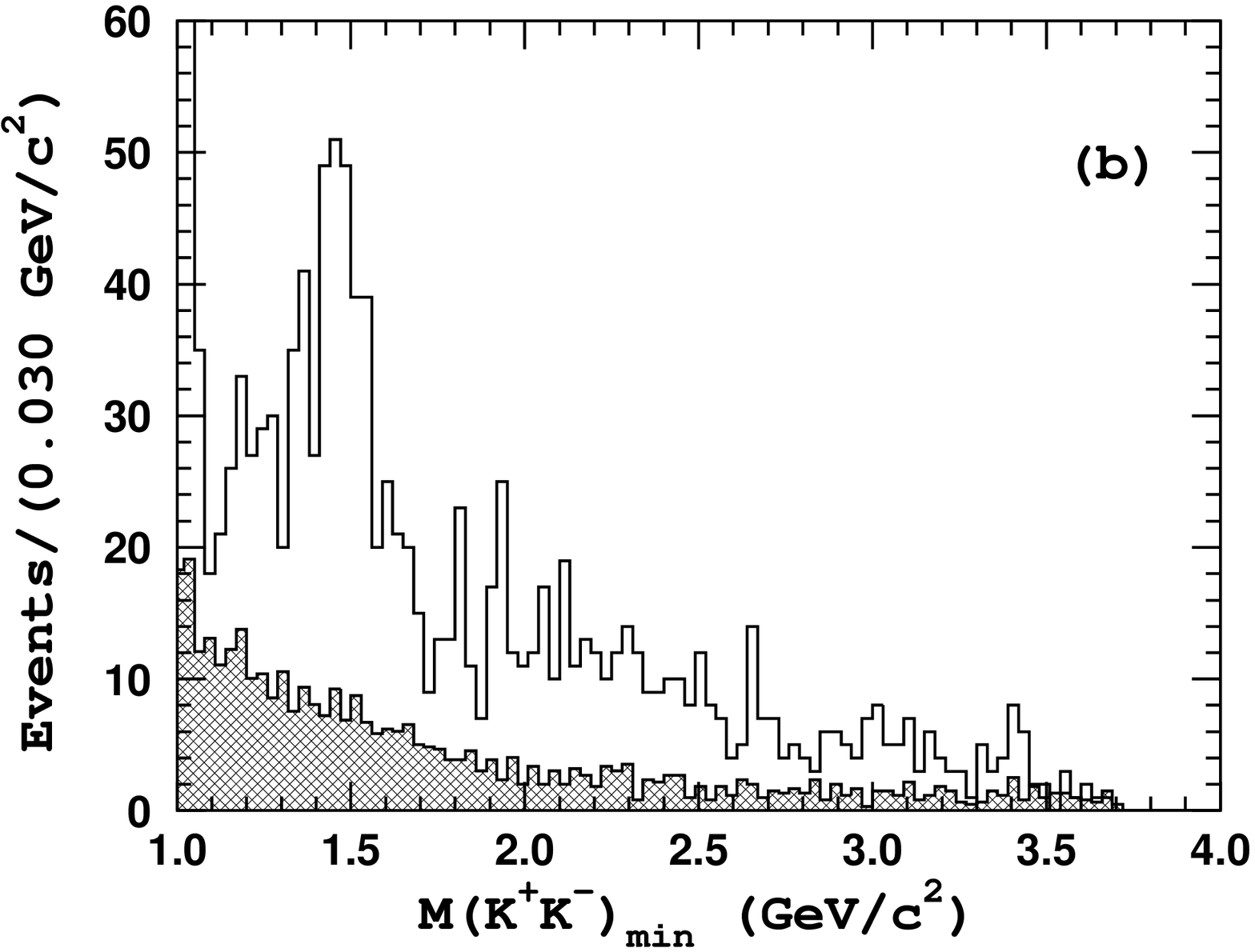} \hfill
  \includegraphics[width=0.32\textwidth]{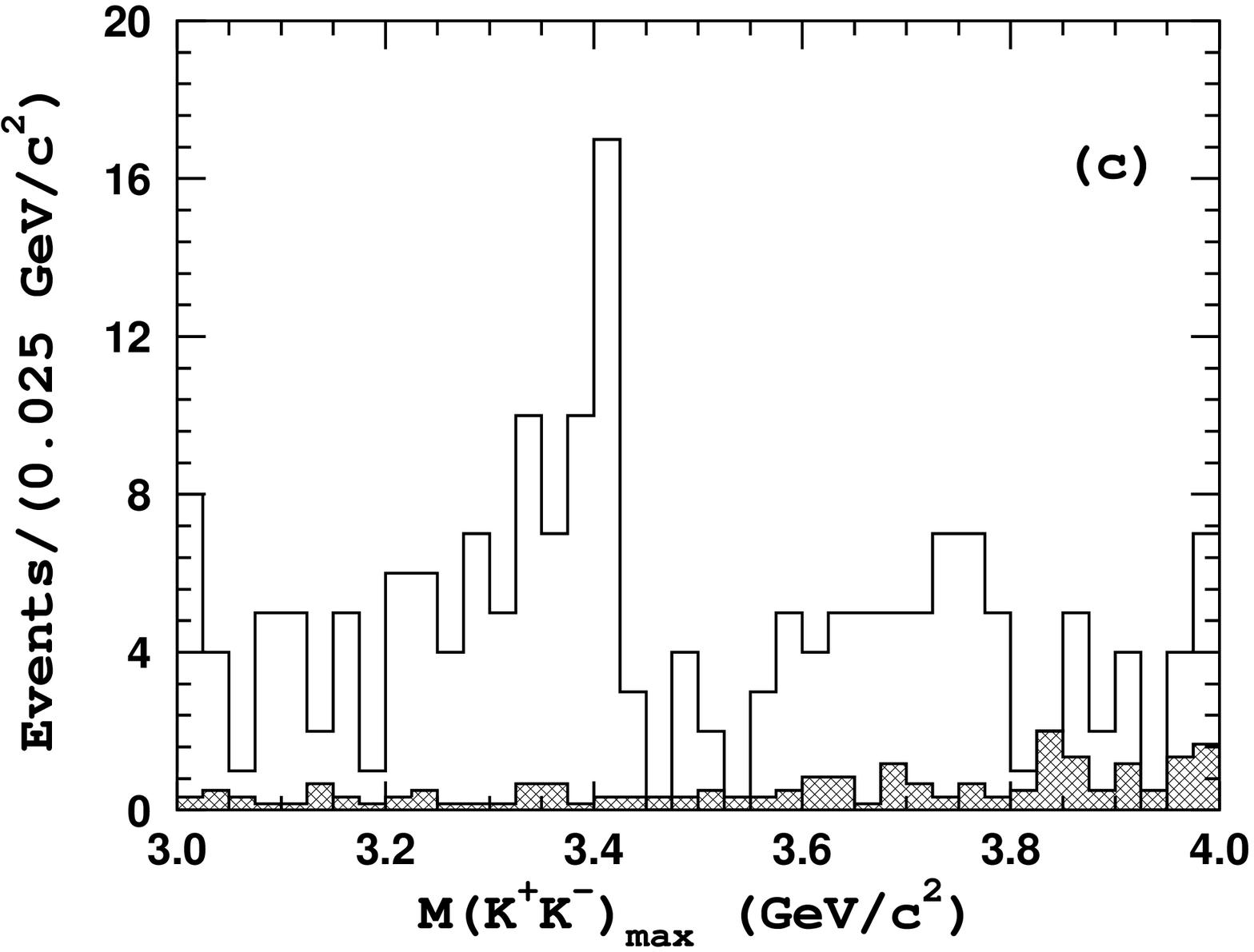}
  \caption{Two-particle invariant mass spectra for  $\bckkk$ 
           candidates in the $B$ signal region (open histograms) and for
           background events in the $\de-\mb$ sidebands (hatched histograms). 
           (a) $\mkkmin$ invariant mass spectrum near the $\phi(1020)$ 
               mass region;
           (b) $\mkkmin$ spectrum in the full range;
           (c) $\mkkmax$ in the $\chic$ mass region with 
               2.0~\Mass~$<\mkkmin<$~3.4~\Mass.}
  \label{fig:kkk_hh}
\end{figure}

The $\kcpi$ and $\pipi$ invariant mass spectra for $\bckpp$ candidate events
in the $\mb-\de$ loose signal region are shown as open histograms in
Fig.~\ref{fig:kpp_hh}. The hatched histograms show the corresponding spectra
for background events in the $\mb-\de$ sidebands, normalized to the estimated
number of background events. To suppress the feed-across between the $\pipi$
and $\kcpi$ resonance states, we require the $\kcpi$ ($\pipi$) invariant mass
to be larger than 2.0 (1.5)~\Mass ~when making the $\pipi$ ($\kcpi$)
projection. The $\kcpi$ invariant mass spectrum is characterized by a narrow
peak around 0.9~GeV/$c^2$, which is identified as the $K^*(892)^0$, and a
broad enhancement around 1.4~\Mass. Possible candidates for this
enhancement are the scalar $K^*_0(1430)$ and tensor $K^*_2(1430)$ resonances.
In the $\pipi$ invariant mass spectrum two distinct structures in the low
mass region are observed. One is slightly below 1.0~\Mass ~and is consistent
with the $f_0(980)$ and the other is between 1.0~\Mass ~and 1.5~\Mass. We
cannot identify unambiguously the resonant state that is responsible for such
a structure. Possible candidates for a resonant state in this mass region
might be $f_0(1370)$, $f_2(1270)$ and perhaps $\rho(1450)$~\cite{PDG}. In what
follows, we refer to this structure as $f_X(1300)$. There is also an indication
for the $\rho(770)^0$. Finally, there is a clear signal for the decay
$B^+\to\chic K^+$, $\chic\to\pipi$. Figure~\ref{fig:kpp_hh}(c) shows the
$\pipi$ invariant mass distributions in the $\chic$ mass region.


The $\kpkm$ invariant mass spectra for $\bckkk$ candidate events in the
$\mb-\de$ signal region with loose requirements are shown as open histograms
in Fig.~\ref{fig:kkk_hh}. Since there are two same-charge kaons in the
\mbox{$B^+\to K^+K^+K^-$} decay, we distinguish the $K^+K^-$ combinations
with smaller, $\mkkmin$, and larger, $\mkkmax$, invariant masses. The
$\mkkmin$ spectrum, shown in Fig.~\ref{fig:kkk_hh}(a), is characterized by a
narrow peak at 1.02 GeV/$c^2$ corresponding to the $\phi(1020)$ meson and a
broad structure around 1.5~GeV/$c^2$, shown in Fig.~\ref{fig:kkk_hh}(b).
Possible candidates for a resonant state in this mass region are the
$f_0(1370)$, $f_0(1500)$ or $f_2'(1525)$ \cite{PDG}. In what follows, we
refer to this structure as $f_X(1500)$. Figure~\ref{fig:kkk_hh}(c) shows
the $\mkkmax$ invariant mass distribution in the $\chic$ mass region. A
clear enhancement is observed at 3.4~\Mass, where the $\chic$ is expected.
Some enhancement of signal events over the expected background level is also
observed in the full mass range shown in Fig.~\ref{fig:kkk_hh}(c). As the
$\chic$ meson has a significant natural width (about 15~\mass)~\cite{PDG}, the
amplitude that is responsible for the $B^+\to\chic K^+$ decay may interfere
with a charmless amplitude. As a result of the interference between these
two amplitudes, the lineshape of the $\chic$ resonance can be distorted.
In our previous analysis of $B$ meson decays to three-body charmless hadronic
final states~\cite{garmash,garmash2}, we imposed a requirement on the
invariant mass of the $\pipi$ and $\kpkm$ combination to veto the
$B^+\to\chic K^+$ signal. In this analysis we do not apply such a requirement.

From these qualitative considerations it is apparent that an amplitude analysis
is required for a more complete understanding of the individual quasi-two-body
channels that contribute to the observed three-body $\bckpp$ and $\bckkk$
signals.


\section{Amplitude Analysis}
\label{sec:aa}

In the preceding Section we found that a significant fraction of the signals
observed in $\bckpp$ and $\bckkk$ decays can be assigned to quasi-two-body
intermediate states. These resonances will cause a non-uniform  distribution
of events in phase space that can be analyzed using the technique pioneered
by Dalitz~\cite{dalitz}. Multiple resonances that occur nearby in phase space
will interfere and provide an opportunity to measure both the amplitudes and
relative phases of the intermediate states. This in turn allows us to deduce
their relative branching fractions. Amplitude analyses of various three-body
$D$ meson decays have been successfully performed by a number of
groups~\cite{d_dp_ana}. From their results we can learn that this kind of
analysis requires, in general, high statistics (of the order of a few thousand
signal events, at least). In contrast to the analysis of $D$ meson three-body
decays, where the level of the combinatorial background is usually quite small,
the signal and background levels in charmless three-body decays of $B$ mesons
are comparable. This complicates the analysis, requiring careful study of the
distribution of background events over the phase space. Finally, independent
of the statistics, the choice of the model (that is the set of quasi-two-body
intermediate states) to fit the data is often not unique. This unavoidably
introduces some model dependence into the determination of quasi-two-body
branching fractions. This is especially true for three-body charmless decays of
$B$ mesons where experimental statistics is quite limited while the available
phase space is large.


\subsection{Formalism}

  Since we are studying the decay of a spin-zero particle to three  
spin-zero daughters $B\to h_1h_2h_3$, only two degrees of freedom are 
required to completely describe the kinematics. There are three invariant 
masses that can be formed by considering all possible pairs of final state 
particles: $\sfs\equiv M^2(h_1h_2)$, $\sft\equiv M^2(h_1h_3)$ and 
$\sst\equiv M^2(h_2h_3)$. Only two of them are independent, however, since
energy and momentum conservation results in the additional constraint
\begin{equation}
M^2 + m^2_1 + m^2_2 + m^2_3 = \sfs + \sft + \sst,
\end{equation}
where $M^2$ is the mass of the initial particle, and $m_i$ are masses of the
daughter particles. In what follows we use $\sft$ and $\sst$ as the two
independent variables.

The density of signal events on the Dalitz plot is described by the matrix 
element $\cal M$ as
\begin{equation} 
d\Gamma = \frac{|{\cal M}|^2}{256\pi^3M^3}d\sft d\sst,
\end{equation}
which in turn depends on the decay dynamics.


\begin{table}[t]
\caption{Blatt-Weisskopf penetration form factors. $p_r$ is the momentum of
         either daughter in the meson rest frame. $p_s$ is the momentum of
         either daughter in the candidate rest frame (same as $p_r$ except the
         parent mass used is the two-track invariant mass of the candidate
         rather than the mass of the meson). $R$ is the meson radial 
         parameter.}
  \medskip
\label{tab:bwff}
\begin{tabular}{lc}  \hline \hline
Spin $J$ \hspace*{50mm} & Form Factor $ F^{(J)}_R $ \\ \hline
~~~0 & 1 \\
~~~1 & $\frac{\sqrt{1 + R^2 p^2_r}}{\sqrt{1+R^2 p^2_s}} $
\vspace*{1mm} \\
~~~2 & $\frac{\sqrt{9 + 3 R^2 p^2_r + R^4 p^4_r}}
             {\sqrt{9 + 3 R^2 p^2_s + R^4 p^4_s}} $ \\ 
\hline \hline
\end{tabular}
\end{table}

The amplitude for $B$ decay to a three-body final state via an intermediate
resonance state $R$ is given by
\begin{equation}
{\cal A}_J = F_B F^{(J)}_R BW_J T_J,
\label{eq:a_general}
\end{equation}
where $F_B$ and $F^{(J)}_R$ are form factors which, in general, are unknown
functions. For $F^{(J)}_R$ we use the Blatt-Weisskopf penetration
factors~\cite{blatt} given in Table~\ref{tab:bwff}. These factors depend on
a single parameter, $R$, which is the ``radius'' of the meson. For all
intermediate resonances we set this parameter to $R=1.5$~GeV$^{-1}$. Form
factors are normalized to unity at the nominal meson mass.
$F_B$ is parametrized in a single pole approximation~\cite{ffactor}
\begin{equation}
 F_B = \frac{1}{1-\frac{s}{M_{\rm pole}^2}}
\label{eq:FB}
\end{equation}
where we use the mass of the $B^*$ meson~\cite{PDG} as a pole mass
$M_{\rm pole}$.

The Breit-Wigner function $BW_J$ in Eq.~(\ref{eq:a_general}) is given by the
expression
\begin{equation}
BW_J(s) = \frac{1}{M^2_R - s - iM_R\Gamma^{(J)}_R(s)},
\label{eq:bw-func}
\end{equation}
where $M_R$ is the nominal mass of the resonance, and  $\Gamma^{(J)}_R(s)$
is the ``mass dependent width''. In the general case, $\Gamma^{(J)}_R(s)$ is
expressed as~\cite{pilkuhn}:
\begin{equation}
\Gamma^{(J)}_R(s) = \Gamma_R \left(\frac{p_s}{p_r}\right)^{2J+1}
              \left( \frac{M_R}{s^{1/2}} \right) F_R^2,
\label{eq:gs1}
\end{equation}
where $p_r$ is the momentum of either daughter in the resonance rest frame,
calculated with the resonance mass equal to the nominal $M_R$ value, $p_s$ is
the momentum of either daughter in the resonance rest frame when the resonance
mass is equal to $s^{1/2}$, $J$ is the spin of the resonance, and $\Gamma_R$
is the width of the resonance.

The function $T_J$ in Eq.~(\ref{eq:a_general}) describes the angular
correlations between the $B$ decay products. We distinguish the following
three cases:

1) {\it \underline{Scalar-Pseudoscalar ($J=0$) decay}}

If $R$ is a scalar state, the decay amplitude Eq.~(\ref{eq:a_general}) takes
the simplest form with $T_0\equiv 1$. We treat the scalar $f_0(980)$ as a
special case, for which we try two parametrizations for the $s$-dependent
width $\Gamma_{f_0}(s)$: by Eq.~(\ref{eq:gs1}), and following the
parametrization by Flatt\'e~\cite{Flatte}
\begin{equation}
\Gamma_{f_0}(s) =  \Gamma_\pi(s)+\Gamma_K(s),
\end{equation}
where
\begin{equation}
\Gamma_\pi(s) = g_\pi\sqrt{s/4 - m^2_\pi},~~
\Gamma_K(s)   = \frac{g_K}{2}\left(\sqrt{s/4 - m^2_{K^+}} +
                                \sqrt{s/4 - m^2_{K^0}}\right),
\label{eq:a_f0}
\end{equation}
and $g_\pi$ and $g_K$ are coupling constants for $f_0(980)\to\pi\pi$ and
$f_0(980)\to KK$, respectively.

2) {\it \underline{Vector-Pseudoscalar ($J=1$) decay}}

In the case of a pseudoscalar-vector decay of the $B$ meson, the
Lorentz-invariant expression for $T_1$ is given by
\begin{equation}
T_1(h_1h_2h_3|R_{23}) = \sfs-\sft+\frac{(M^2-m^2_1)(m^2_3-m^2_2)}{\sst},
\label{eq:a_vector}
\end{equation}
where $R_{23}$ is an intermediate resonance state decaying to $h_2h_3$ final
state.

3) {\it \underline{Tensor-Pseudoscalar ($J=2$) decay}}

For a pseudoscalar-tensor decay, $T_2$ takes the form
$$
T_2(h_1h_2h_3|R_{23}) = \left(\sft-\sfs+\frac{(M_B^2-m_1^2)(m_2^2-m_3^2)} 
                                       {\sst}\right)^2
$$
\begin{equation}
-\frac{1}{3} \left(\sst - 2 M_B^2 - 2 m_1^2 + \frac{( M_B^2 - m_1^2)^2}
 {\sst}\right) \left(\sst - 2 m_2^2 - 2 m_3^2 + 
 \frac{( m^2_2 - m^2_3)^2}{\sst}\right).
\label{eq:a_tenzor}
\end{equation}
We do not consider resonant states of higher spin in our analysis.

There is also the possibility of a so-called ``non-resonant'' amplitude.
In the Dalitz analysis of $D$ meson decays to three-body final states the
non-resonant amplitude is often parametrized as a complex constant.
In the case of $B$ meson decays, where the available phase space is much
larger, it is rather unlikely that the non-resonant amplitude will have a
constant value over the entire phase space; some form factors should be
introduced. Unfortunately, at the moment there is no theoretical consensus
on the properties of non-resonant $B$ meson decays. In our analysis we use
an empirical parametrization that in the case of the $\kpp$ final state is
\begin{equation}
  \Am_{\rm nr}(\kpp) = a^{\rm nr}_1e^{-\alpha \sft}e^{i\delta^{\rm nr}_1} 
                     + a^{\rm nr}_2e^{-\alpha \sst}e^{i\delta^{\rm nr}_2},
  \label{eq:kpp-non-res}
\end{equation}
where $\sft\equiv M^2(\kcpi)$, $\sst\equiv M^2(\pipi)$, and $a_{1,2}^{\rm nr}$,
and $\delta_{1,2}^{\rm nr}$ and $\alpha$ are fit parameters. In a certain
limit this parametrization is equivalent to a constant. Several alternative
parametrizations (mentioned below) are also considered to estimate the model
dependence.

An important feature that should be taken into account in the construction of
the matrix element for the decay $\bckkk$ is the presence of the two identical
kaons in the final state. This is achieved by symmetrizing the matrix element
with respect to the interchange of the two kaons of the same charge, that is
$\sft\leftrightarrow\sst$. Due to symmetrization the non-resonant amplitude for
the $\kkk$ final state becomes
\begin{equation}
{\cal A}_{\rm nr}(\kkk) =
    a^{\rm nr}(e^{-\alpha \sft} + e^{-\alpha{\sst}})e^{i\delta^{\rm nr}},
  \label{eq:kkk-non-res}
\end{equation}
where $\sft\equiv M^2(K^+_1K^-)$, $\sst\equiv M^2(K^+_2K^-)$.

Given the amplitude for each decay type, the overall matrix elements can
be written as a coherent sum
\begin{equation}
{\cal M} = \sum_j a_je^{i\delta_j}{\cal A}^j + {\cal A}_{\rm nr},
\label{eq:m_tot}
\end{equation}
where the index $j$ denotes the quasi-two-body intermediate state, $a_j$ and
$\delta_j$ are the amplitude and relative phase of the $j$-th component.
Since we are sensitive only to the relative phases and amplitudes, we are free
to fix one phase and one amplitude in Eq.~(\ref{eq:m_tot}). The fraction $f_l$
of the total three-body signal attributed to a particular quasi-two-body
intermediate state can be calculated as
\begin{equation}
 f_l = \frac{\int |a_l{\cal A}^l|^2\, d\sft d\sst}
            {\int |{\cal M}|^2\, d\sft d\sst}.
\label{eq:fraction}
\end{equation}
The sum of the fit fractions for all components is not necessarily unity
because of interference.

  The amplitude analysis of $B$ meson three-body decays reported here is
performed by means of an unbinned maximum likelihood fit which minimizes the
function
\begin{equation}
{\cal{F}} = -2\sum_{\rm events}\ln P(\sft,\sst;\xi),
\end{equation}
where the function $P(\sft,\sst;\xi)$ describes the density of experimental
events over the Dalitz plot; $\xi$ is a vector of parameters.

An important question that arises in an unbinned analysis is the estimation
of the goodness-of-fit. As
the unbinned maximum likelihood fitting method does not provide a direct way
to estimate the quality of the fit, we need a measure to assess how well any
given fit represents the data. To do so the following procedure is applied.
We first subdivide the entire Dalitz plot into  1~\Masssq$\times$1~\Masssq~
bins. If the number of events in the bin is smaller than $N_{\rm min}=16$ it
is combined with the adjacent bins until the number of events exceeds
$N_{\rm min}$. After completing this procedure, the entire Dalitz plot is
divided into a set of bins of varying size, and a $\chi^2$ variable for the
multinomial distribution can be calculated as
\begin{equation}
   \chi^2 = -2\sum^{N_{\rm bins}}_{i=1}n_i\ln\left(\frac{p_i}{n_i}\right),
\end{equation}
where $n_i$ is the number of events observed in the $i$-th bin, and $p_i$ is
the number of predicted events from the fit. For a large number of events
this formulation becomes equivalent to the usual one.
Since we are minimizing the unbinned likelihood function, our ``$\chi^2$''
variable does not asymptotically follow a $\chi^2$ distribution but it is
bounded by a $\chi^2$ variable with ($N_{\rm bins}-1$) degrees of freedom
and a $\chi^2$ variable with ($N_{\rm bins}-k-1$) degrees of
freedom~\cite{kendal}, where $k$ is the number of fit parameters. Because
it is bounded by two $\chi^2$ variables, it should be a useful statistic
for comparing the relative goodness of fits for different models.


\subsection{Efficiency, Detector Resolution and Background}

  Several effects should be taken into account when fitting the experimental
data. The reconstruction efficiency can vary significantly over the Dalitz
plot area and distort the initial distribution of signal events. In addition,
there is also some fraction of background that fakes the signal. As is evident
in Fig.~\ref{fig:khh-DE}, the background level in the signal region
is comparable to that of the signal. Thus, understanding the distribution of
background events over the Dalitz plot is important for an amplitude analysis.
Finally, the detector resolution produces some smearing of the Dalitz plot
boundaries so that the phase space for the reconstructed $B$ candidates
exceeds the kinematically allowed area. To correct for this effect, three-body
combinations are kinematically fit to the nominal $B$ mass. As the intermediate
resonances in general have large widths, we neglect the effect of detector
resolution on the resonance shapes in most cases. In the case of narrow
resonant states (for example, the $\phi$ meson in the $\kkk$ final state or
the $\chi_{c0}$ in both $\kkk$ and $\kpp$ final states), we take the detector
resolution into account by convolving the signal probability density function
with a two-dimensional Gaussian resolution function. The widths of the
two-dimensional resolution function depend on the position in the Dalitz plot
and are determined from the MC simulation.

To account for the background events and non-uniform reconstruction efficiency
the event density function, $P(\sft,\sst;\xi)$ can be written as
\begin{equation}
P(\sft,\sst;\xi)=
\frac{N_s\varepsilon(\sft,\sst)S(\sft,\sst,\xi)+
n_b b(\sft,\sst)}{N_s \int \varepsilon(\sft,\sst)
S(\sft,\sst,\xi)d\sft d\sst+n_b},
\label{eq:density_1}
\end{equation}
where $N_s$ is the {\it initial} number of signal events distributed over the
the Dalitz plot according to the signal density function $S(\sft,\sst,\xi)$ ,
$\varepsilon(\sft,\sst)$ is the reconstruction efficiency as a function
of the position on the Dalitz plot, $n_b$ is the  expected number
of the {\it observed} background events distributed with the density 
$b(\sft,\sst)$, and a vector of parameters $\xi$ (masses, widths and
relative amplitudes and phases) is to be determined from the minimization.
Equation (\ref{eq:density_1}) can be written in terms of the expected number of
the observed signal events 
$n_s = N_s\varepsilon_s 
     = N_s\int\varepsilon(\sft,\sst)S(\sft,\sst,\xi)d\sft d\sst$
and the background density function 
$B(\sft,\sst) = \varepsilon_bb(\sft,\sst)/\varepsilon(\sft,\sst)$:
\begin{equation}
P(\sft,\sst,\xi)=
\varepsilon(\sft,\sst)
\frac{n_sS(\sft,\sst,\xi)/\varepsilon_s+
      n_bB(\sft,\sst)/\varepsilon_b}{n_s+n_b},
\label{eq:density_2}
\end{equation}
where the overall efficiencies $\varepsilon_s$ and $\varepsilon_b$ are 
determined from the MC simulation:
\begin{equation}
\varepsilon_s = 
\int \varepsilon(\sft,\sst)S(\sft,\sst,\xi)d\sft d\sst =
\frac{\Delta}{N_{\rm gen}}\sum_{\rm MC} S(\sft,\sst,\xi); \nonumber
\label{eq:eff_sig}
\end{equation}
\begin{equation}
\varepsilon_b =
\int \varepsilon(\sft,\sst)B(\sft,\sst)d\sft d\sst =
\frac{\Delta}{N_{\rm gen}}\sum_{\rm MC} B(\sft,\sst).
\label{eq:eff_bac}
\end{equation}
The sum $\sum_{\rm MC}$ is calculated from a set of MC events generated with
a uniform distribution over the Dalitz plot, passed through the full detector
simulation and subjected to all the event selection requirements;
$N_{\rm gen}$ is the number of generated events; $\Delta$ is the Dalitz plot
area.

The likelihood function to be minimized can be written as
\begin{eqnarray}
{\cal F}=&-&\sum_{\rm events}2\ln{
\left (F \frac{S(\sft,\sst,\xi)}{\sum_{\rm MC}S(\sft^{\rm MC},\sst^{\rm MC},\xi)}
   +(1-F)\frac{B(\sft,\sst)}{\sum_{\rm MC}B(\sft^{\rm MC},\sst^{\rm MC})}\right )}\nonumber\\
  &-& \sum_{\rm events}2\ln\varepsilon(\sft,\sst) 
  ~+~ \frac{(F-F_0)^2}{\sigma^2_{F_0}},
\label{eq:like}
\end{eqnarray}
where 
signal and background density functions are  normalized to satisfy the
requirement
\begin{equation}
\int S(\sft,\sst,\xi)d\sft d\sst = 1; ~~~\int B(\sft,\sst)d\sft d\sst = 1,
\label{eq:norma}
\end{equation}
$F=n_s/(n_s+n_b)$ is the relative fraction of signal events in the
data sample and $F_0$ is the estimated fraction from the fit to the $\de$
distribution. The third term takes into account the uncertainty in our
knowledge of the background contribution. As the second term in
Eq.~(\ref{eq:like}) does not depend on the fit parameters $\xi$, it is constant
for a given set of experimental events and, thus, can be omitted.
In Eq.~(\ref{eq:like}) we assume there is no interference between signal and
background processes. The background density function $B(\sft,\sst)$ is
determined from the unbinned likelihood fit to the experimental events in the
$\mb-\de$ sidebands.



\begin{figure}[t]
\includegraphics[width=0.48\textwidth]{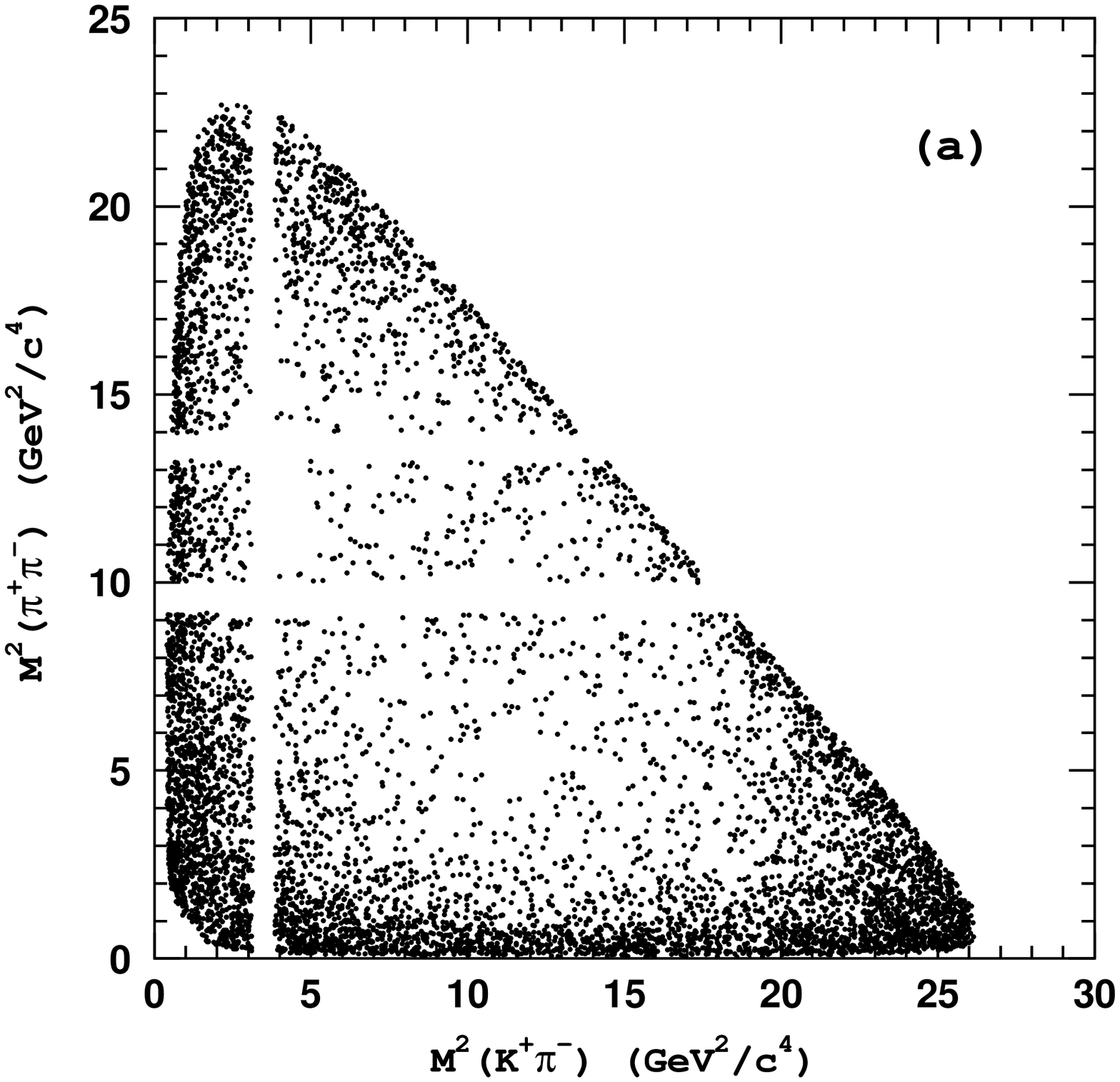} \hfill
\includegraphics[width=0.48\textwidth]{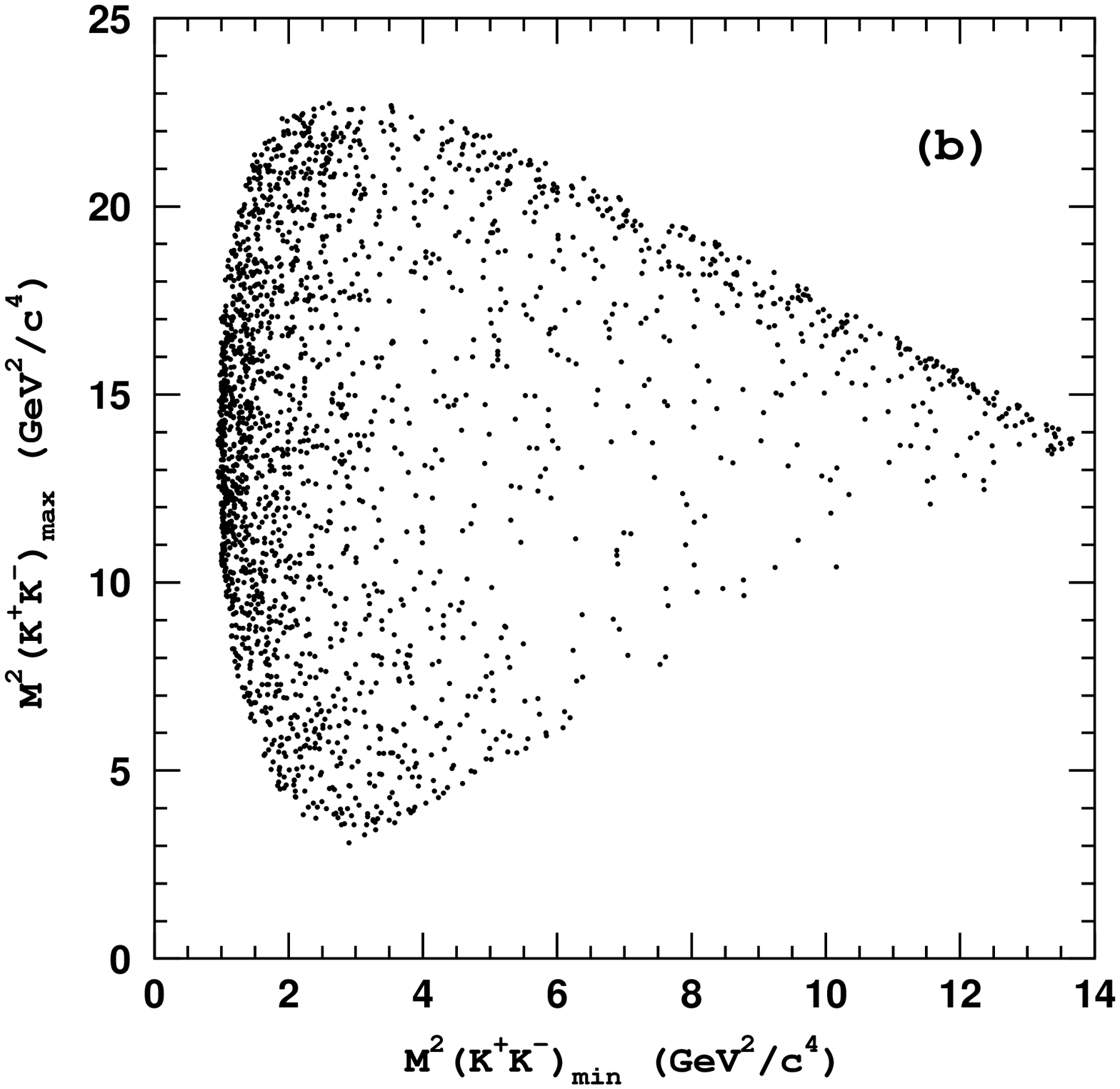}
  \caption{Dalitz plots for events in the $\de-\mb$ sidebands for the
           (a) $\kpp$ and (b) $\kkk$ (right) final states.}
  \label{fig:khh-dp-bac}
\end{figure}

\subsection{Fitting the Background Shape}
\label{sec:khh-bac}

The definition of the $\mb-\de$ sideband region is shown in
Fig.~\ref{fig:dE-Mbc}. Figure~\ref{fig:khh-dp-bac} shows the
Dalitz distributions for events in these sidebands; we find 7360 and 2176
events for the $\kpp$ and $\kkk$ final states, respectively. This is about
seven times the estimated number of background events in the corresponding
signal region.


\begin{figure}[t]
  \centering
  \includegraphics[width=0.48\textwidth]{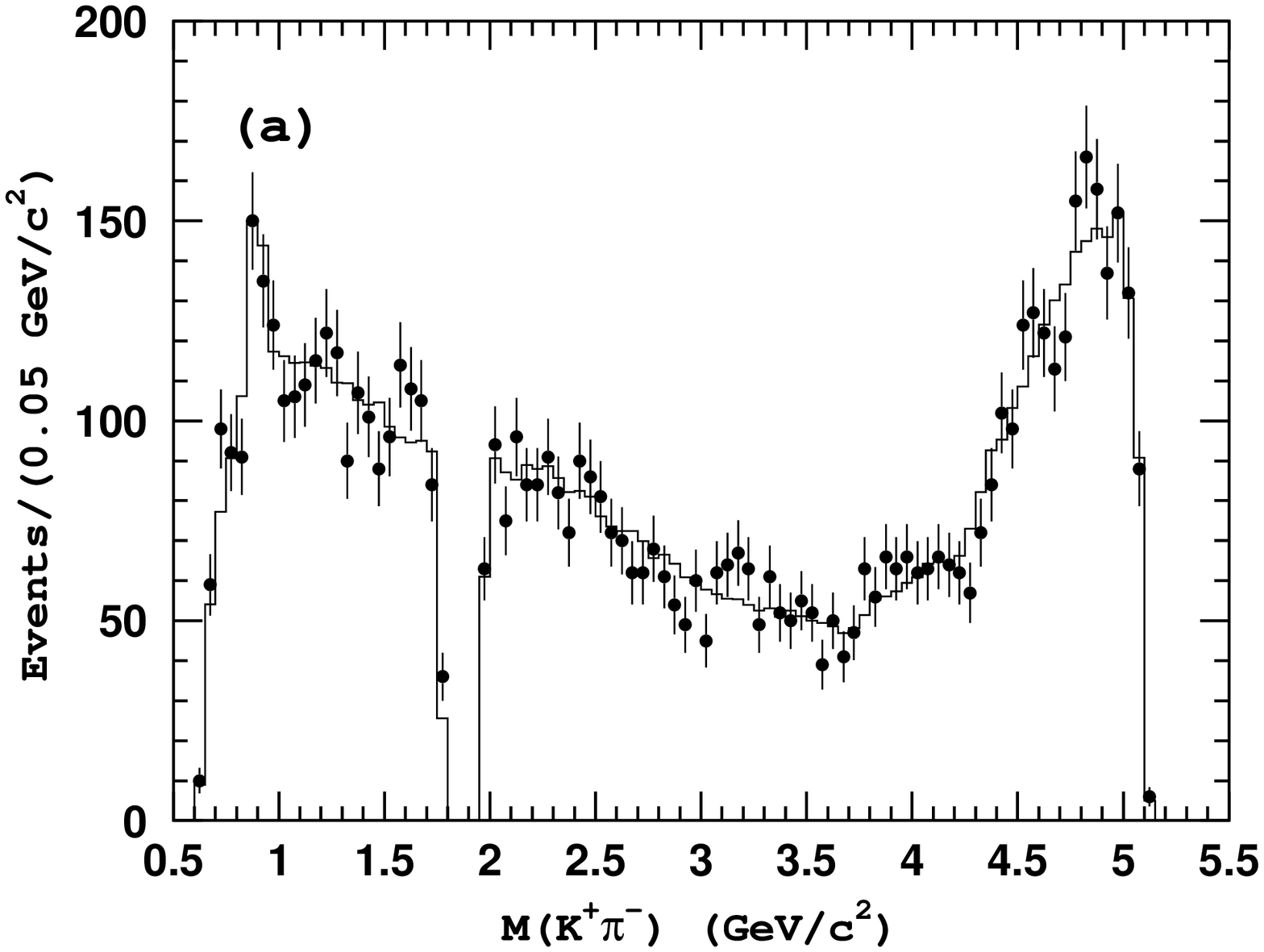} \hfill
  \includegraphics[width=0.48\textwidth]{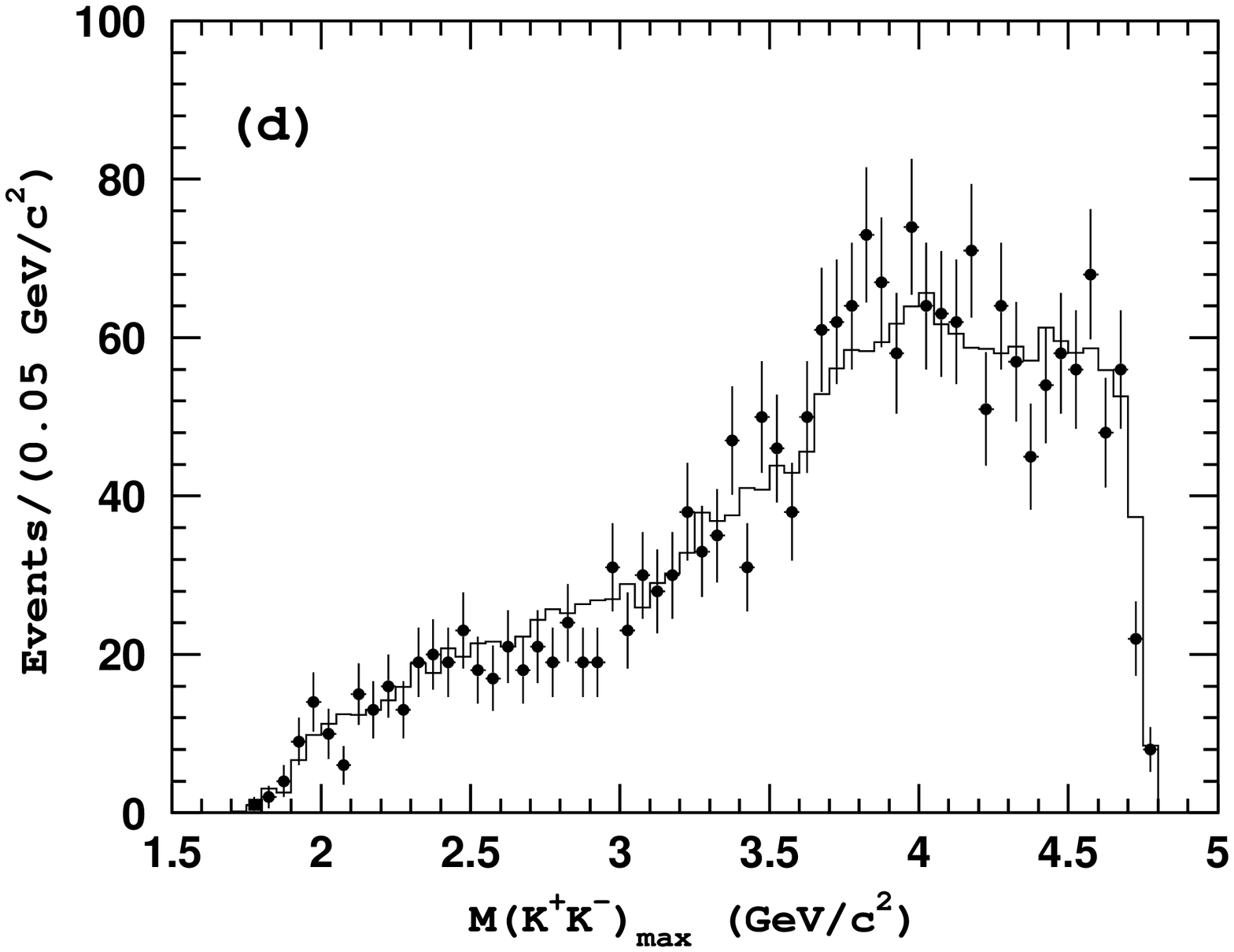} \vspace*{-4mm}\\
  \includegraphics[width=0.48\textwidth]{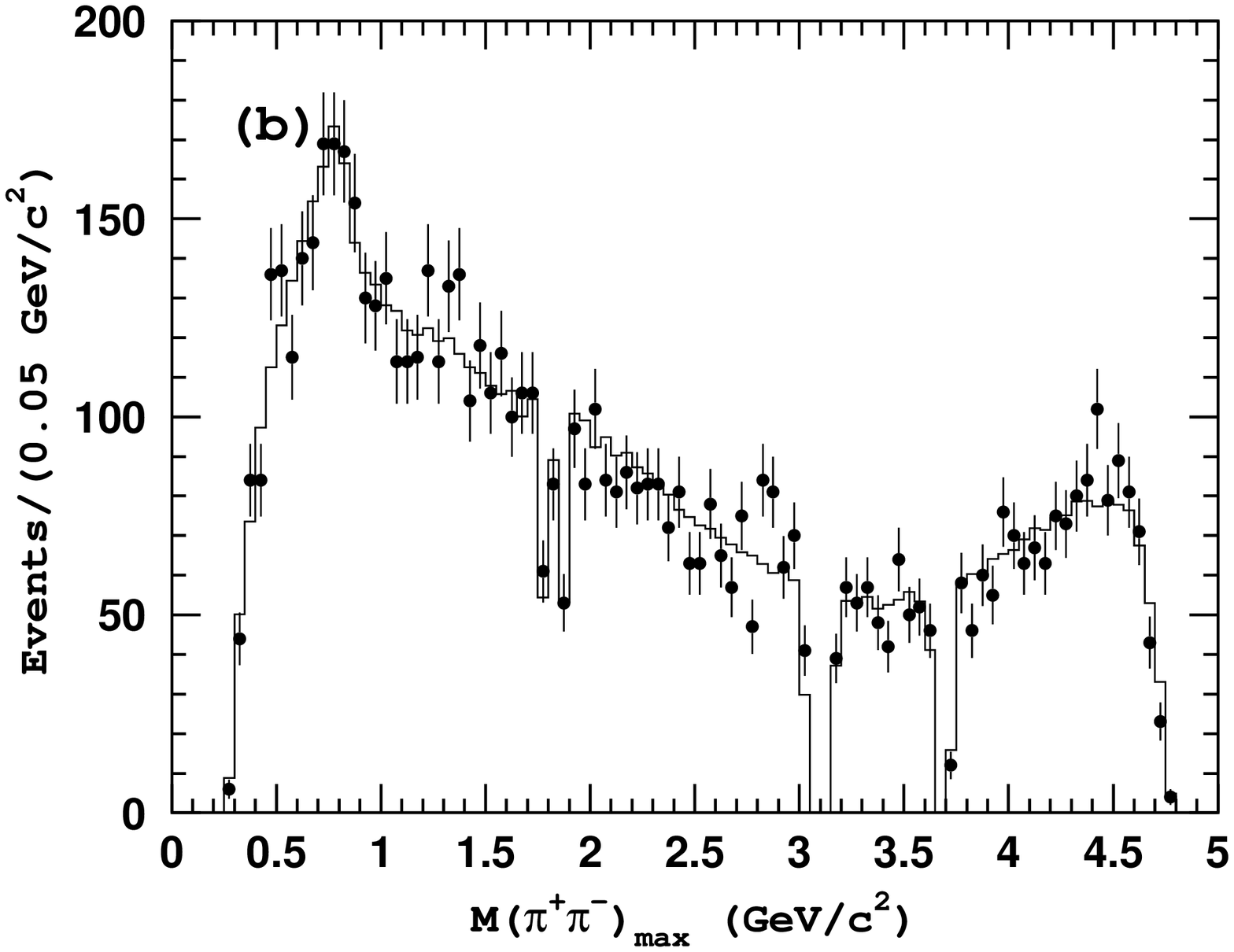} \hfill
  \includegraphics[width=0.48\textwidth]{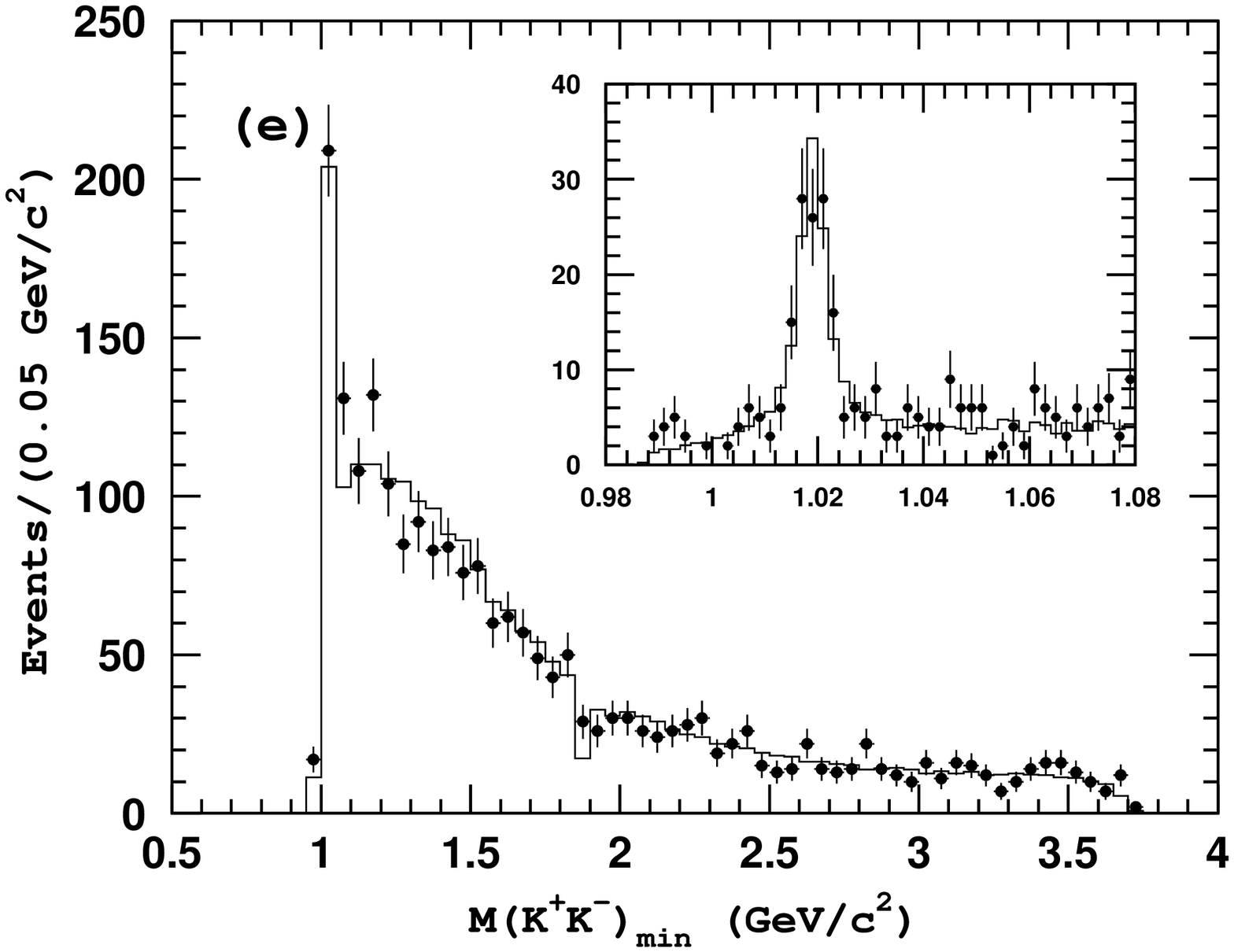} \vspace*{-4mm}\\
  \includegraphics[width=0.48\textwidth]{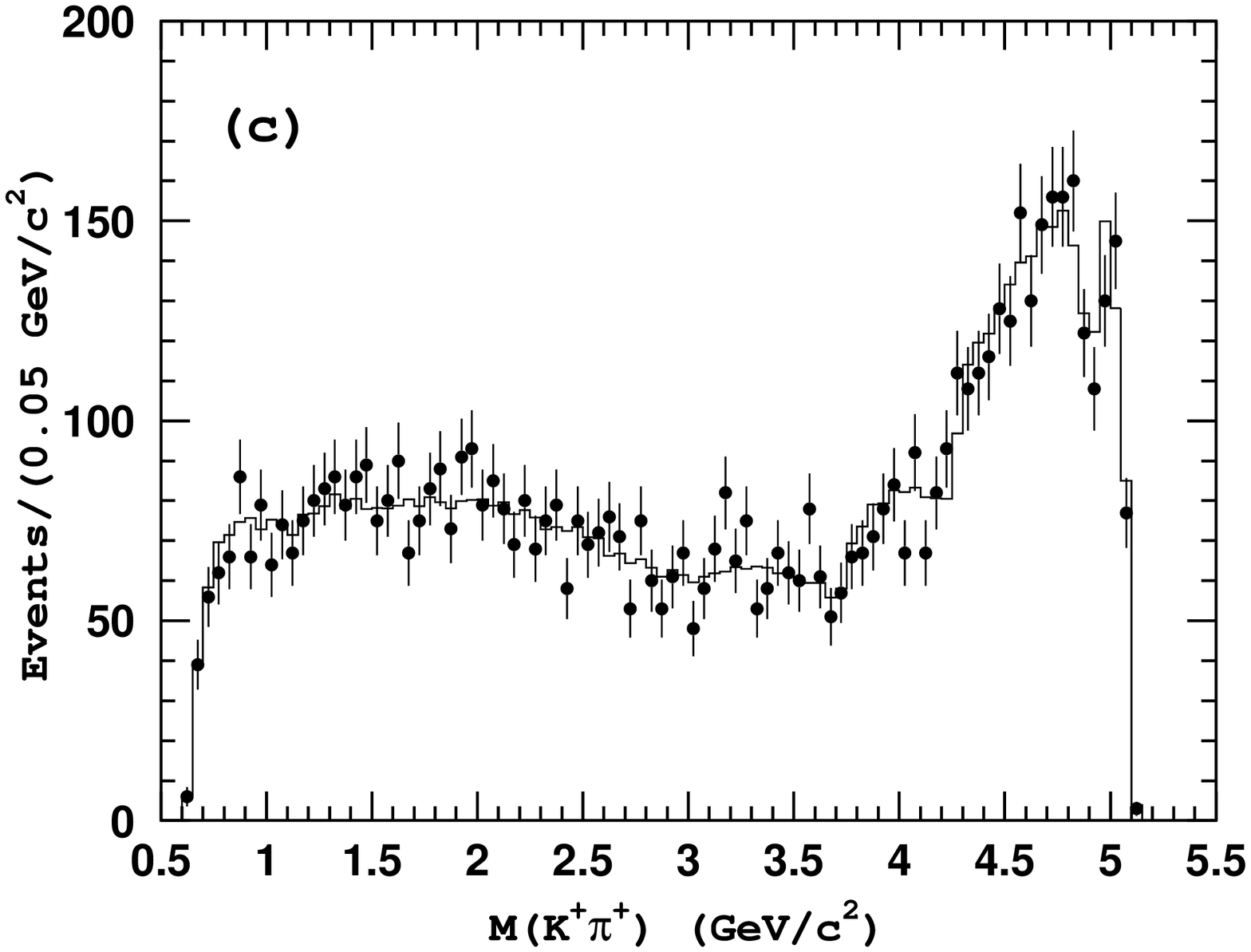} \hfill
  \includegraphics[width=0.48\textwidth]{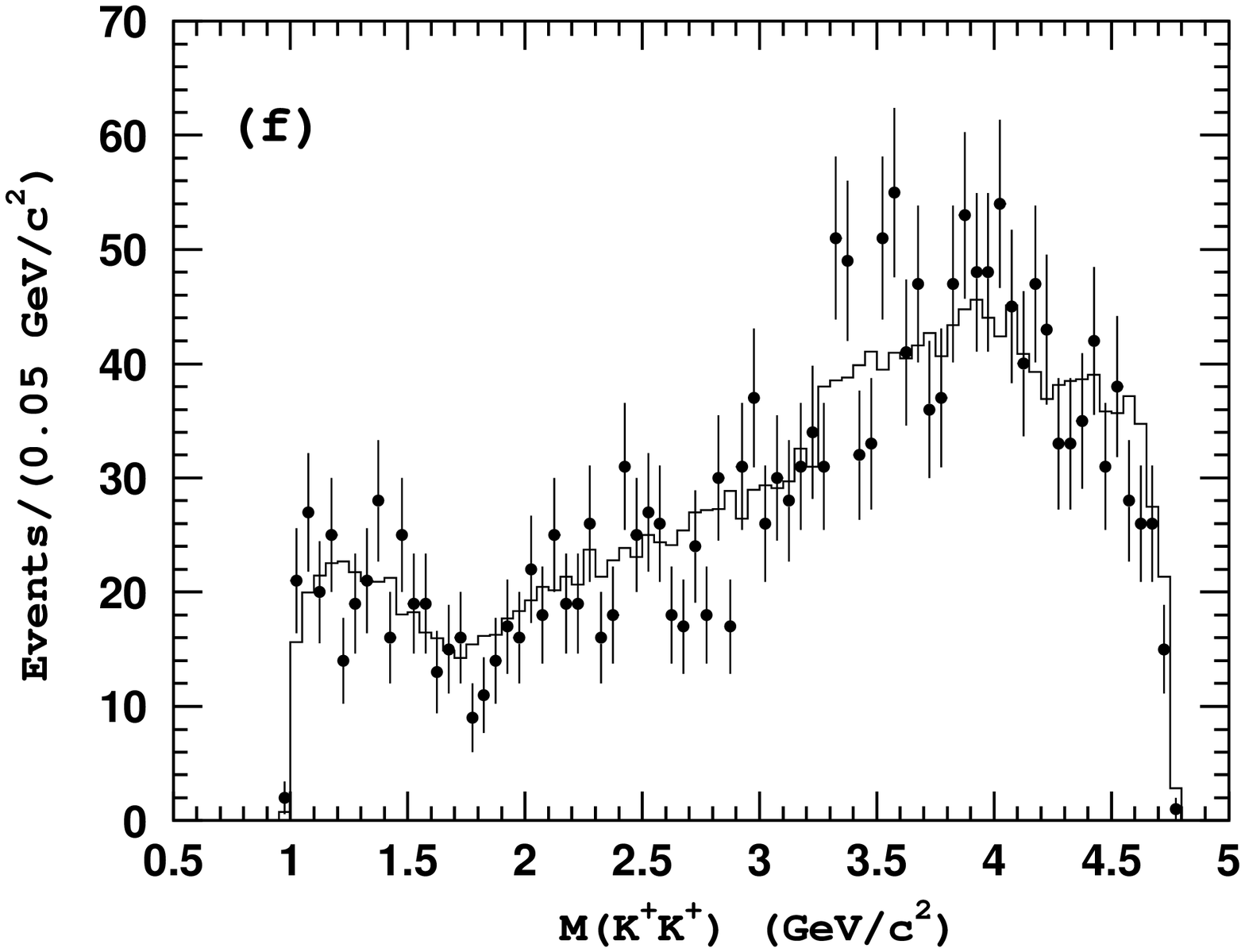} \vspace*{-4mm}\\
  \caption{Results of the best fit to the $\kpp$ (left column) and $\kkk$
           (right column) events in the $\de-\mb$ 
           sidebands shown as projections onto two-particle invariant mass
           variables. Points with error bars are data; histograms
           are fit results.  The inset in (e) shows the $\phi(1020)$ mass
           region in 2 MeV/$c^2$ bins.}
\label{fig:khh_back}
\end{figure}

We use the following empirical parametrization to describe the distribution
of background events over the Dalitz plot in the $\kpp$ final state
\begin{eqnarray}
B_{\Kpp}(\sft,\sst) &=& \alpha_1e^{-\beta_1\sft}
                       ~+~ \alpha_2e^{-\beta_2\sst}
                       ~+~ \alpha_3e^{-\beta_3\sfs} \nonumber \\
                       &+& \alpha_4e^{-\beta_4(\sft+\sst)}
                       ~+~ \alpha_5e^{-\beta_5(\sft+\sfs)}
                       ~+~ \alpha_6e^{-\beta_6(\sst+\sfs)} \nonumber \\
            &+& \gamma_1|BW_1(K^*(892))|^2 ~+~ \gamma_2|BW_1(\rho(770))|^2,
\label{eq:kpp_back}
\end{eqnarray}
where $\sft \equiv M^2(\kcpi)$, $\sst\equiv M^2(\pipi)$ and $\alpha_i$
($\alpha_1\equiv 1.0$), $\beta_i$ and $\gamma_i$ are fit parameters.
The first three terms in Eq.~(\ref{eq:kpp_back}) are introduced to describe
the background enhancement in the two-particle low invariant mass regions.
This enhancement originates mainly from $e^+e^-\to\qqbar$ continuum
events. Due to the jet-like structure of this background, all three particles
in a three-body combination have almost collinear momenta. Hence, the
invariant mass of at least one pair of particles is in the low mass region.
In addition, it is often the case that two high momentum particles are
combined with a low momentum particle to form a $B$ candidate. In this
case there are two pairs with low invariant masses and one pair with high
invariant mass. This results in even  stronger enhancement of the background
in the corners of the Dalitz plot. This is taken into account by terms $4-6$
in Eq.~(\ref{eq:kpp_back}). To account for the contribution from real
$K^*(892)^0$ and $\rho(770)^0$ mesons, we introduce two more terms in
Eq.~(\ref{eq:kpp_back}), that are (non-interfering) squared Breit-Wigner
amplitudes (as in Eq.~(\ref{eq:bw-func})), with masses and widths fixed at
world average values~\cite{PDG}. For the $\kkk$ final state the following,
somewhat more complicated, parametrization is used
\begin{eqnarray} 
B_{KKK}(\sft,\sst) &=& 
    \alpha_1(1-\alpha_2(\sst-\delta)^2)e^{-\beta_1(\sqrt{\sft}-\sqrt{\sft^0})}
~+~ \alpha_3(1-\alpha_4(\sst-\delta)^2)e^{-\beta_3(\sqrt{\sfs}-\sqrt{\sfs^0})} \nonumber \\
                   &+&
    \alpha_5e^{-\beta_5(\sqrt{\sft}+\sqrt{\sfs})}
~+~ \alpha_6e^{-\beta_6(\sqrt{\sft}+\sqrt{\sst})} \nonumber \\
                   &+& \gamma_1|BW(\phi)|^2,
\label{eq:kkk_back}
\end{eqnarray}
where $\sft\equiv {\rm min}\{M^2(K^+_1K^-),M^2(K^+_2K^-)\}$,
$\sst\equiv {\rm max}\{M^2(K^+_1K^-),M^2(K^+_2K^-)\}$, $\sfs\equiv\mkkssks$
and $\sft^0$ ($\sfs^0$) is the minimal possible value (determined by
phase space) for $\sft$ ($\sfs$), given the value of $\sst$.


\begin{table}[t]
  \caption{Parameters of the background density functions 
           determined from the fit to events in the $\mb-\de$ sidebands.}
  \medskip
  \label{tab:khh_back_pars}
\centering
  \begin{tabular}{lccccccccccccccc} \hline \hline
~Final state~ &
~~$\alpha_1$~~  & ~~$\alpha_2$~~ & ~~$\alpha_3$~~ &
~~$\alpha_4$~~  & ~~$\alpha_5$~~ & ~~$\alpha_6$~~ &
~~$\beta_1$~~   & ~~$\beta_2$~~  & ~~$\beta_3$~~  &
~~$\beta_4$~~   & ~~$\beta_5$~~  & ~~$\beta_6$~~  &
~~$\gamma_1$~~  & ~~$\gamma_2$~~ & ~~$\delta$~~
\\ \hline \hline
\vspace*{-2mm} \\
           $\kpp$  &
           \rotatebox{90}{\hspace*{-10mm}$1.0$ (fixed)}   &
           \rotatebox{90}{\hspace*{-10mm}$0.78\pm0.10$}   &
           \rotatebox{90}{\hspace*{-10mm}$1.22\pm0.27$}   &
           \rotatebox{90}{\hspace*{-10mm}$1.51\pm0.22$}   &
           \rotatebox{90}{\hspace*{-10mm}$2.05\pm0.28$}   &
           \rotatebox{90}{\hspace*{-10mm}$1.98\pm0.36$}   &
           \rotatebox{90}{\hspace*{-10mm}$1.25\pm0.09$}   &
           \rotatebox{90}{\hspace*{-10mm}$1.66\pm0.10$}   &
           \rotatebox{90}{\hspace*{-10mm}$2.17\pm0.23$}   &
           \rotatebox{90}{\hspace*{-10mm}$0.27\pm0.01$}   &
           \rotatebox{90}{\hspace*{-10mm}$0.38\pm0.01$}   &
           \rotatebox{90}{\hspace*{-10mm}$0.27\pm0.02$}   &
           \rotatebox{90}{\hspace*{-10mm}$0.80\pm0.23$}   &
           \rotatebox{90}{\hspace*{-10mm}$2.25\pm0.61$}   &
           \rotatebox{90}{\hspace*{-10mm}$~~~~~~-$}
\vspace*{12mm} \\ \hline
\vspace*{-2mm} \\
  $\kkk$  &
           \rotatebox{90}{\hspace*{-10mm}$1.0$ (fixed)}   &
           \rotatebox{90}{\hspace*{-10mm}$0.0131\pm0.0017$}   &
           \rotatebox{90}{\hspace*{-10mm}$0.51\pm0.10$}   &
           \rotatebox{90}{\hspace*{-10mm}$0.0118\pm0.0031$}   &
           \rotatebox{90}{\hspace*{-10mm}$0.40\pm0.17$}   &
           \rotatebox{90}{\hspace*{-10mm}$3.36\pm1.13$}   &
           \rotatebox{90}{\hspace*{-10mm}$4.09\pm0.32$}   &
           \rotatebox{90}{\hspace*{-10mm}$~~~~~~-$}   &
           \rotatebox{90}{\hspace*{-10mm}$4.83\pm0.69$}   &
           \rotatebox{90}{\hspace*{-10mm}$~~~~~~-$}   &
           \rotatebox{90}{\hspace*{-10mm}$0.89\pm0.13$}   &
           \rotatebox{90}{\hspace*{-10mm}$1.53\pm0.16$}   &
           \rotatebox{90}{\hspace*{-10mm}$2.80\pm0.45$}   &
           \rotatebox{90}{\hspace*{-10mm}$~~~~~~-$}   &
           \rotatebox{90}{\hspace*{-10mm}$14.21\pm0.50$}
\vspace*{15mm} \\
\vspace*{-8mm} \\ \hline \hline
  \end{tabular}
\end{table}

The projections of the data and fits for the background events are shown
in Fig.~\ref{fig:khh_back}. The numerical values of the fit parameters are
given in Table~\ref{tab:khh_back_pars}. The $\chi^2/N_{\rm bins}$ values of
the fits to the Dalitz plots are $213.7/195$ for the $\kpp$ and $57.6/66$ for
the $\kkk$ final state, respectively.



\begin{figure}[t]
\includegraphics[width=0.48\textwidth]{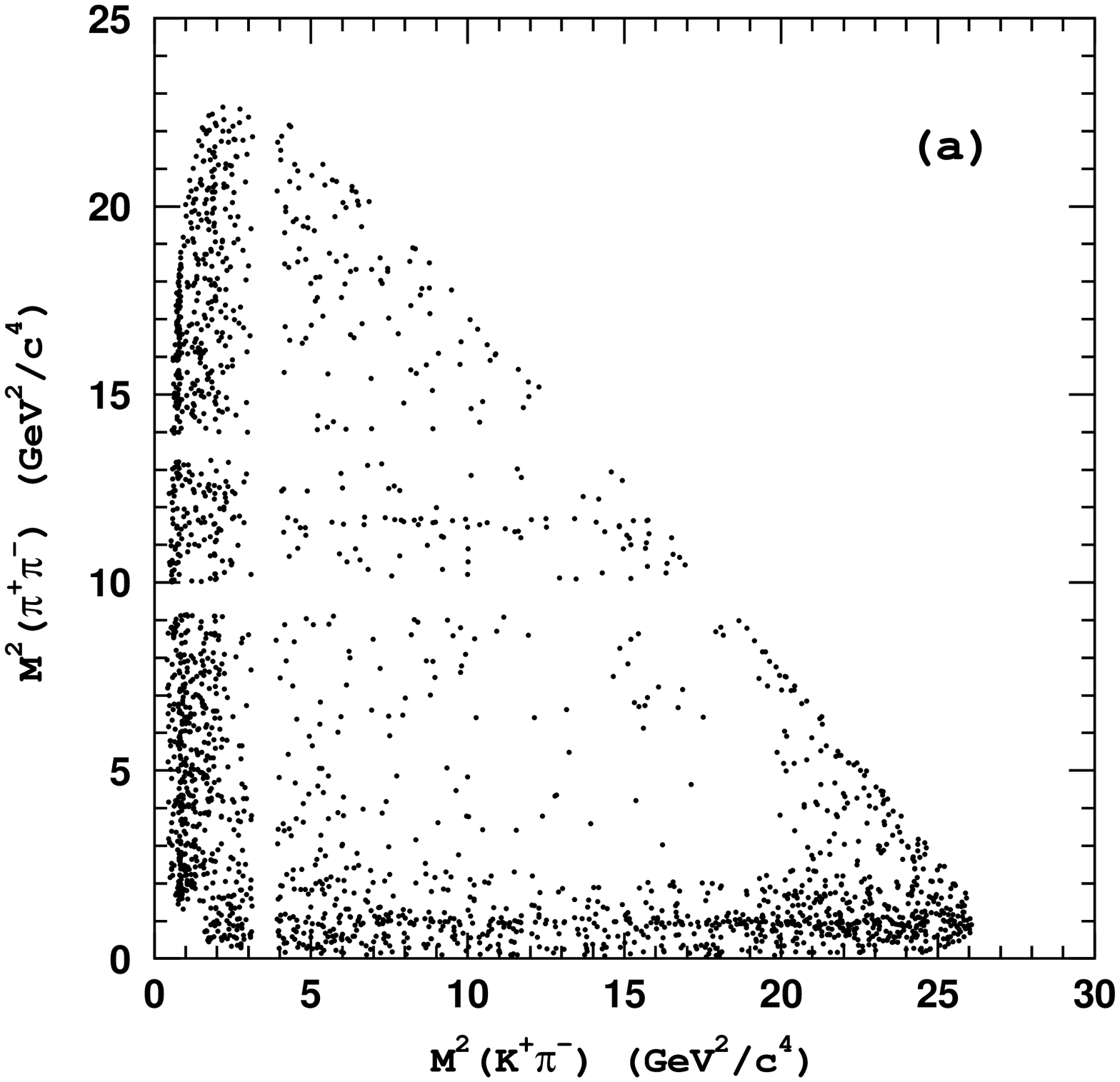} \hfill
\includegraphics[width=0.48\textwidth]{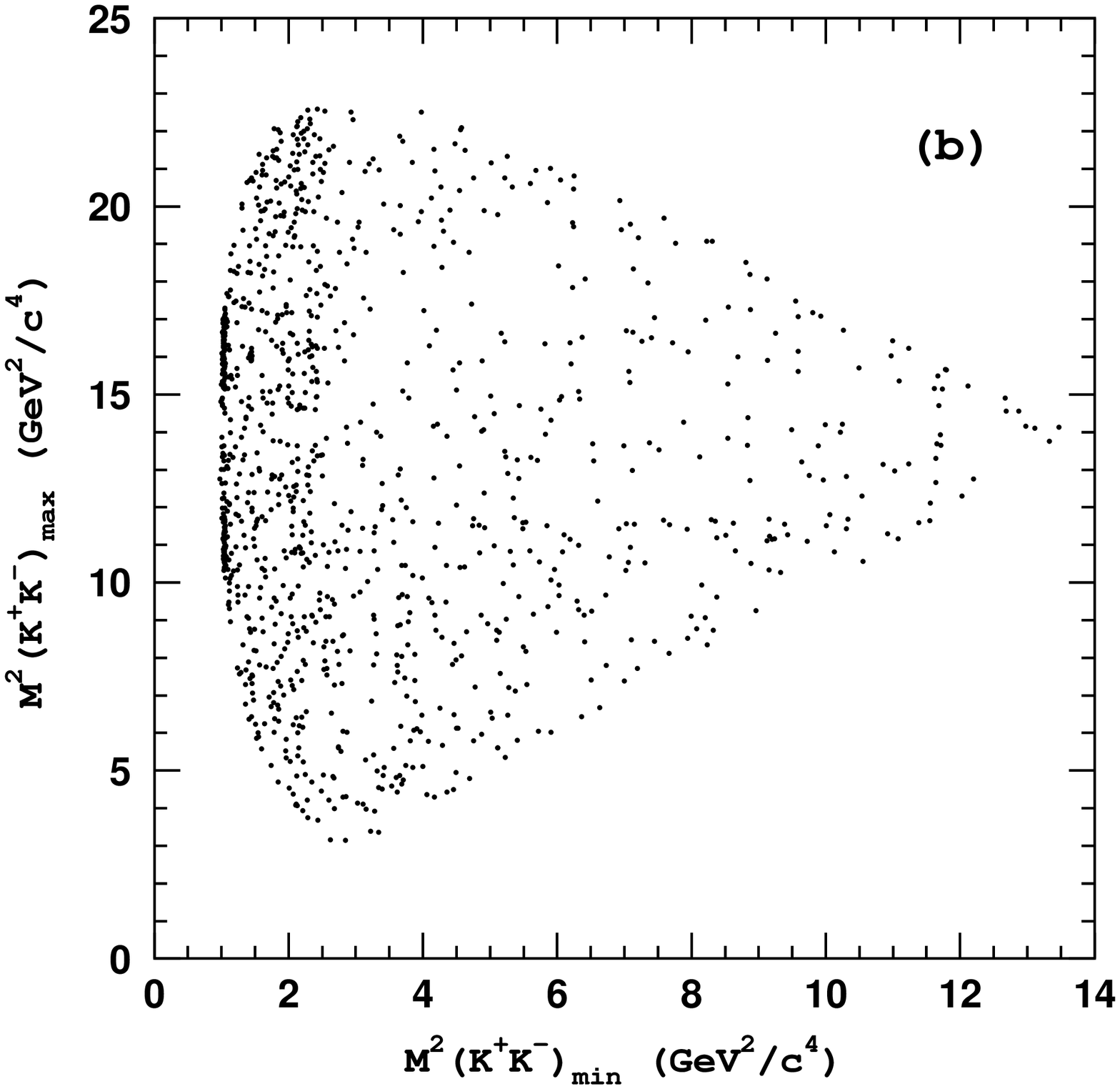}
  \caption{Dalitz plots for events in the signal region for the
           (a) $\kpp$ and (b) $\kkk$ final states.}
  \label{fig:khh-dp-sig}
\end{figure}

\subsection{Fitting the $\bckpp$ Signal}
\label{sec:kpp-sig}

\begin{figure}[t]
  \centering
  \includegraphics[width=0.48\textwidth,height=55mm]{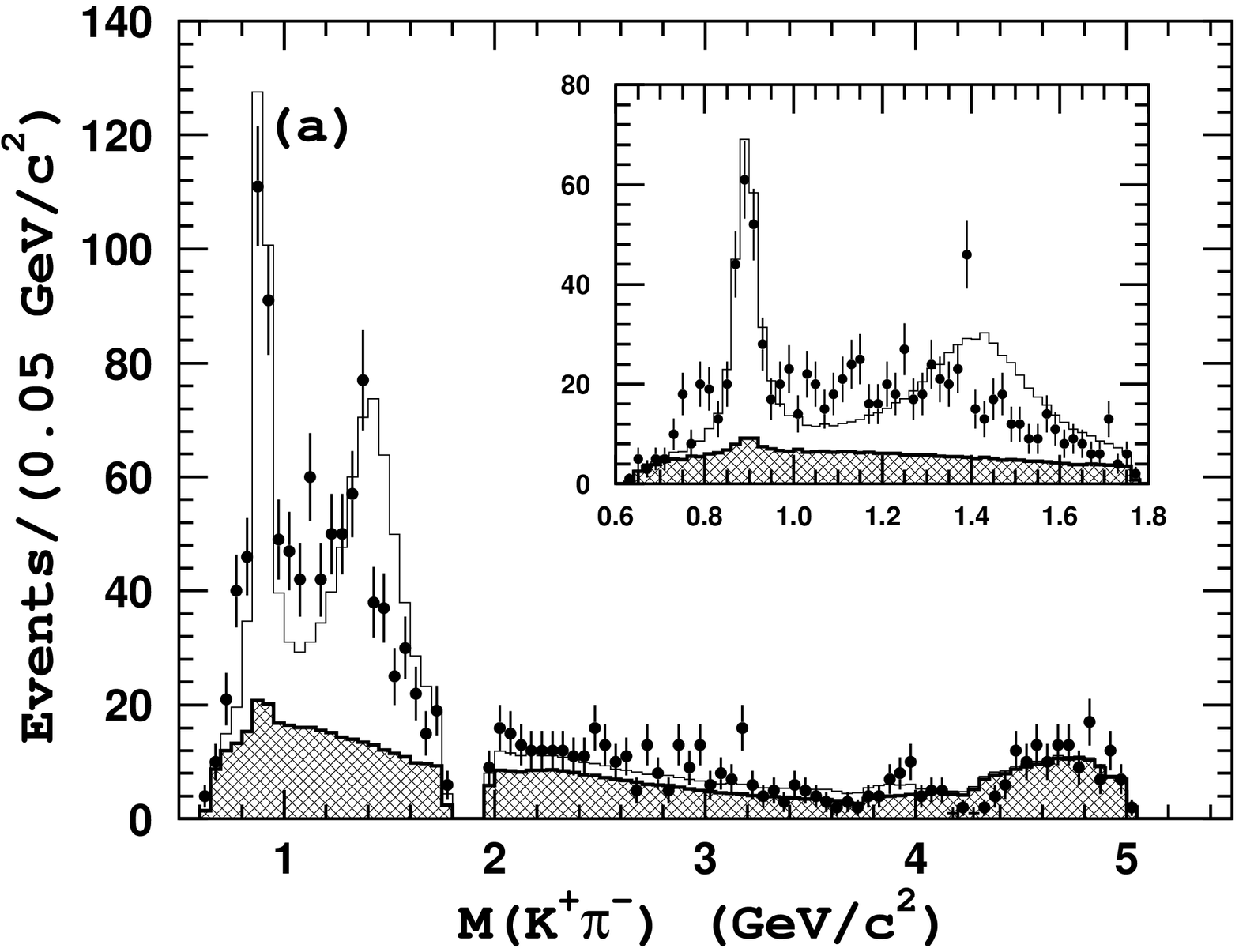} \hfill
  \includegraphics[width=0.48\textwidth,height=55mm]{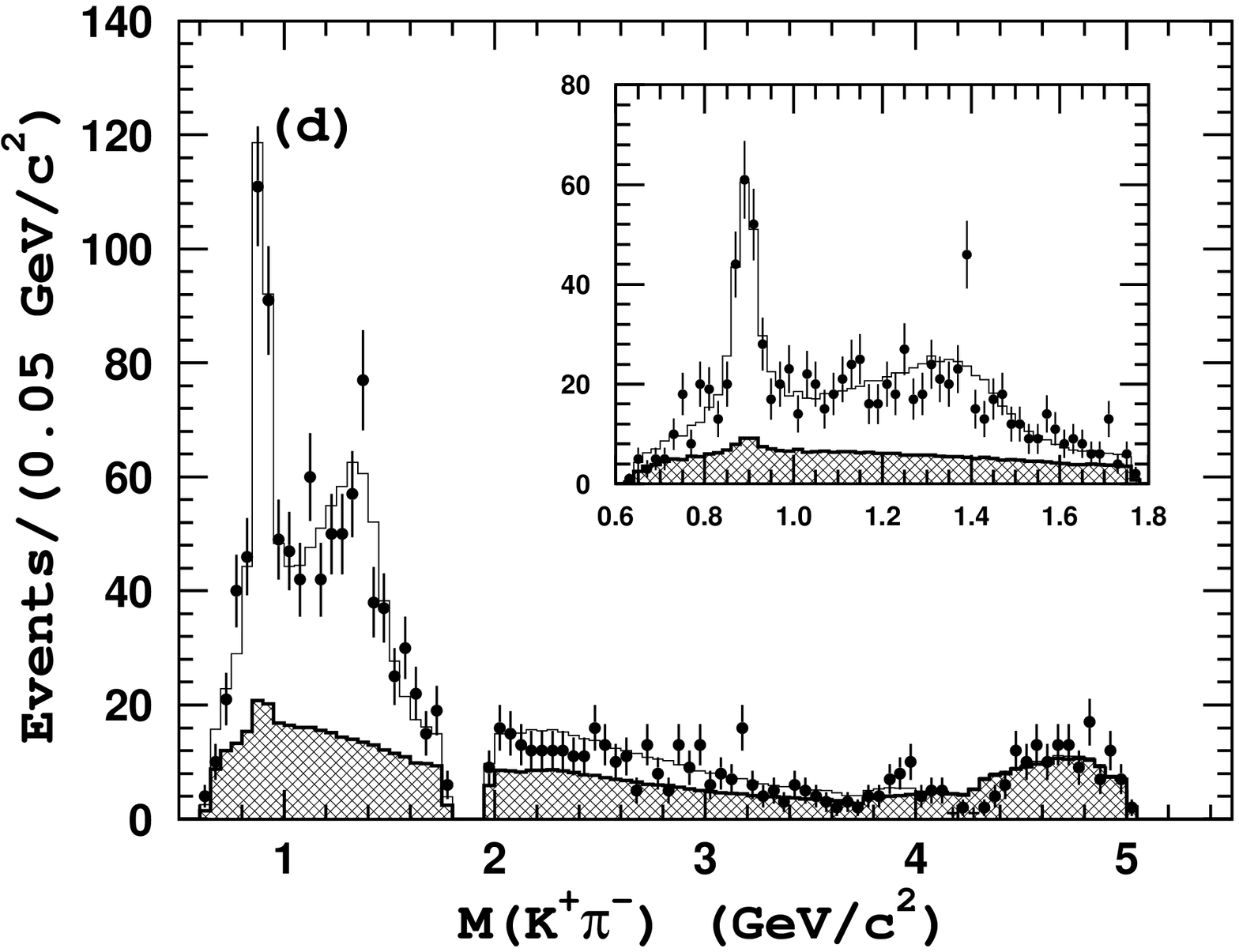}
  \includegraphics[width=0.48\textwidth,height=55mm]{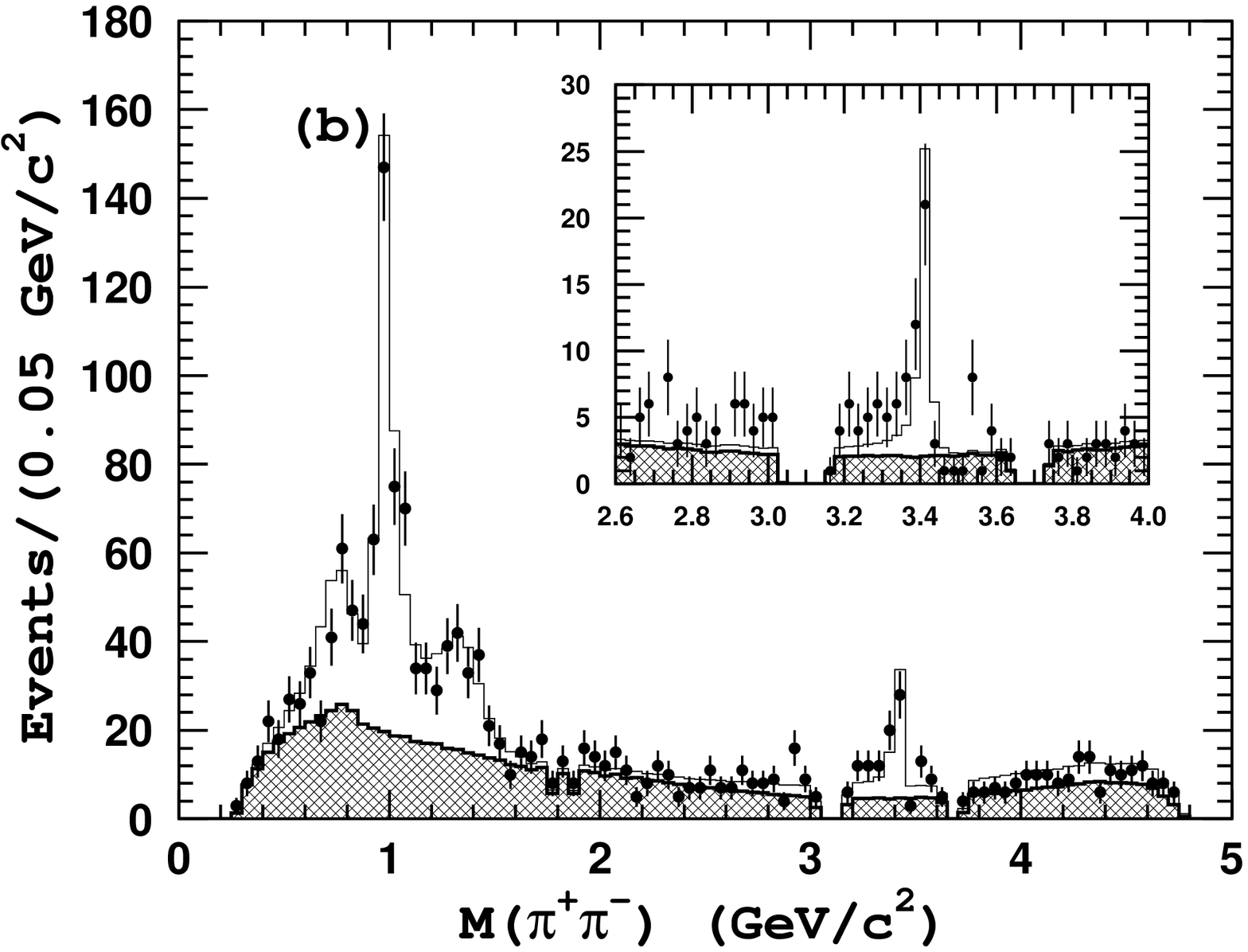} \hfill
  \includegraphics[width=0.48\textwidth,height=55mm]{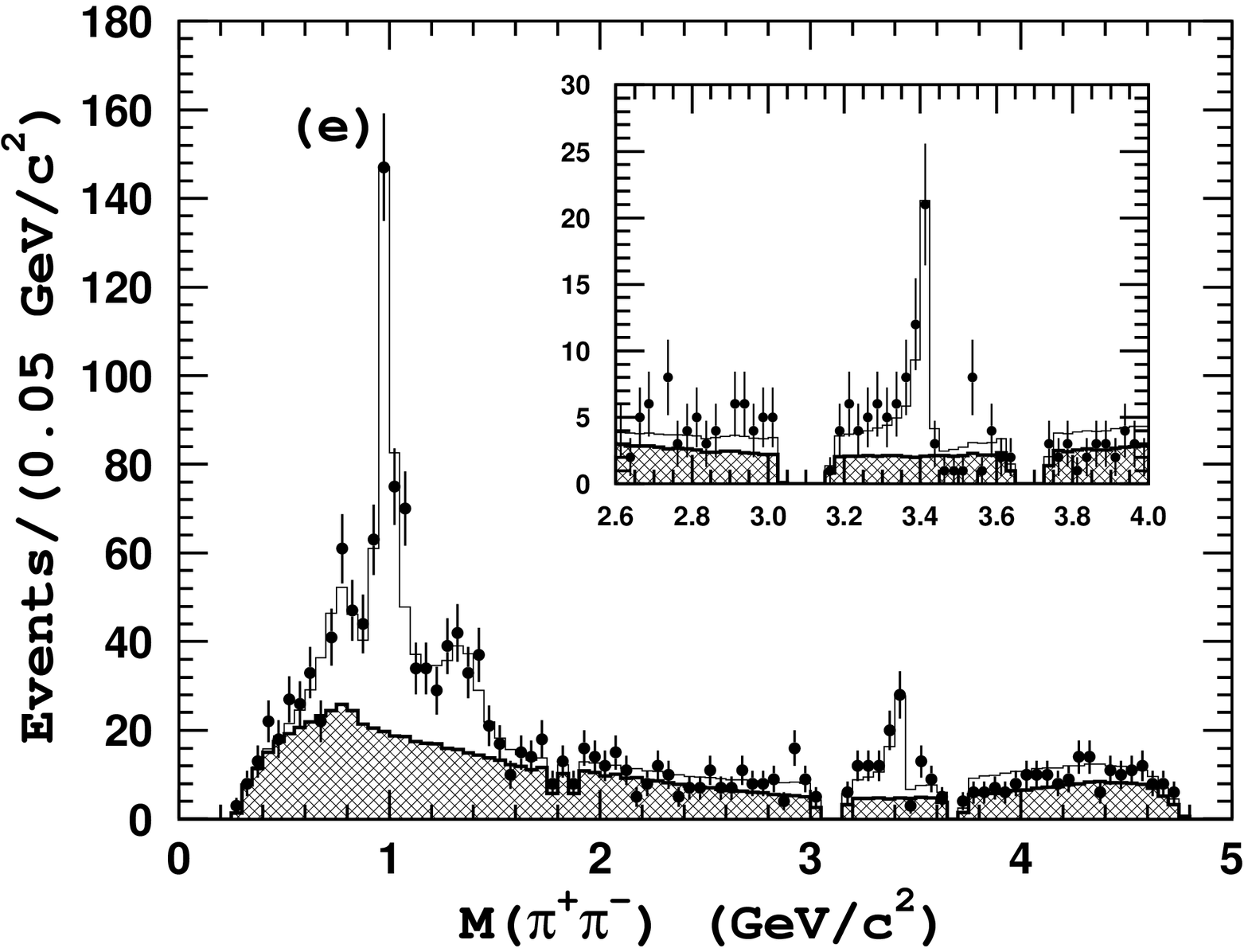}
  \includegraphics[width=0.48\textwidth,height=55mm]{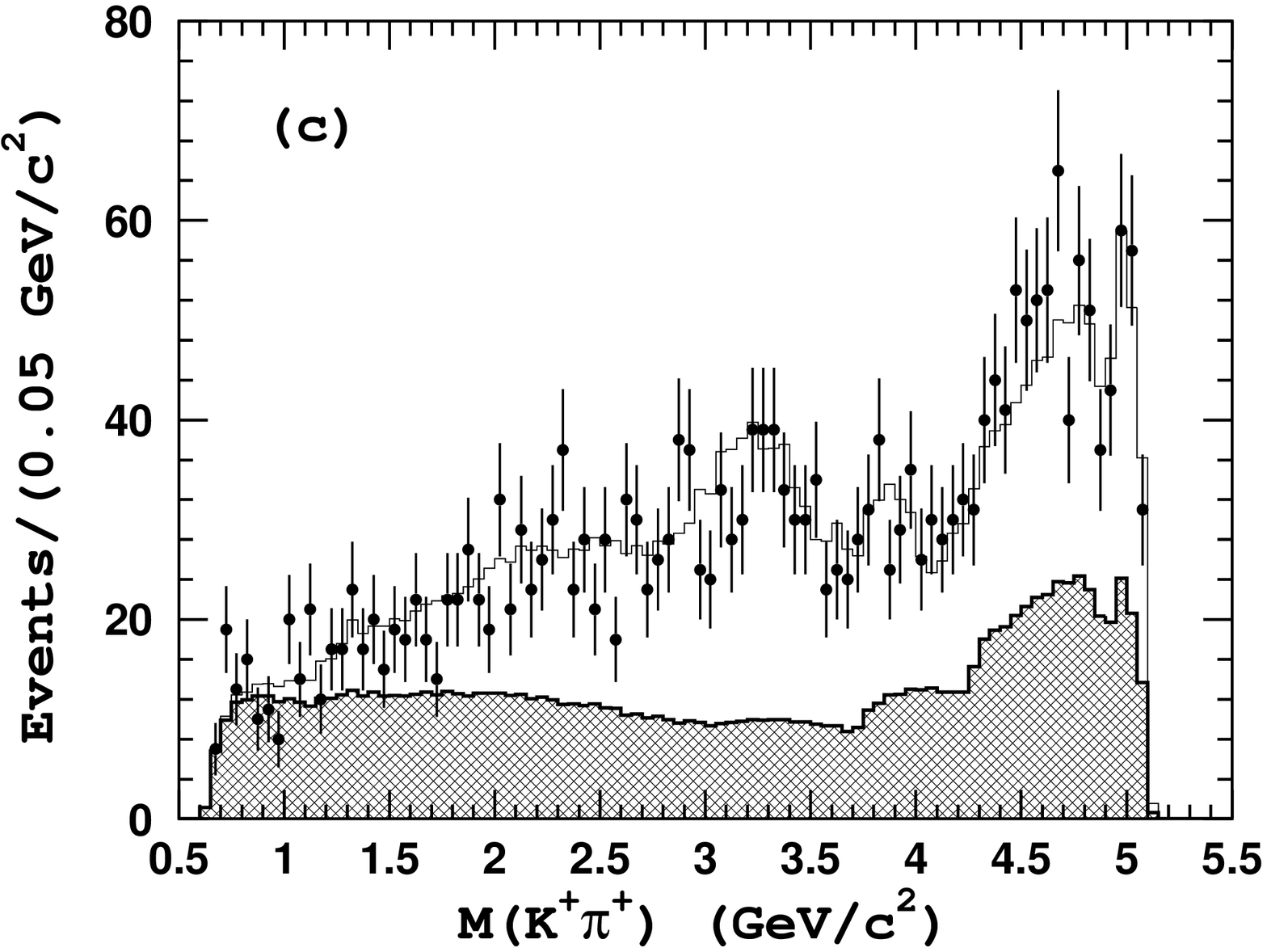} \hfill
  \includegraphics[width=0.48\textwidth,height=55mm]{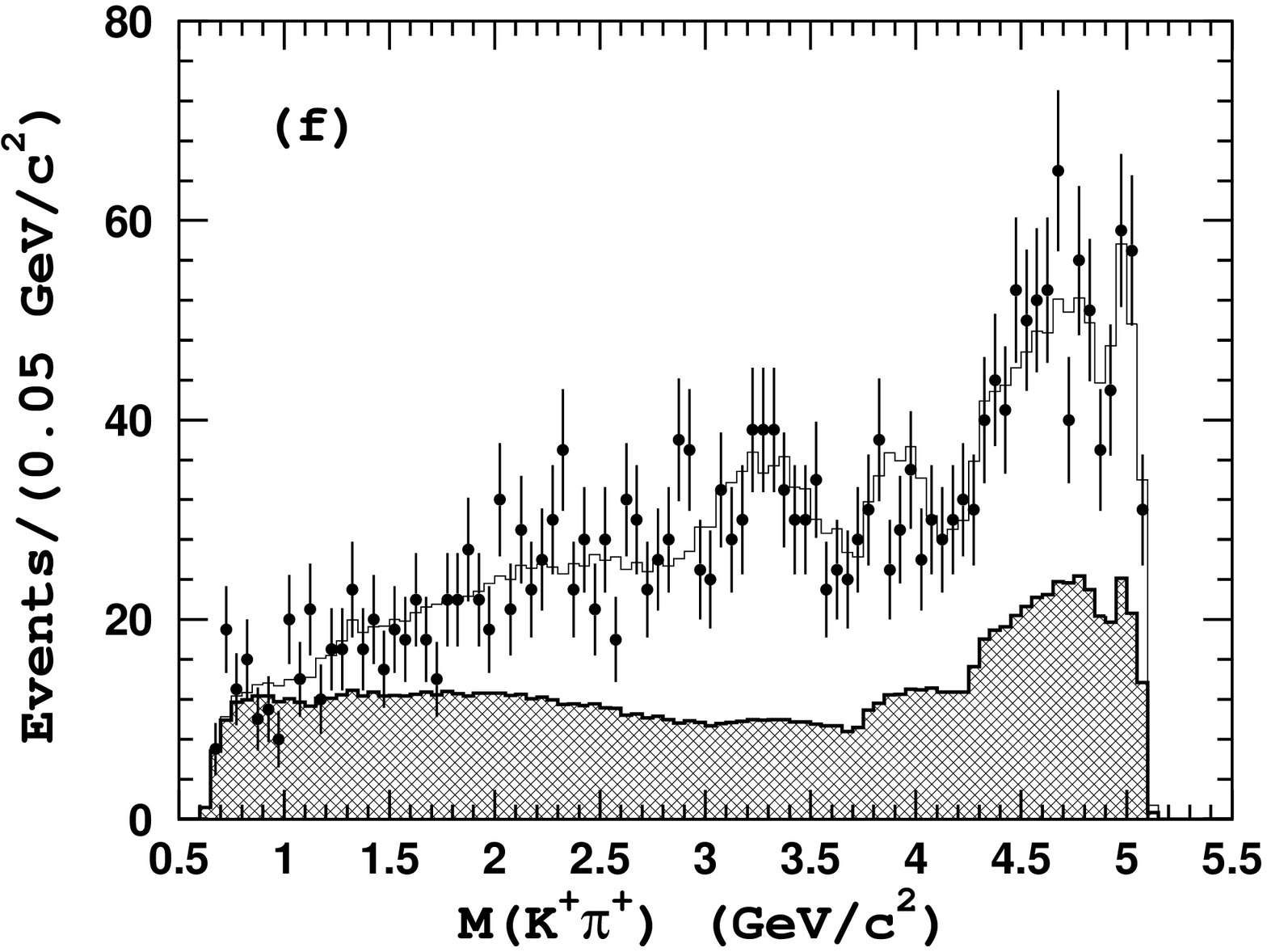}
  \caption{Results of the fit to $\kpp$ events in the signal region with
           model~$\Kpp-$A$_0$ (left column) and model~$\Kpp-$C$_0$ (right
           column). Points with error bars are data, the open histograms
           are the fit result and hatched histograms are the background
           components. Insets in (a) and (d) show the $K^*(892)-K_0^*(1430)$
           mass region in 20~\mass~ bins. Insets in (b) and (e) show the
           $\chic$ mass region in 25~\mass~ bins. Note that for plots
           (a) and (d) ((b) and (e)) an additional requirement
           $M(\pipi)>1.5$~GeV/$c^2$ ($M(\kcpi)>1.5$~GeV/$c^2$) is imposed.}
\label{fig:kpp-mods}
\end{figure}

\begin{figure}[t]
  \centering
  \includegraphics[width=0.48\textwidth]{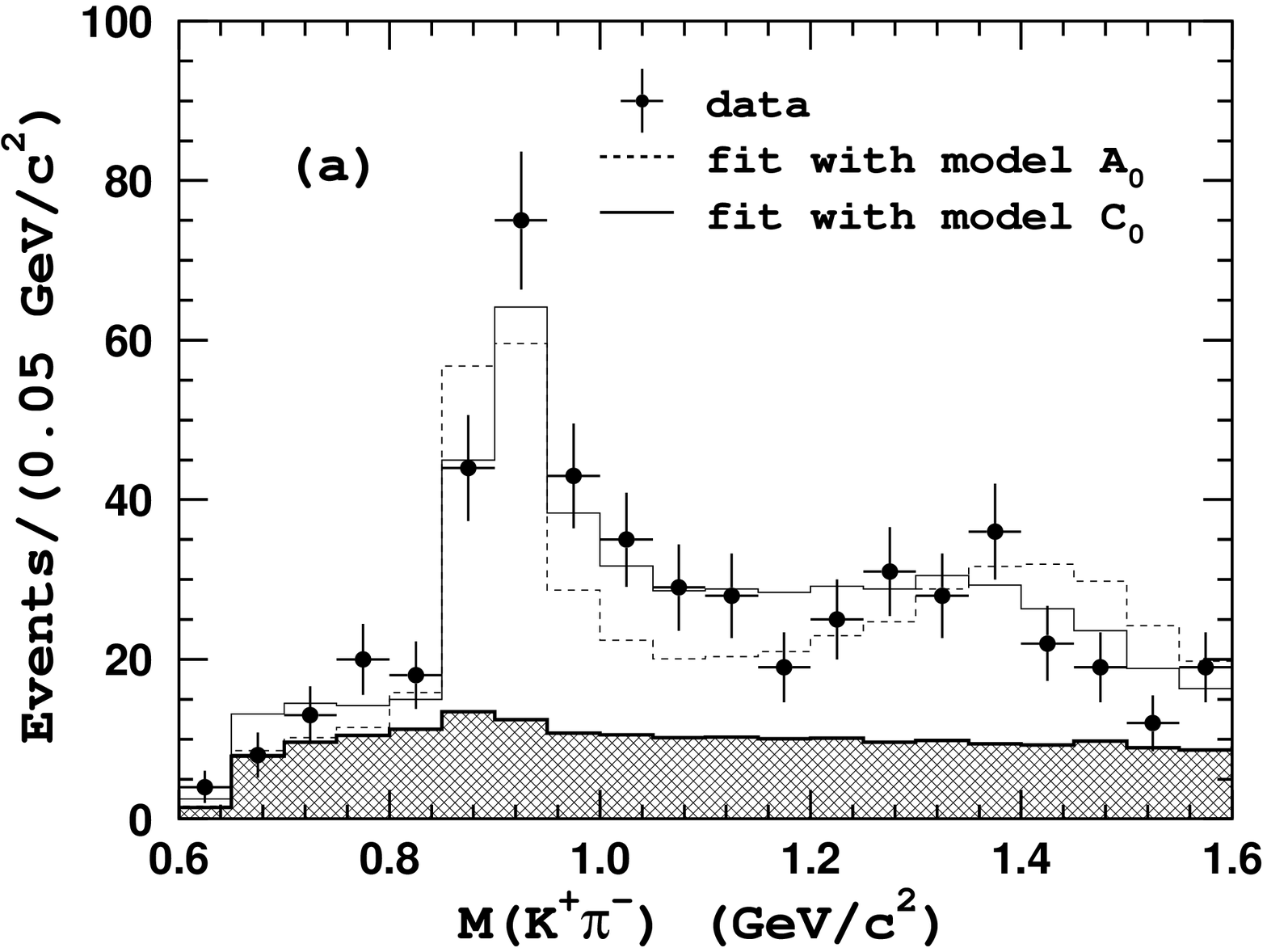} \hfill
  \includegraphics[width=0.48\textwidth]{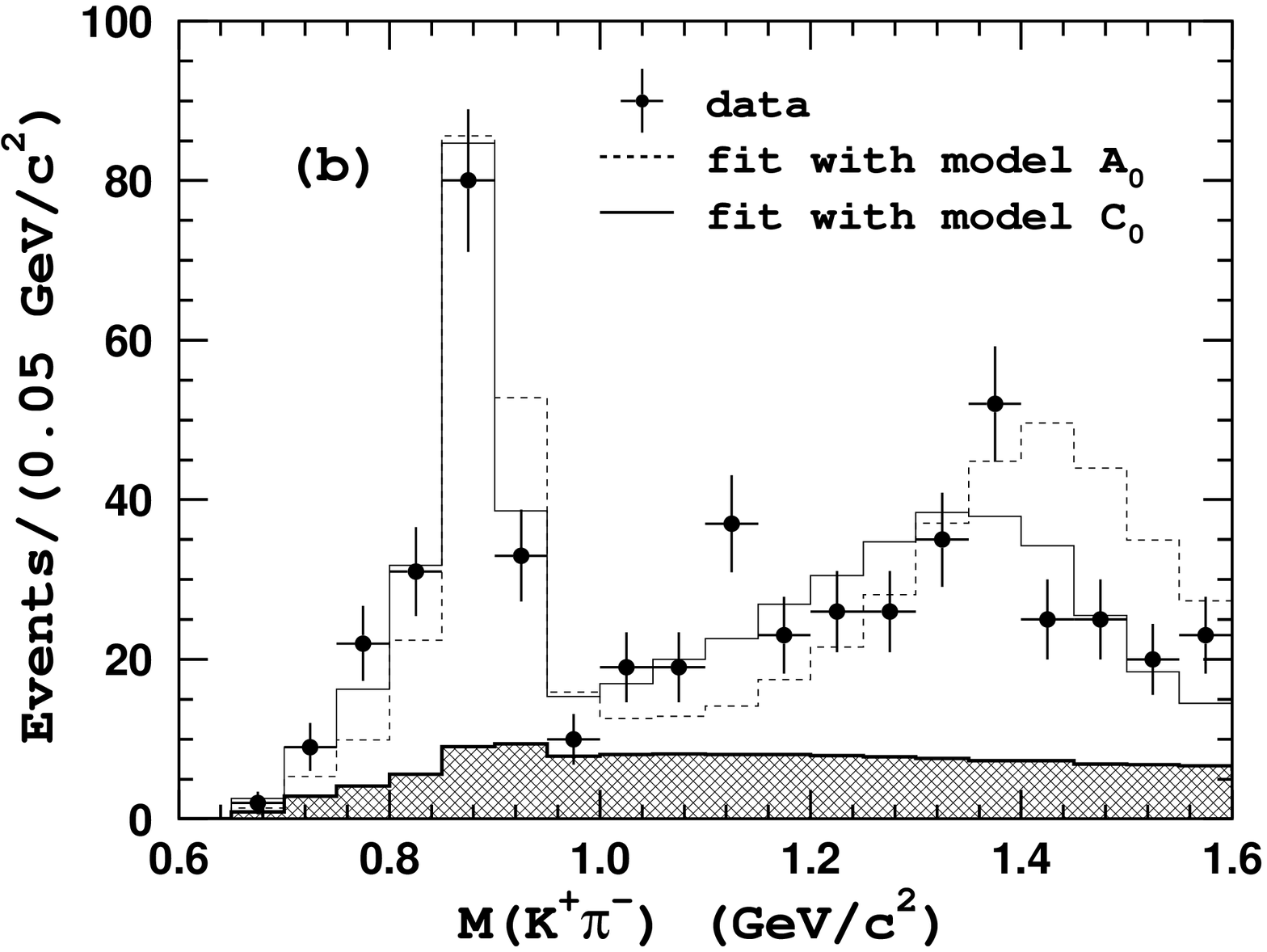}
  \caption{$\kcpi$ invariant mass distributions
           for $\kpp$ events with 
           (a)~$M^2(\pipi)<11~$\Mass~ and
           (b)~$M^2(\pipi)>11~$\Mass.}
\label{fig:kpp-kst-reg}
\end{figure}

The Dalitz plot for $\kpp$ events in the signal region is shown in 
Fig.~\ref{fig:khh-dp-sig}(a). There are 2584 events in the signal region that
satisfy all the selection requirements. In an attempt to describe all the
features of the $\kcpi$ and $\pipi$ mass spectra mentioned in
Section~\ref{sec:khh}, we start with the following minimal matrix element for
the $\bckpp$ decay (referred to as model~$\Kpp-$A$_J$):
\begin{eqnarray}
S_{A_J}(\kpp) & = & a_{K^*}e^{i\delta_{K^*}}\Am_1(\pi^+\kcpi|K^*(892)^0)
   ~+~  a_{K^*_0}e^{i\delta_{K^*_0}}\Am_0(\pi^+\kcpi|K^*_0(1430)^0)\nonumber \\
  & + & a_{\rho}e^{i\delta_{\rho}}\Am_1(\kpp|\rho(770)^0)
   ~+~  a_{f_0}e^{i\delta_{f_0}}\Am_0(\kpp|f_0(980)) \nonumber \\
  & + & a_{f_X}e^{i\delta_{f_X}}\Am_J(\kpp|f_X)
   ~+~  a_{\chic}e^{i\delta_{\chic}}\Am_0(\kpp|\chic),
\label{eq:kpp-modA}
\end{eqnarray}
where the subscript $J$ denotes the unknown spin of the $f_X(1300)$ resonance;
amplitudes $a_i$, relative phases $\delta_i$, masses and widths of the
$f_0(980)$ and $f_X(1300)$ resonances are fit parameters. The masses and widths
of all other resonances are fixed at their world average values~\cite{PDG}.
While fitting the data, we choose the $K^*(892)^0\pi^+$ signal as our reference
by fixing its amplitude and phase ($a_{K^*}\equiv 1$ and
$\delta_{K^*}\equiv 0$). Figures~\ref{fig:kpp-mods}(a,b,c) show the fit
projections with model $\Kpp-$A$_0$ and the data~\cite{footnote-2}.
The numerical values of the fit parameters are given in
Table~\ref{tab:kpp-fit-res}. However, the data are not well represented by
this matrix element, especially in the low $\kcpi$ mass region as shown in
Fig.~\ref{fig:kpp-mods}(a). This is also demonstrated in
Fig.~\ref{fig:kpp-kst-reg}, where the $\kcpi$ invariant mass distributions are
shown for the two $M^2(\pipi)$ regions: $M^2(\pipi)<11~$\Mass~ and
$M^2(\pipi)>11~$\Mass, which approximately correspond to the two helicity angle
regions: $\cos\theta^{K\pi}_H<0$ and $\cos\theta^{K\pi}_H>0$, respectively.
Result of the fit using model $\Kpp-$A$_0$ is shown as a dashed histogram in
Fig.~\ref{fig:kpp-kst-reg}. (The helicity angle is defined as the angle between
the direction of flight of the $\pi^-$ in the $\kcpi$ rest frame and the
direction of $\kcpi$ system in the $B$ rest frame.) The difference in shape of
the $M(\kcpi)$ spectra clearly observed in Figs.~\ref{fig:kpp-kst-reg}(a) and
\ref{fig:kpp-kst-reg}(b) is consistent with what is expected in the case of
interference of vector and scalar amplitudes. The scalar amplitude introduced
by the $K^*_0(1430)^0$ state is found to be insufficient to reproduce this
pattern. Thus, we modify the matrix element
(Eq.~\ref{eq:kpp-modA}) by introducing an additional scalar amplitude. First,
we construct model $\Kpp-$B$_J$, that is model $\Kpp-$A$_J$ with an additional
scalar $\kcpi$ resonance. A candidate for such a state could be the so-called
$\kappa$ resonance. An indication of the presence of the $\kappa$ in
$D^+\to K^-\pi^+\pi^+$ decay with $M(\kappa)=797\pm19\pm43$~MeV/$c^2$
and $\Gamma(\kappa)=410\pm43\pm87$~MeV/$c^2$ was reported by the E791
collaboration~\cite{e791-kappa}. Results of the fit with model $\Kpp-$B$_0$
are summarized in Table~\ref{tab:kpp-fit-res}. The agreement with the data
is somewhat improved as compared to the model $\Kpp-$A$_0$. However, if
the mass and the width of the $\kappa$ are allowed to float, the fit finds
\mbox{$M($''$\kappa$''$)=1.23\pm0.07$~GeV/$c^2$} and
\mbox{$\Gamma($''$\kappa$''$)=2.41\pm0.26$~GeV/$c^2$}. Both the mass and the
width differ from those for the $\kappa$ state measured by the E791
collaboration.  On the other hand, a scalar amplitude with such a large width
could be an indication of the presence of a non-resonant amplitude. To check
this hypothesis, we construct model $\Kpp-$C$_J$, that is model $\Kpp-$A$_J$
plus a non-resonant amplitude parametrized by Eq.~(\ref{eq:kpp-non-res}).
Results of the fit with model $\Kpp-$C$_0$ are given in
Table~\ref{tab:kpp-fit-res} and shown in Figs.~\ref{fig:kpp-mods}(d,e,f).


\begin{table}[p]
\caption{Summary of fit results to $\kpp$ events in the signal region.
         The two values given for model $\Kpp-$C$_0$ correspond to two
         solutions (see text for details).}
\medskip
\label{tab:kpp-fit-res}
\centering
  \begin{tabular}{lr|ccc} \hline \hline
   \multicolumn{2}{c}{Parameter}    &   \multicolumn{3}{|c}{Model}   \\
 &
 & \hspace*{5mm} $\Kpp-$A$_0$ \hspace*{5mm}
 & \hspace*{5mm} $\Kpp-$B$_0$ \hspace*{5mm}
 & \hspace*{5mm} $\Kpp-$C$_0$ \hspace*{5mm} \\
 & & & & Solution 1/Solution 2
\\ \hline \hline  
$K^*(892)^0\pi^+$        &
    fraction, \%~~       &  $18.0\pm1.4$ &   $14.1\pm1.3$ & $13.7\pm1.1$/$12.6\pm1.3$
\\
&    phase,~$^\circ$~~   &  \multicolumn{3}{|c}{ $0~~$ (~f~i~x~e~d~)}                          
\\ \hline  
$K_0^*(1430)^0\pi^+$        &
    fraction, \%~~       &  $42.1\pm3.7$ &   $48.6\pm3.4$ & $58.4\pm2.7$/$10.7\pm2.8$
\\
&    phase,~$^\circ$~~   &   $11\pm8$    &       $73\pm9$ &   $36\pm7$/$-11\pm9$
\\ \hline  
 $\rho(770)^0K^+$        &
    fraction, \%~~       &  $11.2\pm1.4$ &  $9.85\pm1.20$ & $10.0\pm1.5$/$8.18\pm0.92$
\\
&    phase,~$^\circ$~~   &   $-17\pm18$  &      $25\pm25$ &   $-52\pm18$/$47\pm25$
\\ \hline  
 $f_0(980)K^+$           &
    fraction, \%~~       &  $16.5\pm1.5$ &   $17.4\pm1.7$ &   $15.8\pm2.5$/$14.0\pm1.4$
\\
&    phase,~$^\circ$~~   &   $33\pm19$   &    $74\pm23$   &    $20\pm16$/$94\pm17$
\\
&     Mass, GeV/$c^2$~~  &$0.975\pm0.004$&$0.976\pm0.004$ &$0.976\pm0.004$/$0.975\pm0.003$
\\
&    Width, GeV/$c^2$~~  &$0.063\pm0.009$&$0.065\pm0.009$ &$0.061\pm0.009$/$0.053\pm0.009$
\\ \hline  
 $\chi_{c0}K^+$          &
    fraction, \%~~       & $3.56\pm0.93$ &  $3.09\pm0.87$ & $2.86\pm0.58$/$2.13\pm0.67$
\\
&    phase,~$^\circ$~~   &  $-124\pm16$  &     $-37\pm24$ &    $-29\pm23$/$-15\pm22$
\\ \hline  
 $f_X(1300)K^+$          &
    fraction, \%~~       & $6.70\pm1.42$ &  $6.14\pm1.50$ & $5.47\pm2.47$/$3.75\pm1.70$
\\
&    phase,~$^\circ$~~   &  $160\pm18$   &     $185\pm21$ &    $158\pm18$/$-134\pm22$
\\
&  Mass, GeV/$c^2$~~     &$1.369\pm0.026$&$1.344\pm0.026$ &$1.369\pm0.026$/$1.400\pm0.028$
\\
&  Width, GeV/$c^2$~~    &$0.220\pm0.063$&$0.227\pm0.070$ &$0.185\pm0.052$/$0.165\pm0.048$
\\ \hline  
$\kappa\pi^+$        &
   fraction, \%~~        &      $-$      &   $20.3\pm0.0$ &            $-$  
\\
&   phase,~$^\circ$~~    &      $-$      &     $-139\pm6$ &            $-$  
\\
&    Mass, GeV/$c^2$~~   &      $-$      &$0.797$ (fixed) &            $-$ 
\\
&    Width, GeV/$c^2$~~  &      $-$      &$0.410$ (fixed) &            $-$
\\ \hline  
 Non-Resonant            &
  fraction, \%~~         &      $-$      &       $-$      &    $36.2\pm3.2$/$40.1\pm5.2$
\\
&$a_2^{\rm nr}/a_1^{\rm nr}$~~   & $-$   &       $-$      &   $0.34\pm0.09$/$0.42\pm0.09$
\\
&$\delta^{\rm nr}_1$,~$^\circ$~~ & $-$   &       $-$      &       $-25\pm7$/$8\pm8$
\\
&$\delta^{\rm nr}_2$,~$^\circ$~~ & $-$   &       $-$      &      $140\pm16$/$-146\pm13$
\\
&   $\alpha$~~            &       $-$    &       $-$      & $0.102\pm0.023$/$0.106\pm0.022$
\\ \hline  
 Charmless Total$^{a}$          & 
\\
&    fraction, \%~~       & $97.7\pm0.6$ &   $96.6\pm0.8$ &    $97.5\pm0.7$/$97.6\pm0.6$
\\ \hline \hline  
\multicolumn{2}{c|}{$-2\ln{\cal{L}}$}    &     $-3845.3$  &    $-3966.6$   & $-4041.8$/$-4024.4$
\\ \hline  
\multicolumn{2}{c|}{$\chi^2$}            &      $227.8$   &     $129.0$    & $104.2$/$107.1$
\\
\multicolumn{2}{c|}{$N_{\rm bins}$}      &       $106$    &      $106$     &   $106$    
\\
\multicolumn{2}{c|}{$N_{\rm fit. var.}$} &       $14$     &      $16$      &   $19$
\\ \hline \hline  
\multicolumn{5}{l}{$^{a}$ Here ``Charmless Total'' refers to the
total three-body $\bckpp$ signal excluding the} \\
\multicolumn{5}{l}{contribution from $B^+\to\chic K^+$.}
  \end{tabular}
\vspace*{30mm}
\end{table}

\begin{figure}[t]
  \centering
 \includegraphics[width=0.48\textwidth,height=45mm]{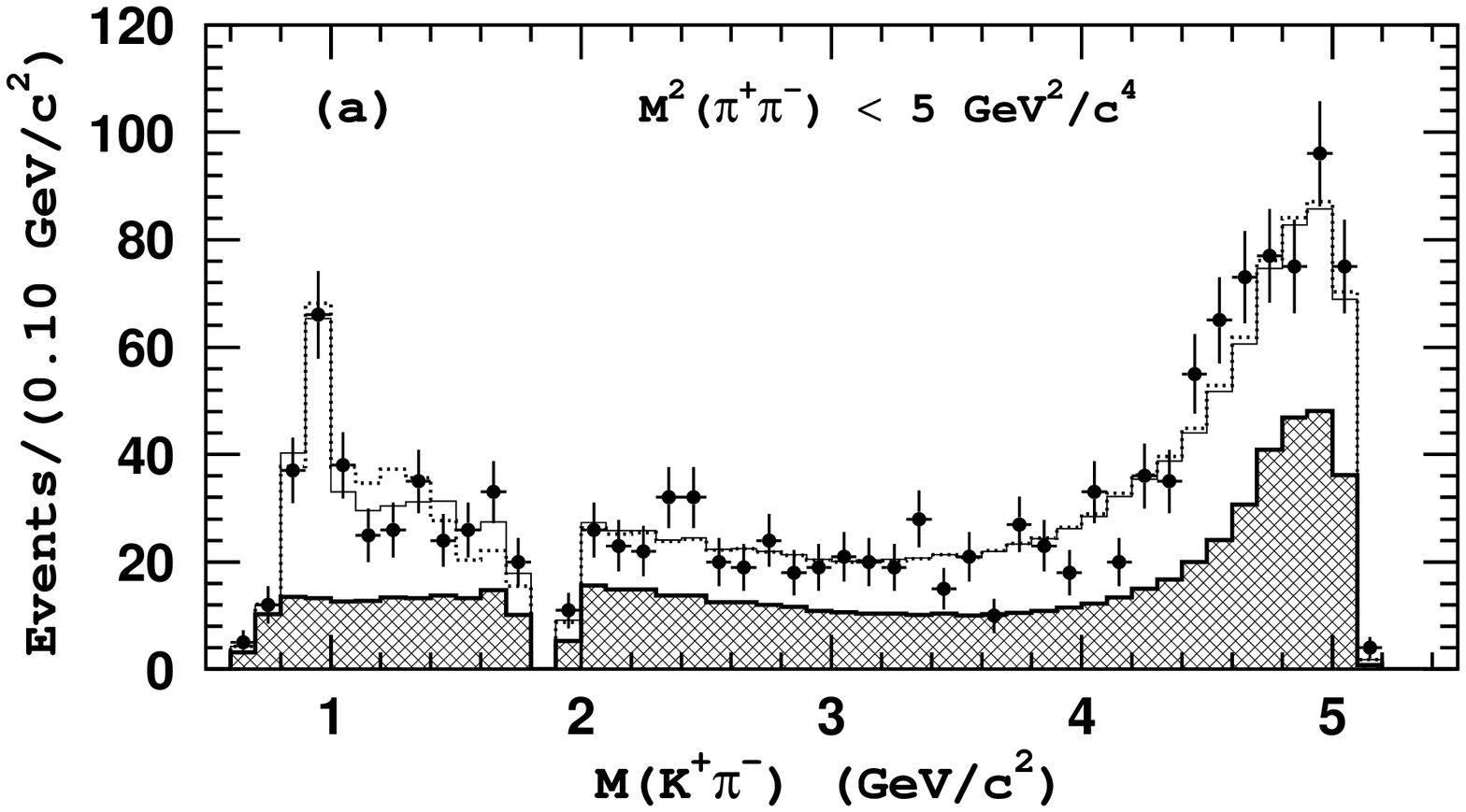} \hfill
 \includegraphics[width=0.48\textwidth,height=45mm]{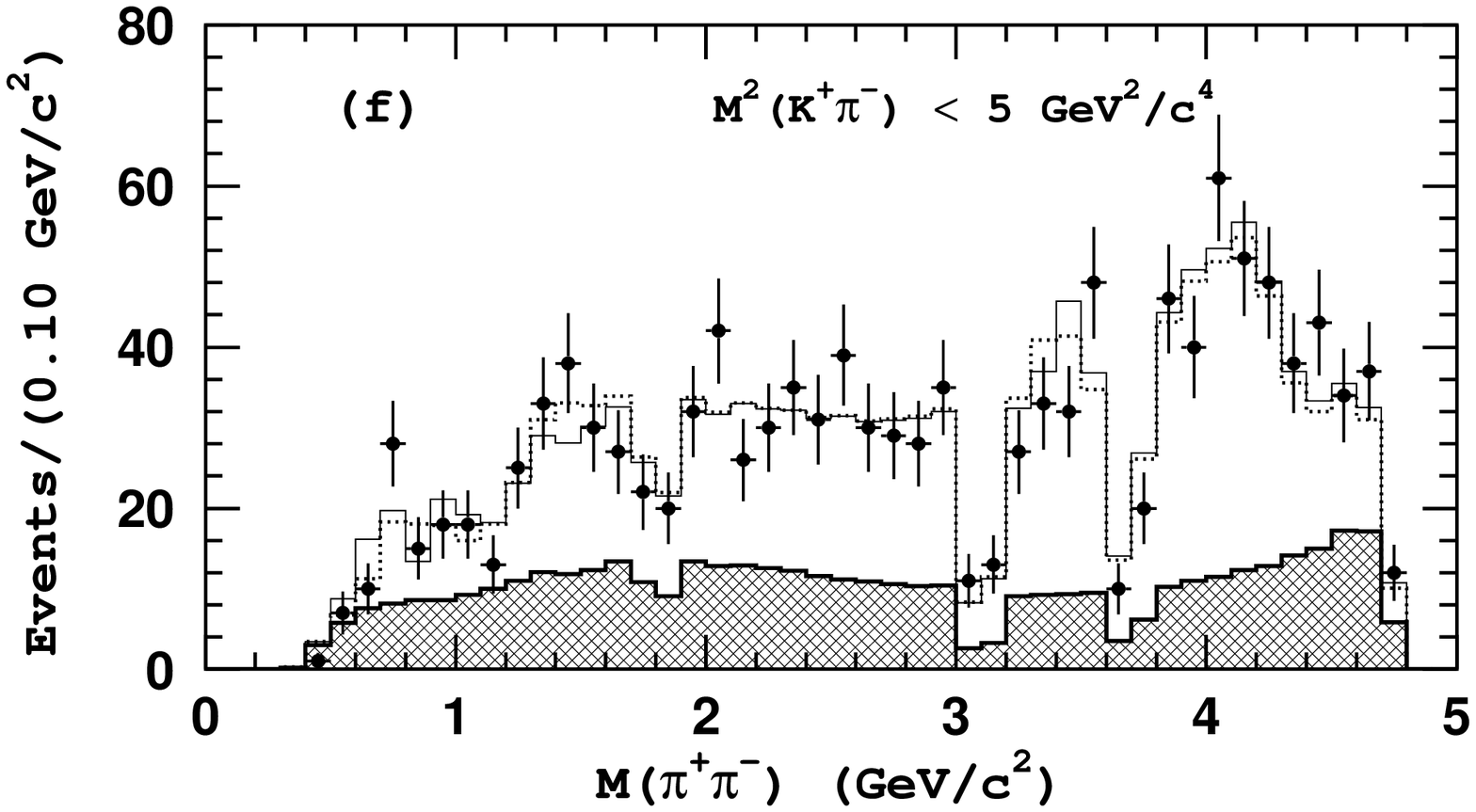} \vspace*{-5mm}\\
 \includegraphics[width=0.48\textwidth,height=45mm]{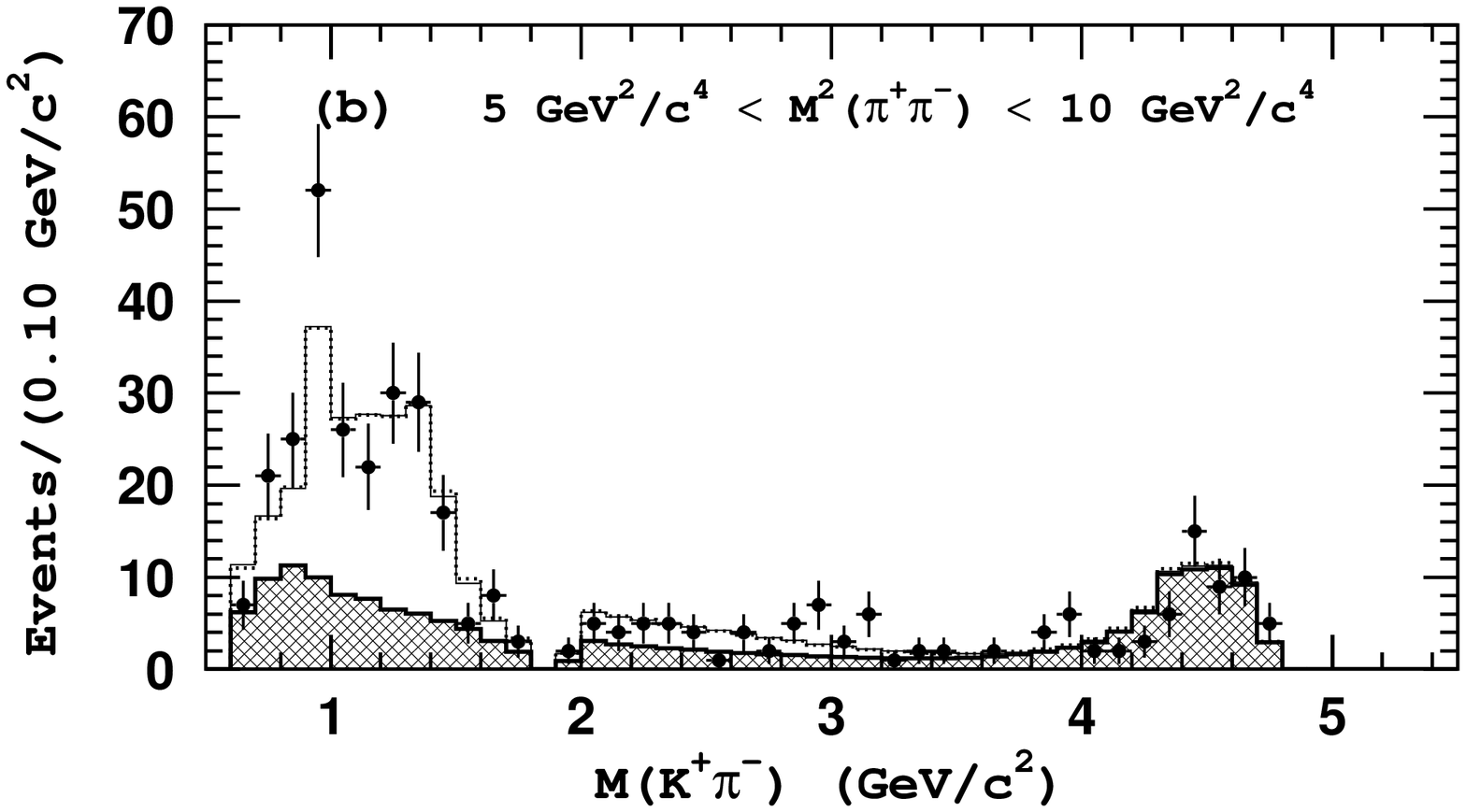} \hfill
 \includegraphics[width=0.48\textwidth,height=45mm]{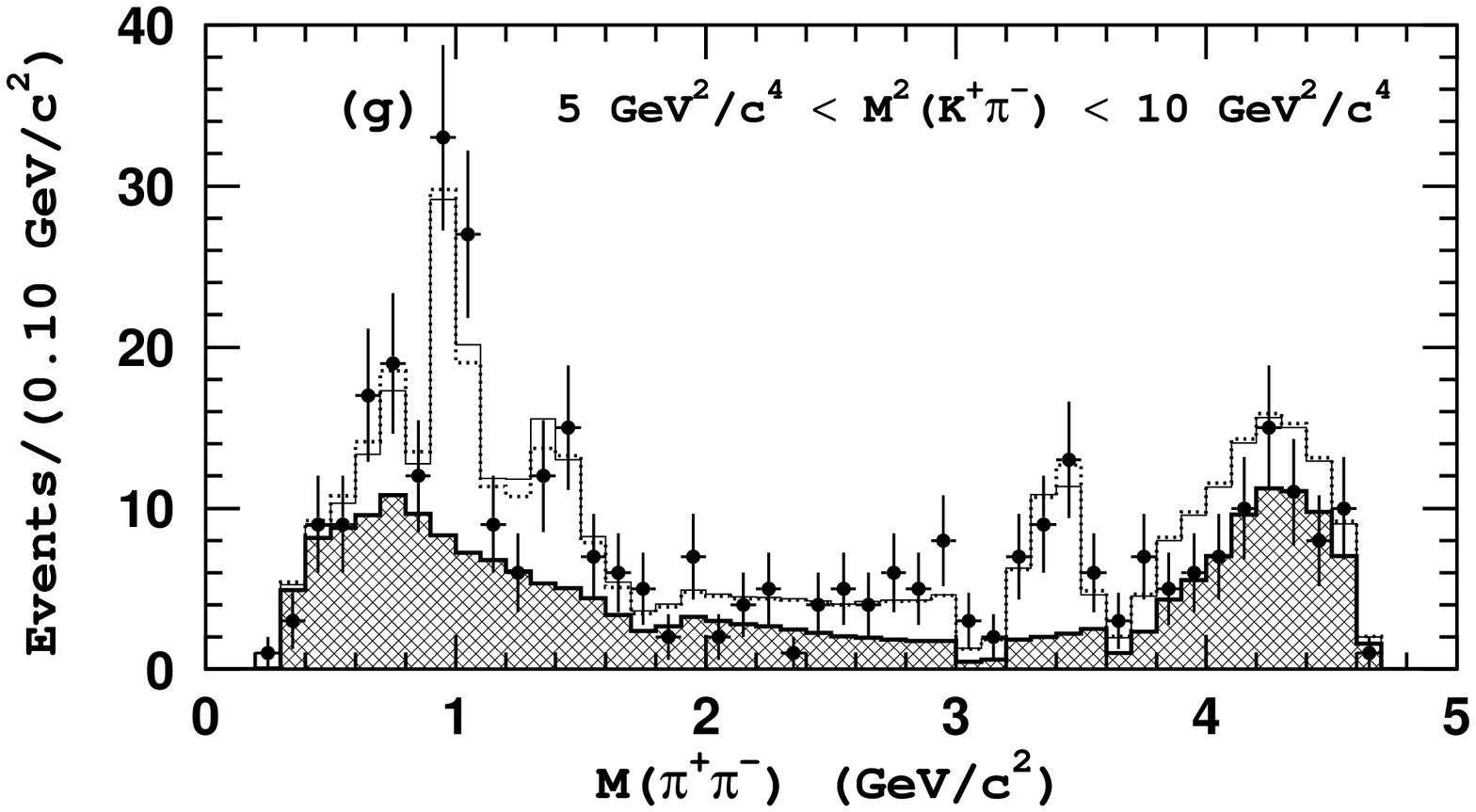} \vspace*{-5mm}\\
 \includegraphics[width=0.48\textwidth,height=45mm]{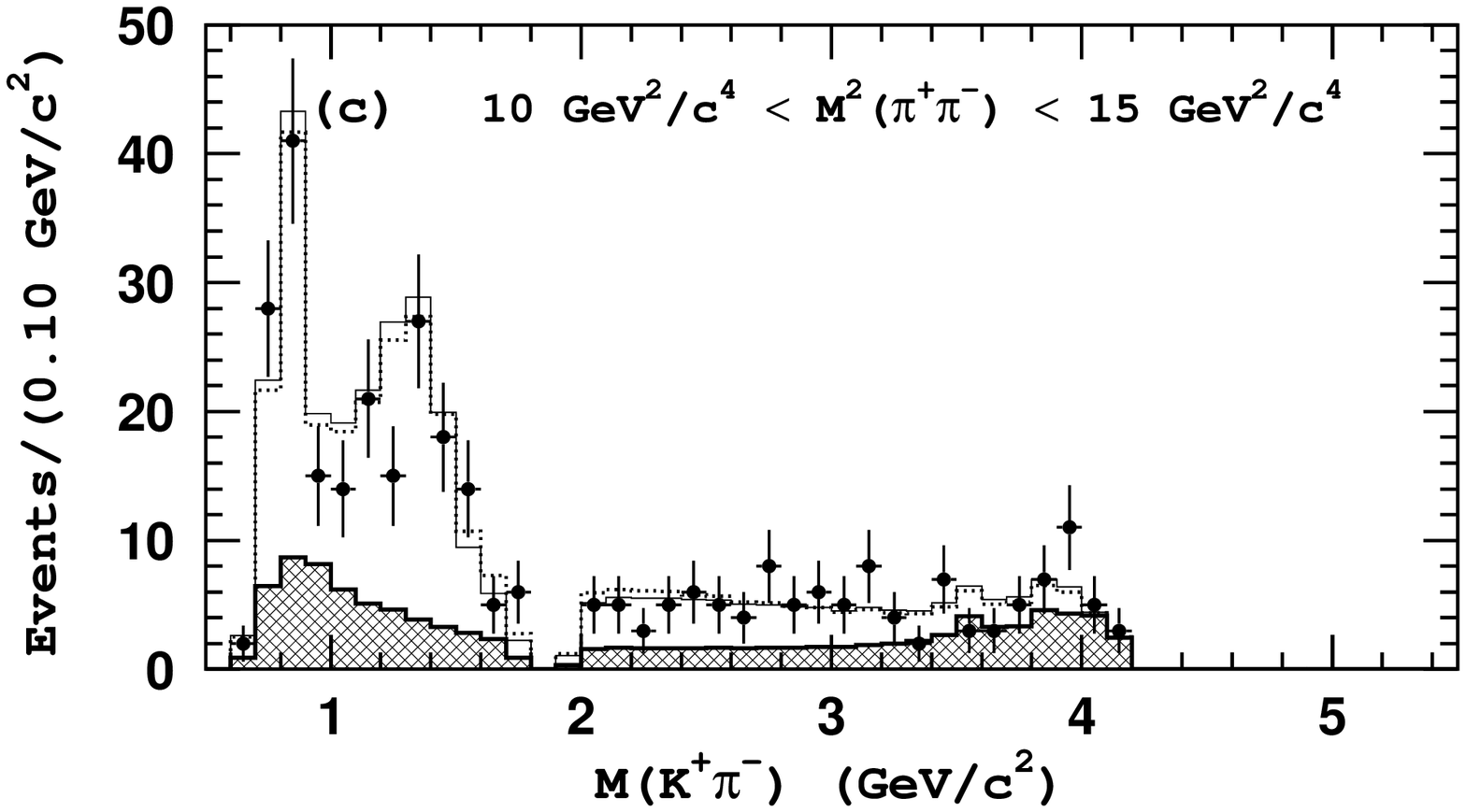} \hfill
 \includegraphics[width=0.48\textwidth,height=45mm]{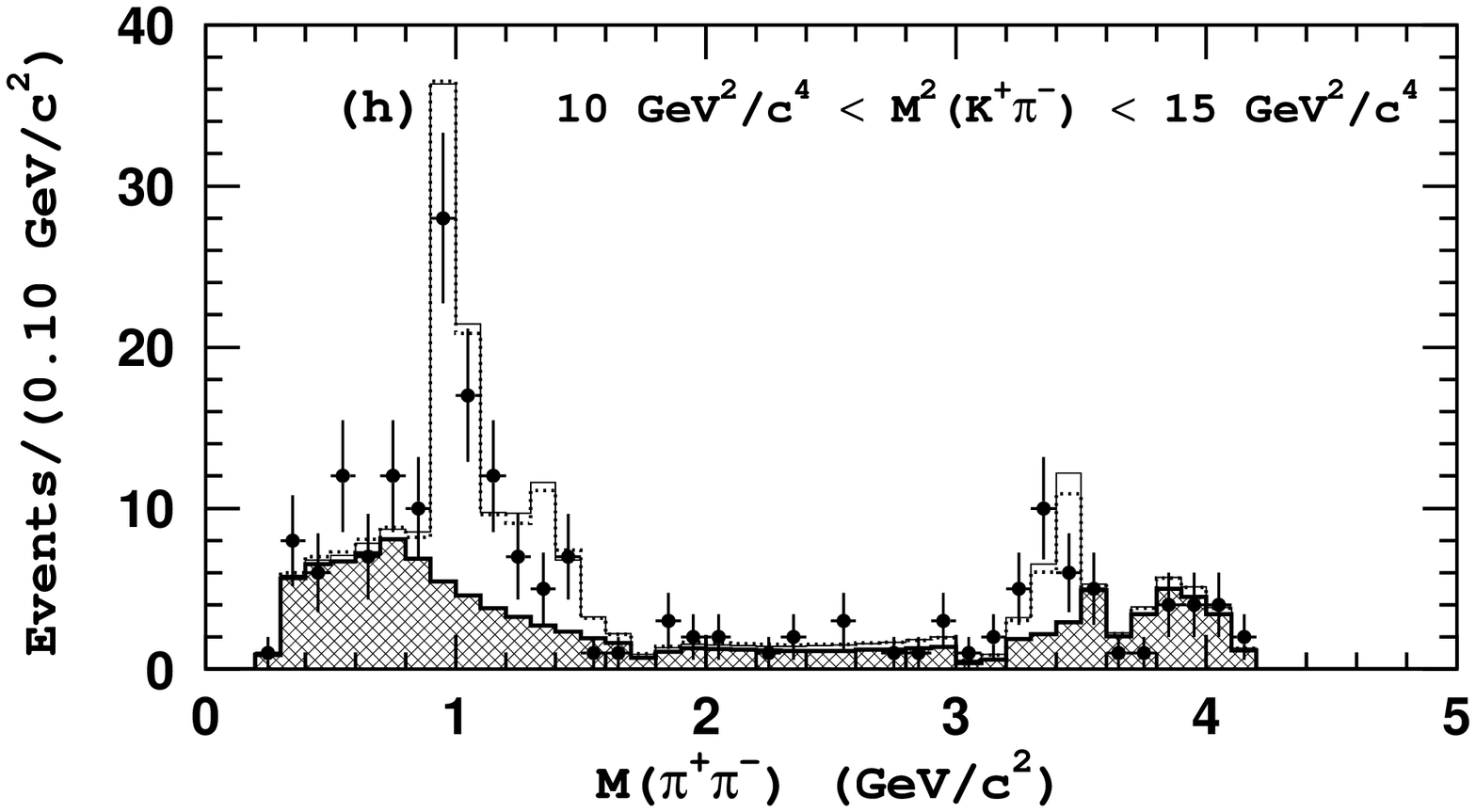} \vspace*{-5mm}\\
 \includegraphics[width=0.48\textwidth,height=45mm]{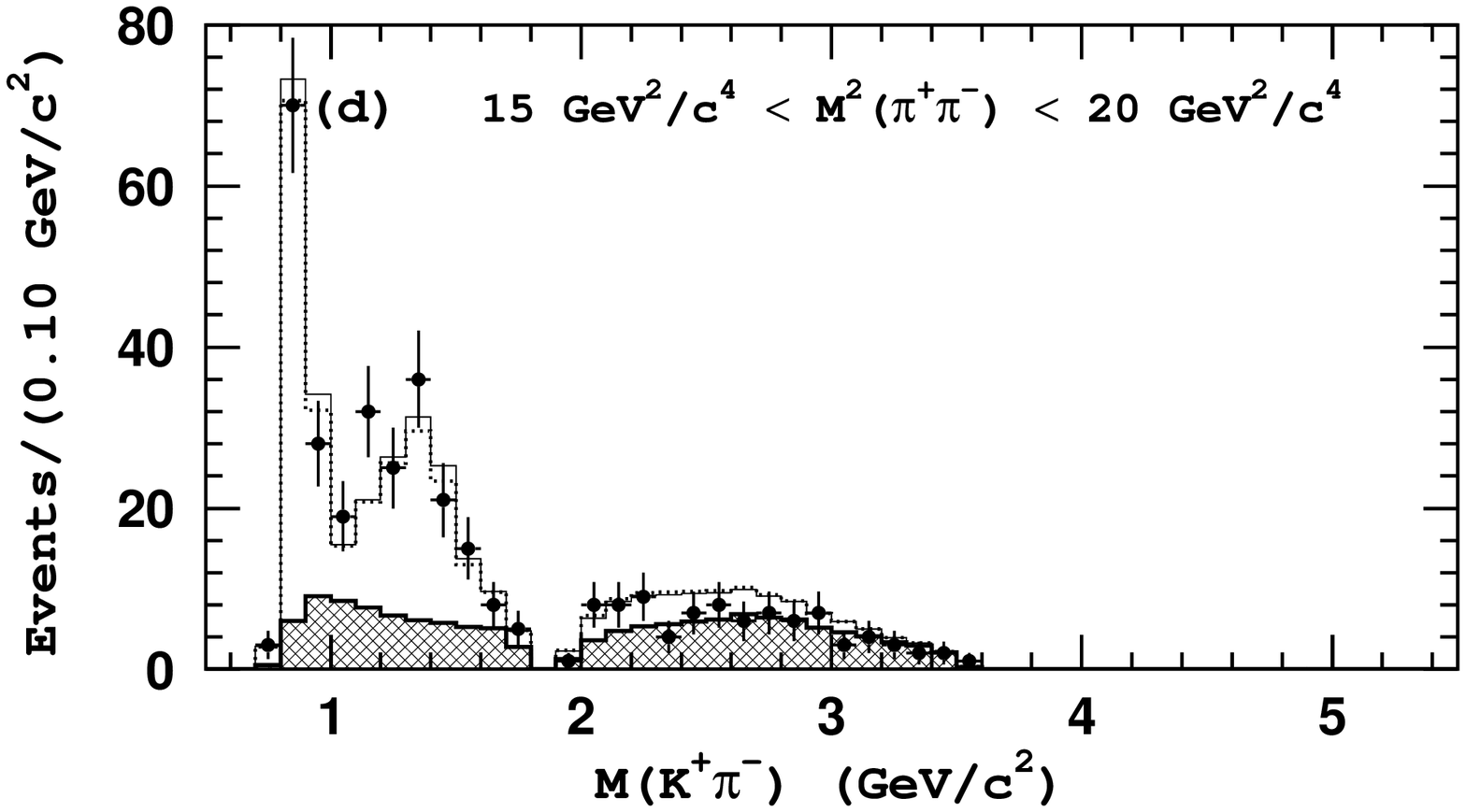} \hfill
 \includegraphics[width=0.48\textwidth,height=45mm]{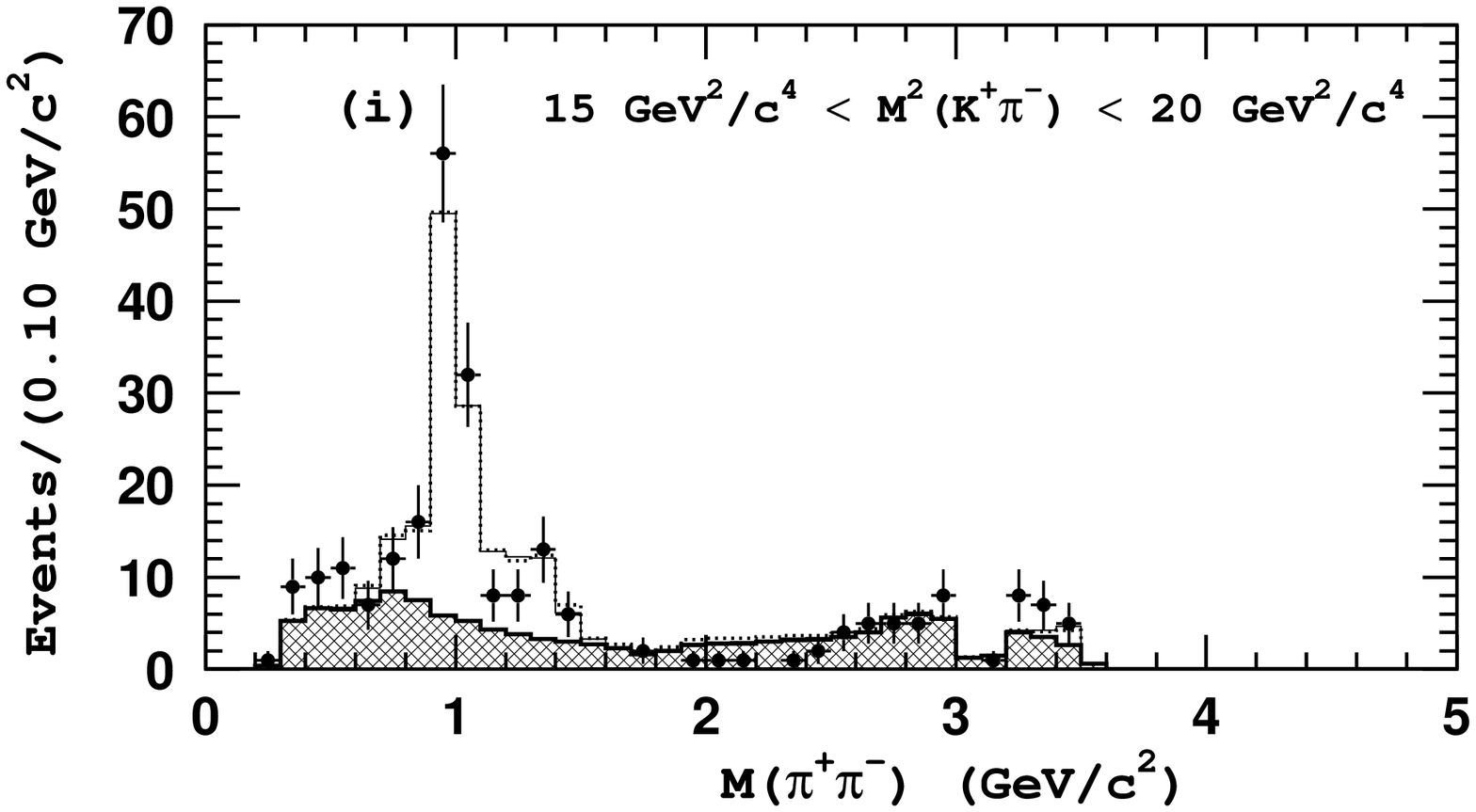} \vspace*{-5mm}\\
 \includegraphics[width=0.48\textwidth,height=45mm]{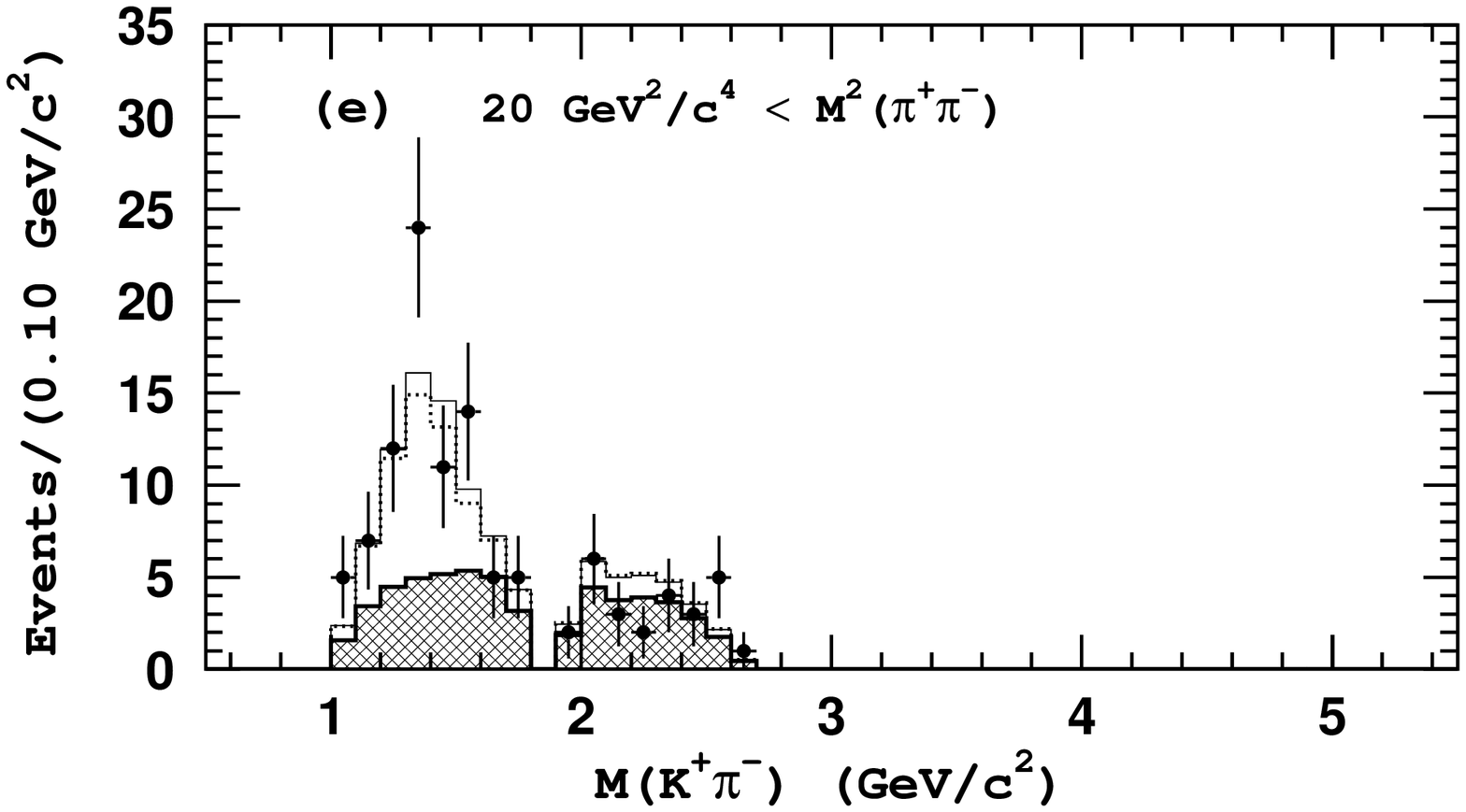} \hfill
 \includegraphics[width=0.48\textwidth,height=45mm]{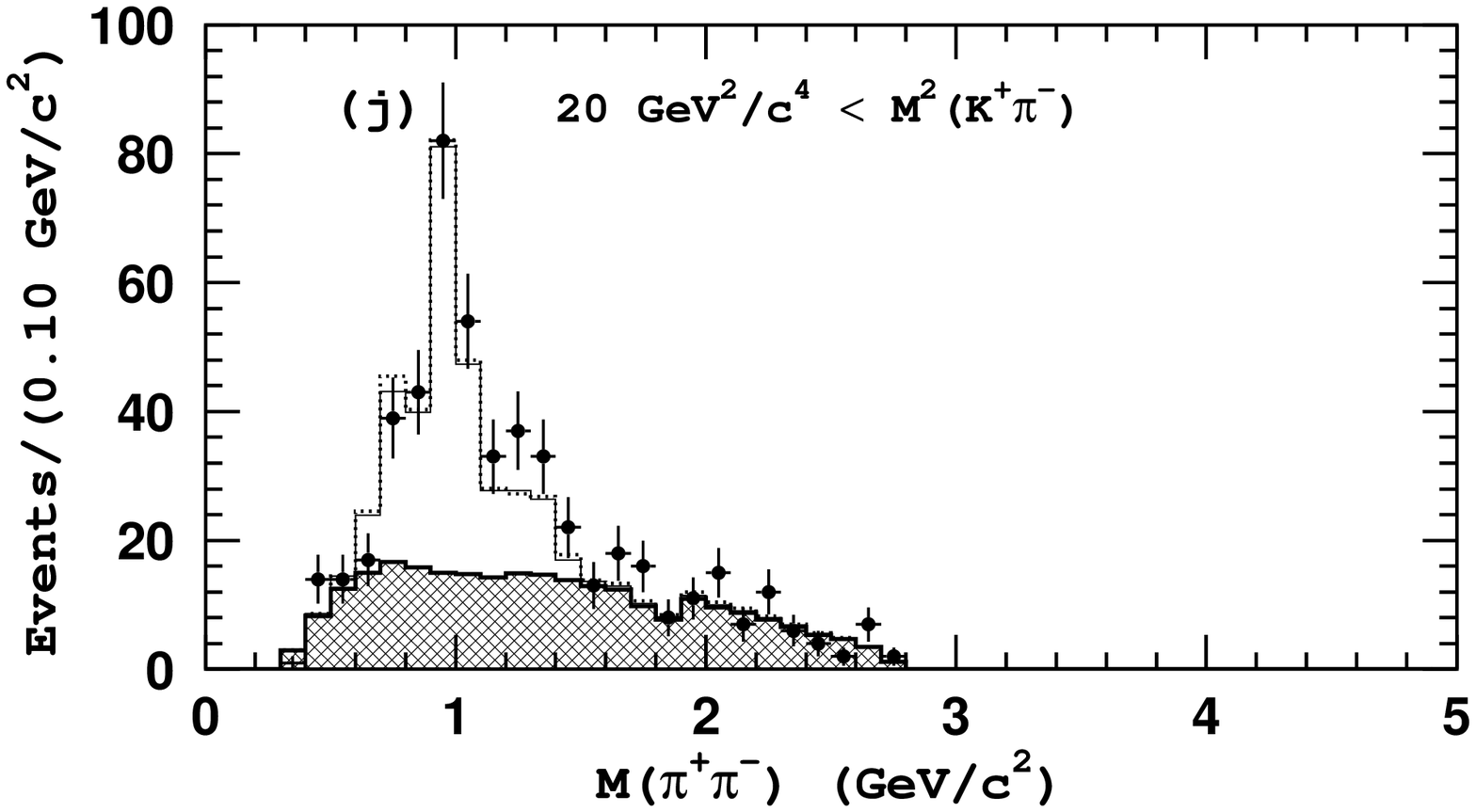} \vspace*{-3mm}
  \caption{$M(\kcpi)$ ($M(\pipi)$) distributions in slices of $M(\pipi)$
           ($M(\kcpi)$).
           Points with error bars are data, the open histograms are the fit
           results with model $\Kpp-$C$_0$ and the hatched histogram is the
           background component. Solid and dotted histograms correspond to
           Solution~1 and Solution~2, respectively 
           (see Table~\ref{tab:kpp-fit-res} and text for details). }
\label{fig:kpp-slices}
\end{figure}

The mass and width of the $f_X(1300)$ state obtained from the fit with 
model $\Kpp-$C$_0$ are consistent with those for the $f_0(1370)$~\cite{PDG}.
If a tensor amplitude is used for the $f_X(1300)$ state (model $\Kpp-$C$_2$),
the fit finds $M(f_X(1300))=1.377\pm0.038$~GeV/$c^2$ and
$\Gamma(f_X(1300))=0.085\pm0.031$~GeV/$c^2$ ($-2\log{\cal{L}}=-4013.0$;
$\chi^2/N_{\rm bin}=103.8/106$), which disagree with the world average
$f_2(1270)$ parameters~\cite{PDG}. In the case of a vector amplitude (model
$\Kpp-$C$_1$), the fit gives $M(f_X(1300))=1.330\pm0.019$~GeV/$c^2$ and
$\Gamma(f_X(1300))=0.210\pm0.048$~GeV/$c^2$ ($-2\log{\cal{L}}=-4048.1$;
$\chi^2/N_{\rm bin}=105.5/106$). Based on this study, we choose model
$\Kpp-$C$_0$ as our default and obtain all the final results for the $\bckpp$
decay using this model. Figure~\ref{fig:kpp-slices} shows the $M(\kcpi)$
($M(\pipi)$) distributions in slices of $M^2(\pipi)$ ($M^2(\kcpi)$) to allow a
more detailed comparison of the  data and fit results with model $\Kpp-$C$_0$.

\begin{figure}[t]
  \centering
  \includegraphics[width=0.48\textwidth]{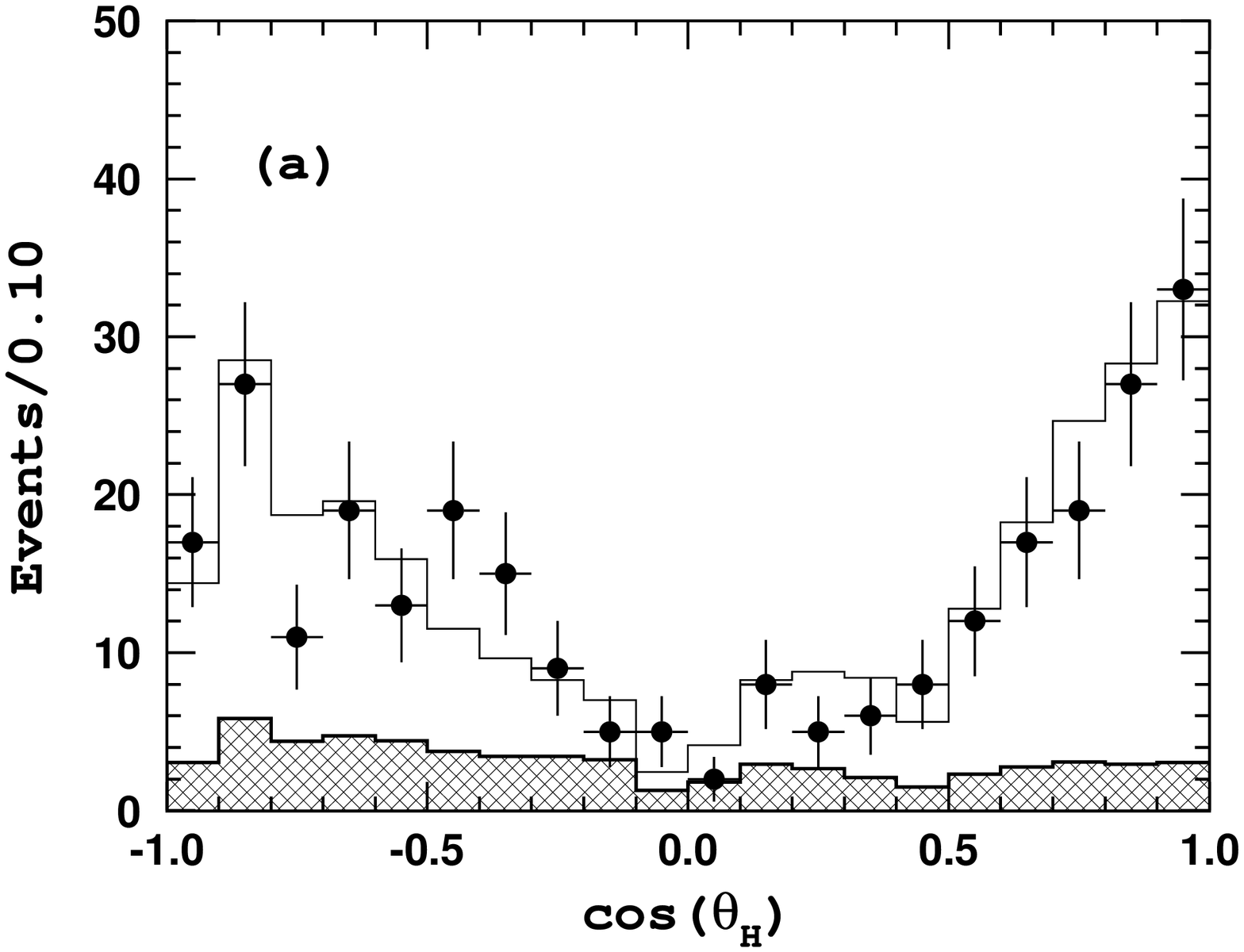} \hfill
  \includegraphics[width=0.48\textwidth]{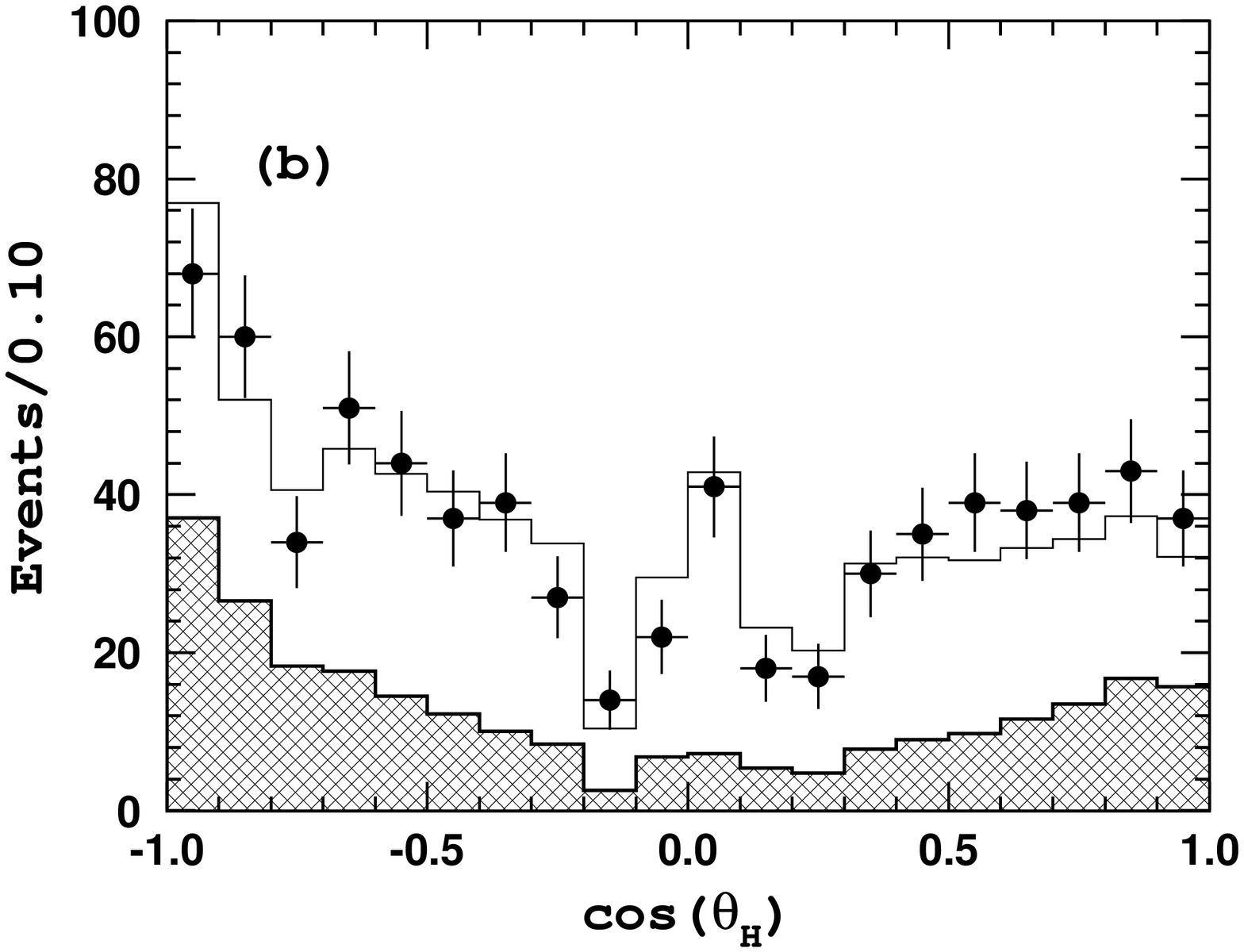} \\
  \includegraphics[width=0.48\textwidth]{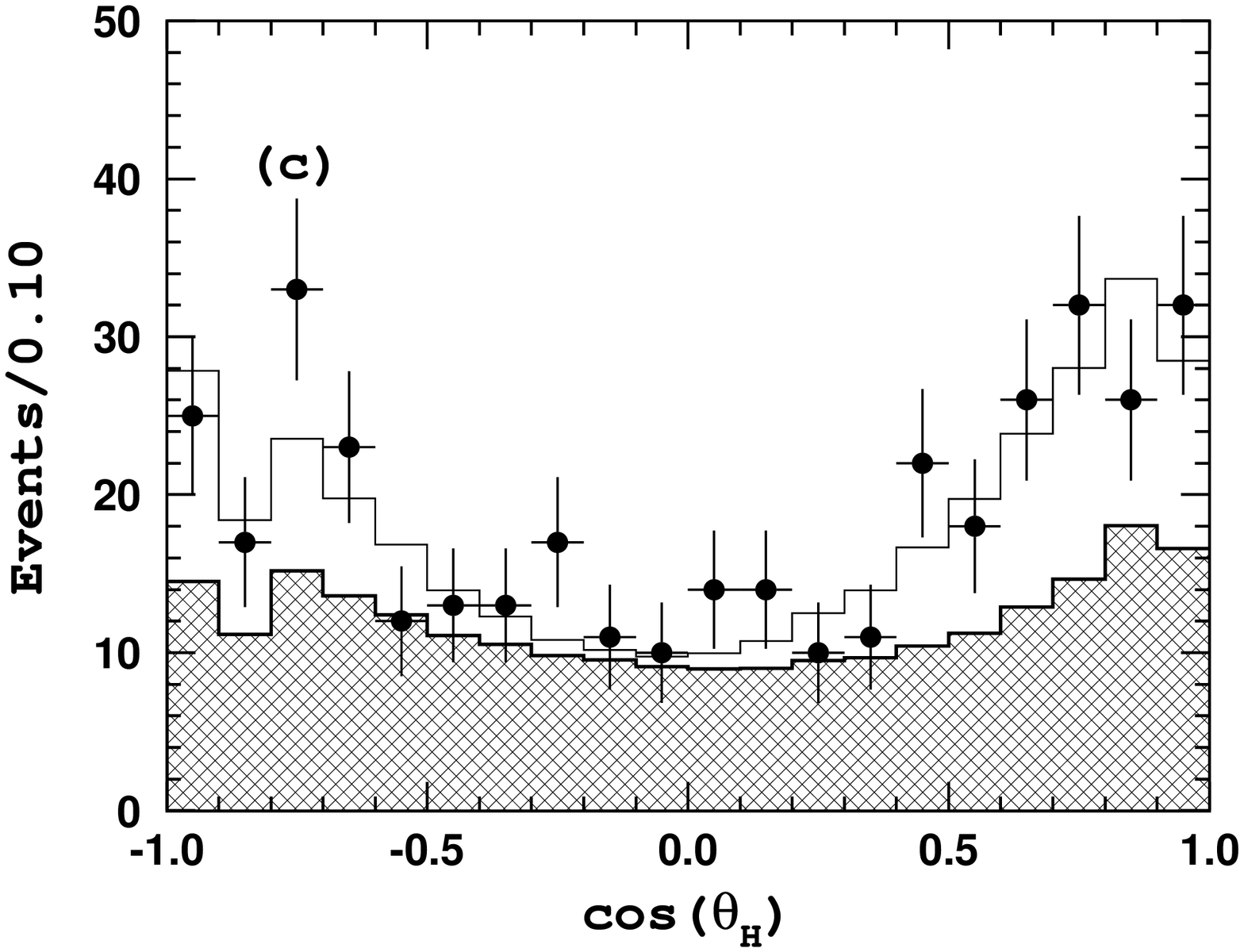} \hfill
  \includegraphics[width=0.48\textwidth]{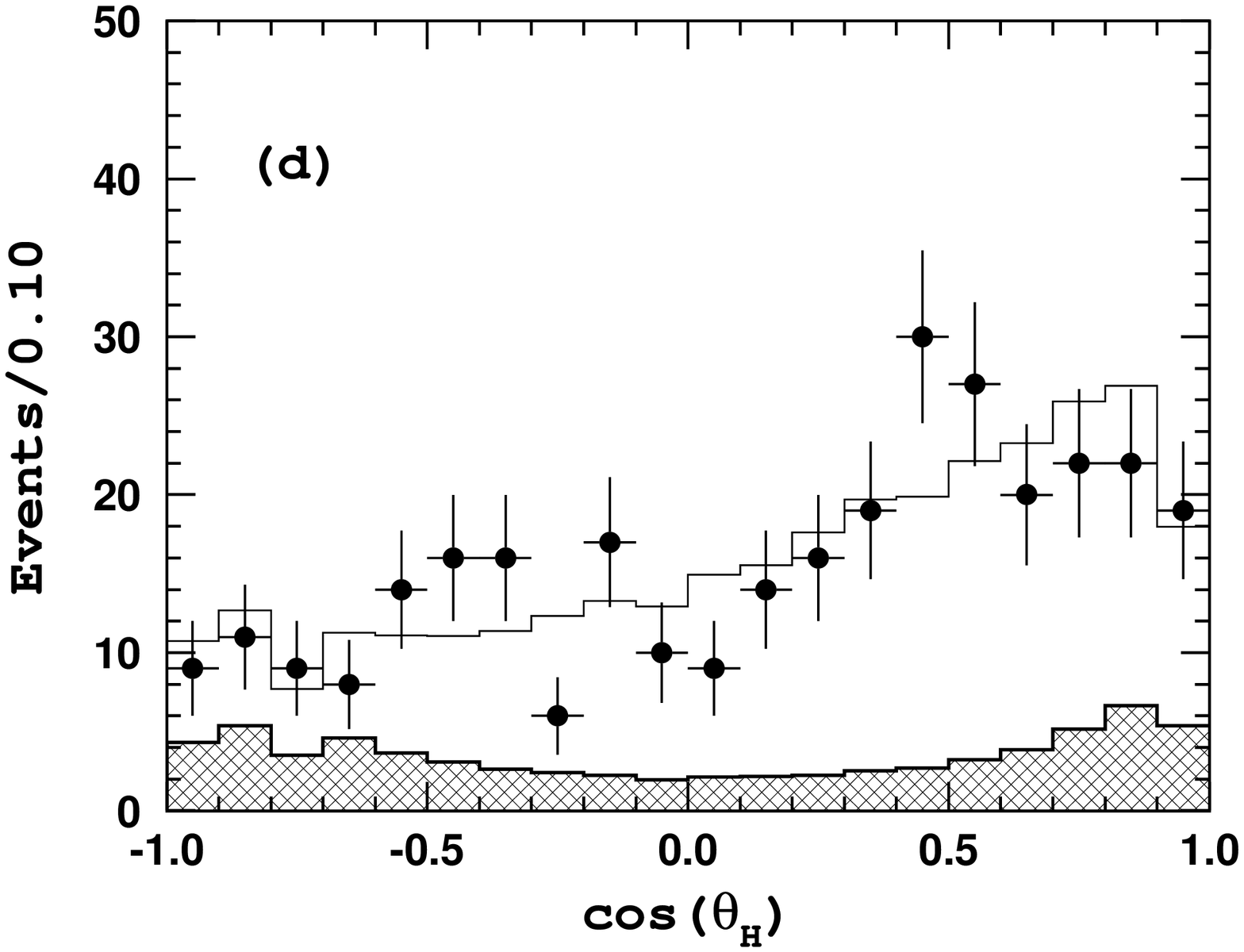}
  \caption{Helicity angle distributions for $\kpp$ events in different regions:
           (a)~$K^*(892)^0$ (0.82~\Mass~$<M(\kcpi)<$~0.97~\Mass);
           (b)~$K^*_0(1430)^0$ (1.0~\Mass~$<M(\kcpi)<$~1.76~\Mass);
           (c)~$\rho(770)^0$ ($M(\pipi)<$~0.90~\Mass) and
           (d)~$f_0(980)$ (0.90~\Mass~$<M(\pipi)<$~1.06~\Mass).
           Points with error bars are data, the open histogram is the fit
           result with model $\Kpp-$C$_0$ and the hatched histogram is the
           background component.  Visible irregularities are due to
           vetoes applied on invariant masses of two-particle combinations.}
\label{fig:kpp-heli}
\end{figure}

In addition to the two-particle invariant mass distributions shown in
Figs.~\ref{fig:kpp-mods} and~\ref{fig:kpp-slices}, the helicity angle
distributions for several regions are shown in Fig.~\ref{fig:kpp-heli}.
(For the $\pipi$ combination the helicity angle is defined in a similar
way as for $\kcpi$ combination.) All plots shown in Fig.~\ref{fig:kpp-heli}
demonstrate good agreement between data and the fit.

To test for the contribution of other possible quasi-two-body intermediate
states such as $K^*(1410)^0\pi^+$, $K^*(1680)^0\pi^+$, $K^*_2(1430)^0\pi^+$
or $f_2(1270)K^+$, we include an additional amplitude of each of these
channels in model $\Kpp-$C$_0$ and repeat the fit to data.
None of these channels have a statistically significant signal.


To estimate the model dependent uncertainty in the branching fractions for
individual quasi-two-body channels, we use the results of fits obtained with
models $\Kpp-$B$_J$ and $\Kpp-$C$_J+R$, where $R$ is one of an additional
resonances mentioned above. We also use different parametrizations of the
non-resonant amplitude to estimate the related uncertainty. We try the
following alternative parametrizations:
\begin{itemize}
 \item{ $\Am_{\rm nr}(\kpp) =
          a^{\rm nr}_1e^{-\alpha \sft }e^{i\delta^{\rm nr}_1}+
          a^{\rm nr}_2e^{-\alpha \sst}e^{i\delta^{\rm nr}_2}+
          a^{\rm nr}_3e^{-\alpha \sfs}e^{i\delta^{\rm nr}_3}$;}
 \item{ $\Am_{\rm nr}(\kpp)=
         \frac{a_1^{\rm nr}}{\sft^\alpha}e^{i\delta_1^{\rm nr}}+
         \frac{a_2^{\rm nr}}{\sst^\alpha}e^{i\delta_2^{\rm nr}}$;}
 \item{ $\Am_{\rm nr}(\kpp)=a^{\rm nr}e^{i\delta^{\rm nr}}$.}
\end{itemize}


While fitting the data with model $\Kpp-$C$_0$, we found that two solutions
with very similar likelihood values exist. The comparison between the two
solutions and the data are shown in Fig.~\ref{fig:kpp-slices}. The main
difference between these two solutions is the relative fractions of the total
$\bckpp$ signal ascribed to the $B^+\to K^*_0(1430)^0\pi^+$ decay: the
fraction of this channel changes by a factor of about five. The reason for
the existence of the second solution is similar behavior of the two amplitudes
(the non-resonant component parametrized by Eq.(\ref{eq:kpp-non-res}) and the
scalar $K^*_0(1430)^0\pi^+$ amplitude) as functions of $M^2(\kcpi)$. Due to the
large width of the $K^*_0(1430)^0$ resonance, these two amplitudes can be, to a
large extent, interchanged providing a nearly identical description of the
data. An even stronger effect is observed in the case of model $\Kpp-$B$_0$
when the mass and width of the ``$\kappa$'' resonance is allowed to float.
In this case the two amplitudes are almost identical. A similar behavior is
observed for all the parametrizations used to describe the non-resonant
amplitude. The existence of secondary maxima of the likelihood function is
confirmed with MC simulation (see Section~\ref{sec:discussion} for the more
detailed discussion).


\subsection{Fitting the $\bckkk$ Signal}
\label{sec:kkk-sig}

\begin{figure}[p]
  \centering
  \includegraphics[width=0.48\textwidth]{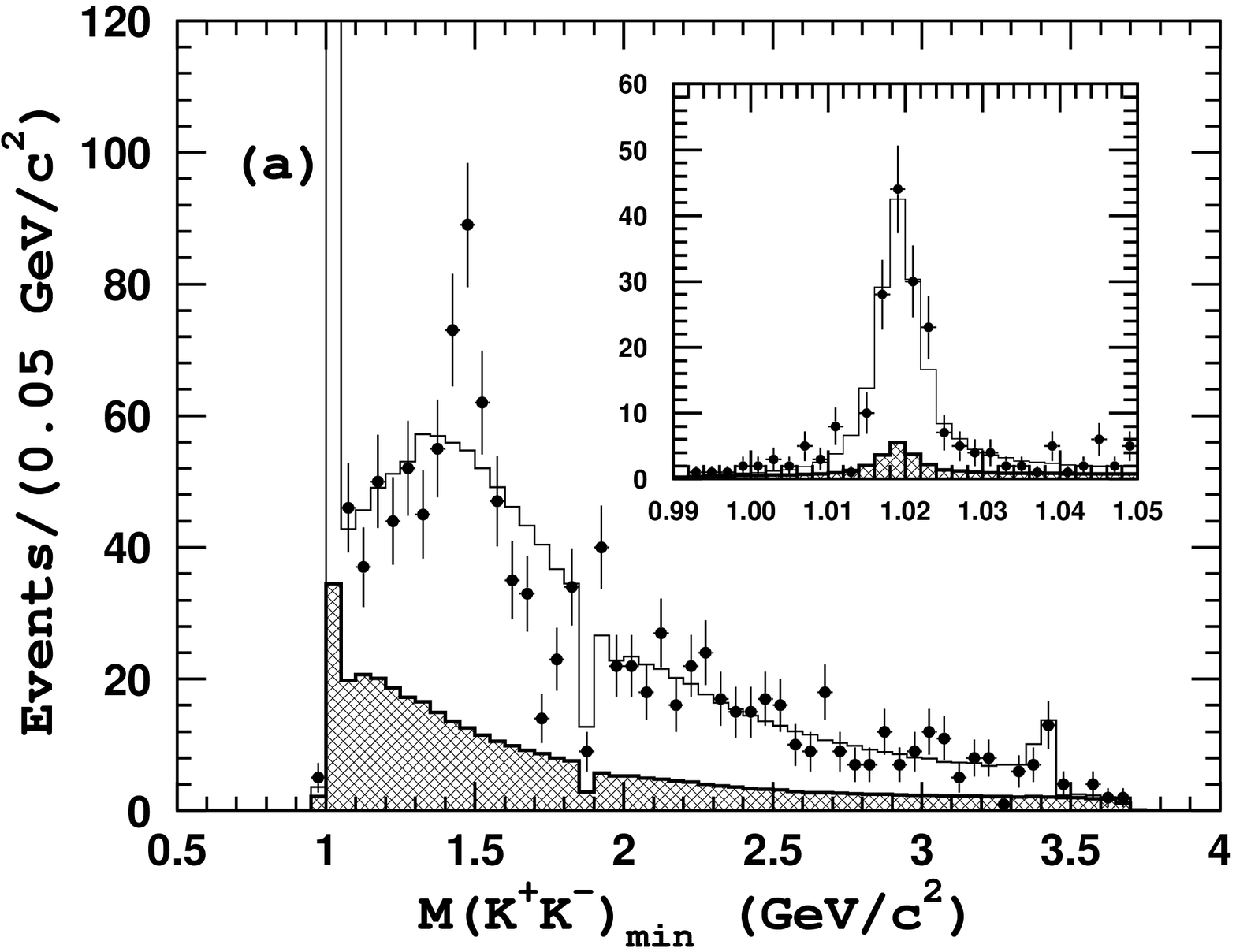} \hfill
  \includegraphics[width=0.48\textwidth]{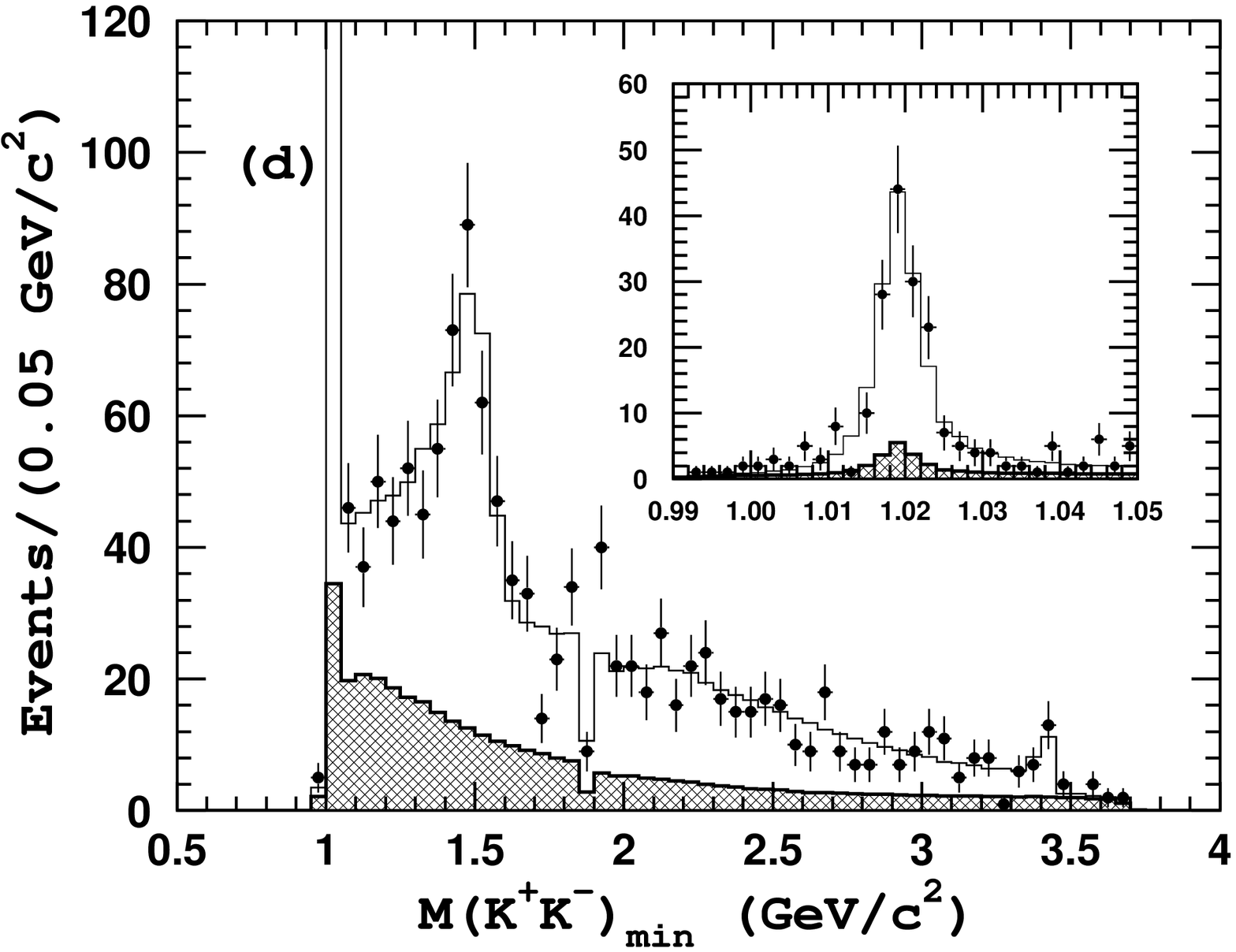} \\
  \includegraphics[width=0.48\textwidth]{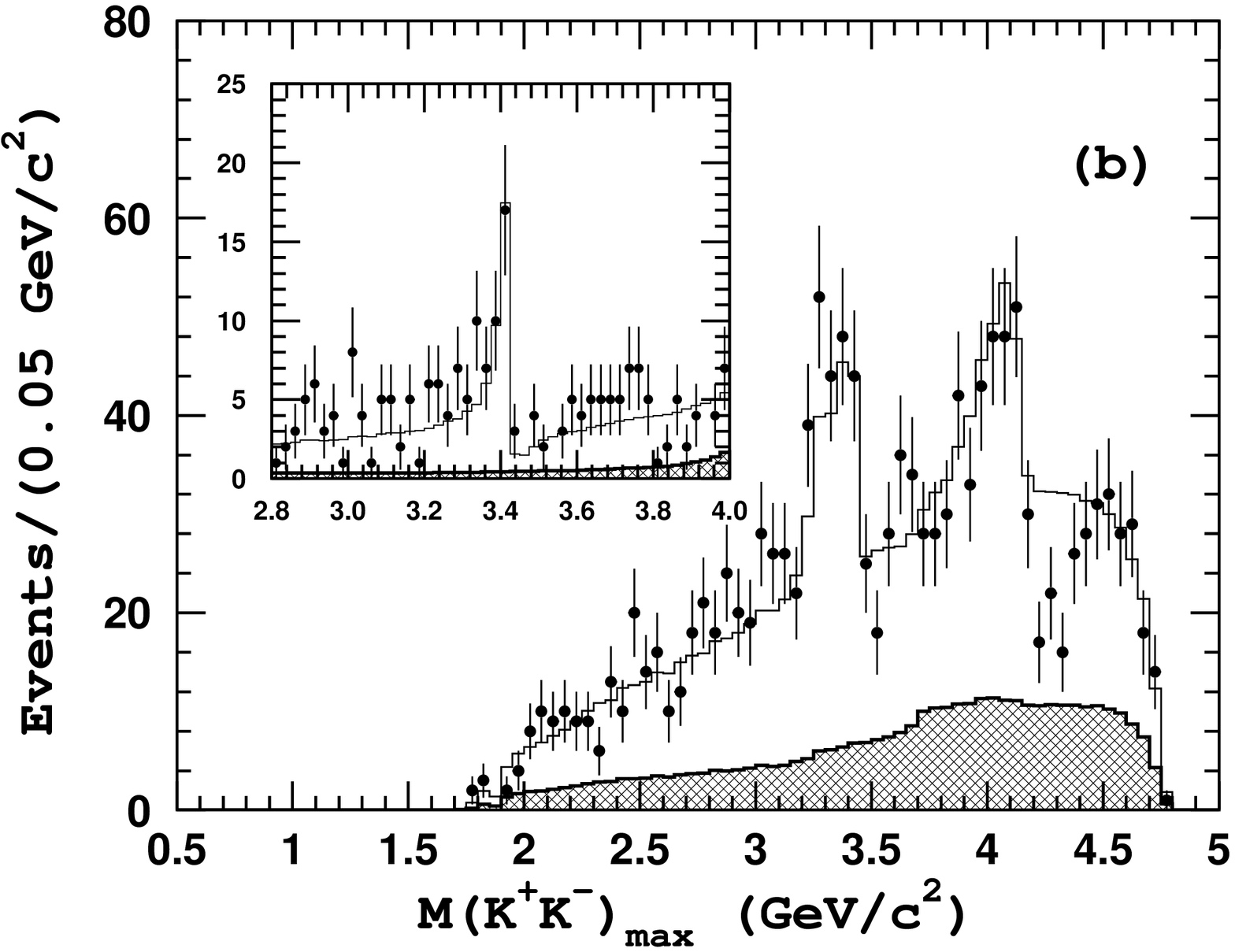} \hfill
  \includegraphics[width=0.48\textwidth]{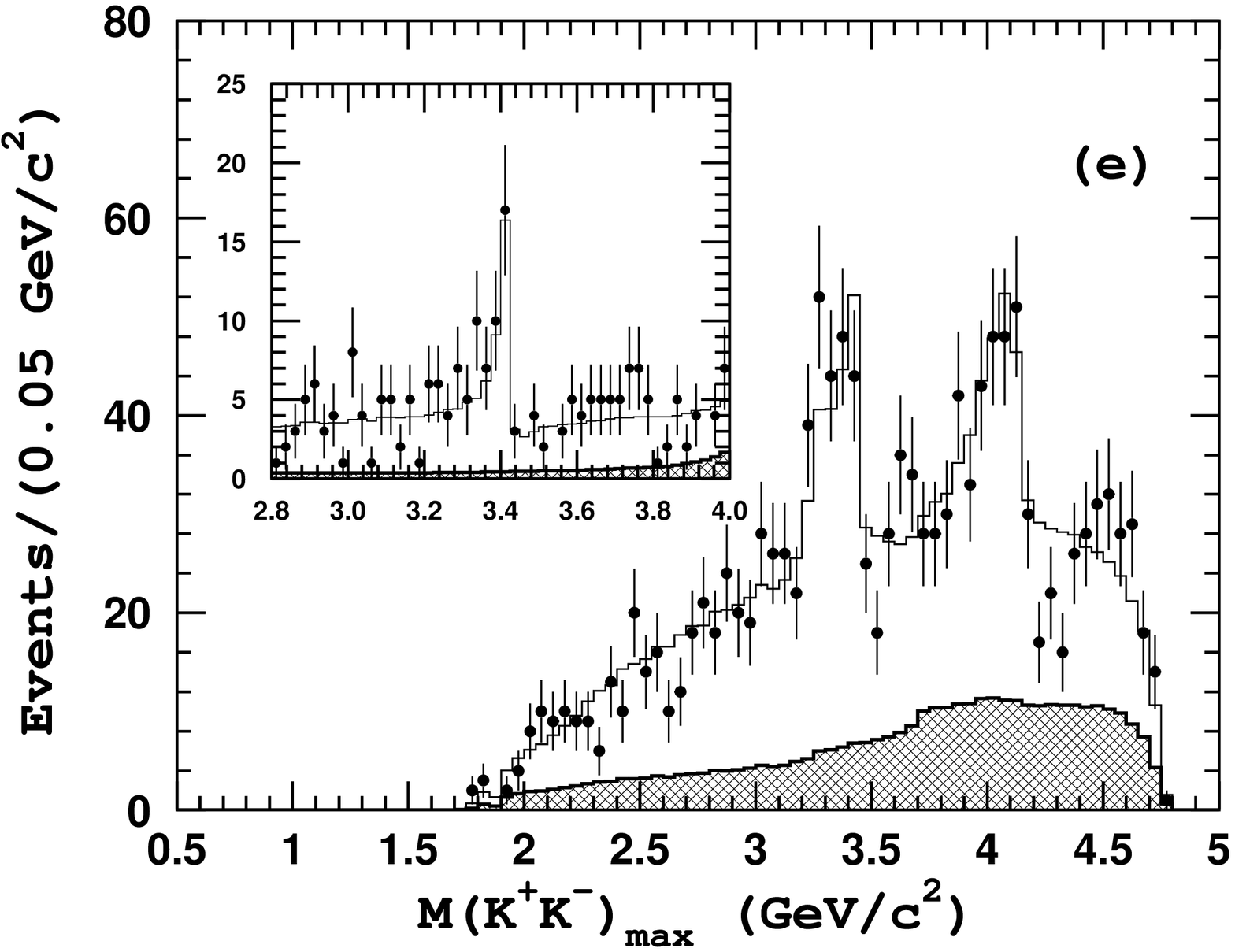} \\
  \includegraphics[width=0.48\textwidth]{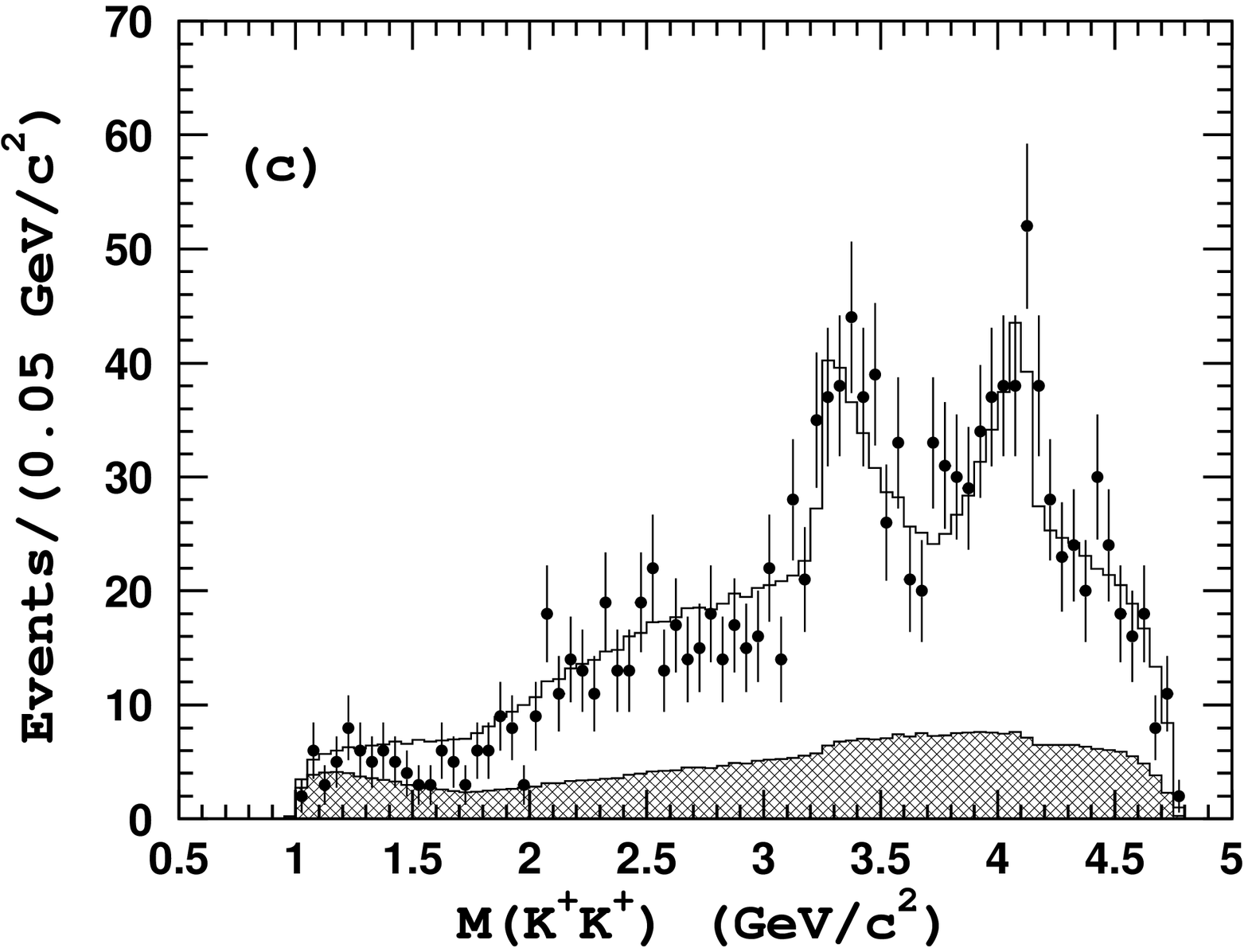} \hfill
  \includegraphics[width=0.48\textwidth]{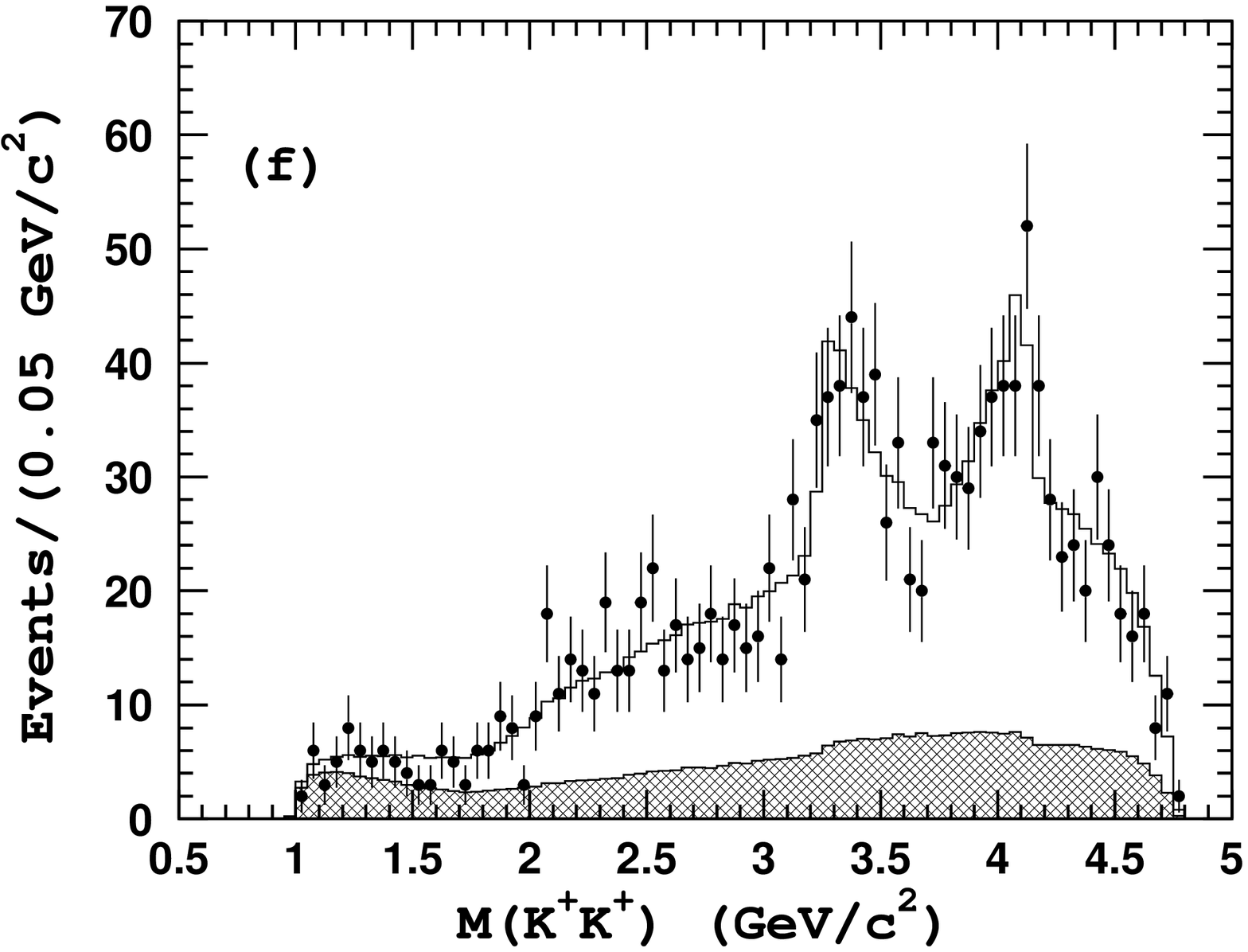}
  \caption{Results of the fit with the model $\KKK-$A$_0$ (left column) and 
           $\KKK-$B$_0$
           (right column) to events in the signal region. Points with
           error bars are data, the open histogram is the fit result and
           hatched histogram is the background component. Insets in (a) and (d)
           show the $\phi(1020)$ mass region in 2~\mass~ bins. Insets
           in (b) and (e) show the $\chic$ mass region in 25~\mass~ bins
           with an additional requirement 2.0~\Mass~$<\mkkmin<$3.4~\Mass.}
\label{fig:kkk-mods}
\end{figure}

\begin{figure}[t]
  \centering
  \includegraphics[width=0.48\textwidth]{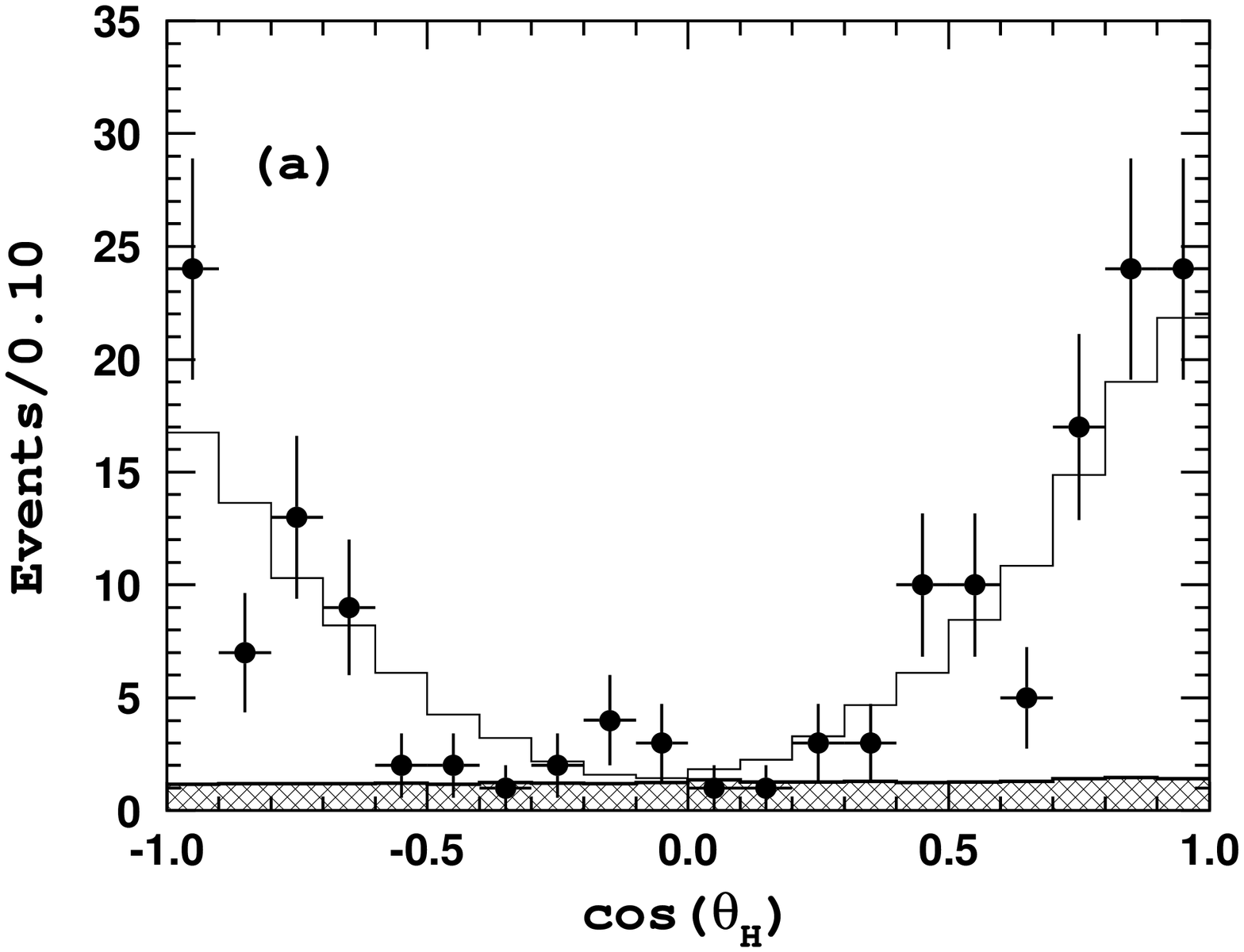} \hfill
  \includegraphics[width=0.48\textwidth]{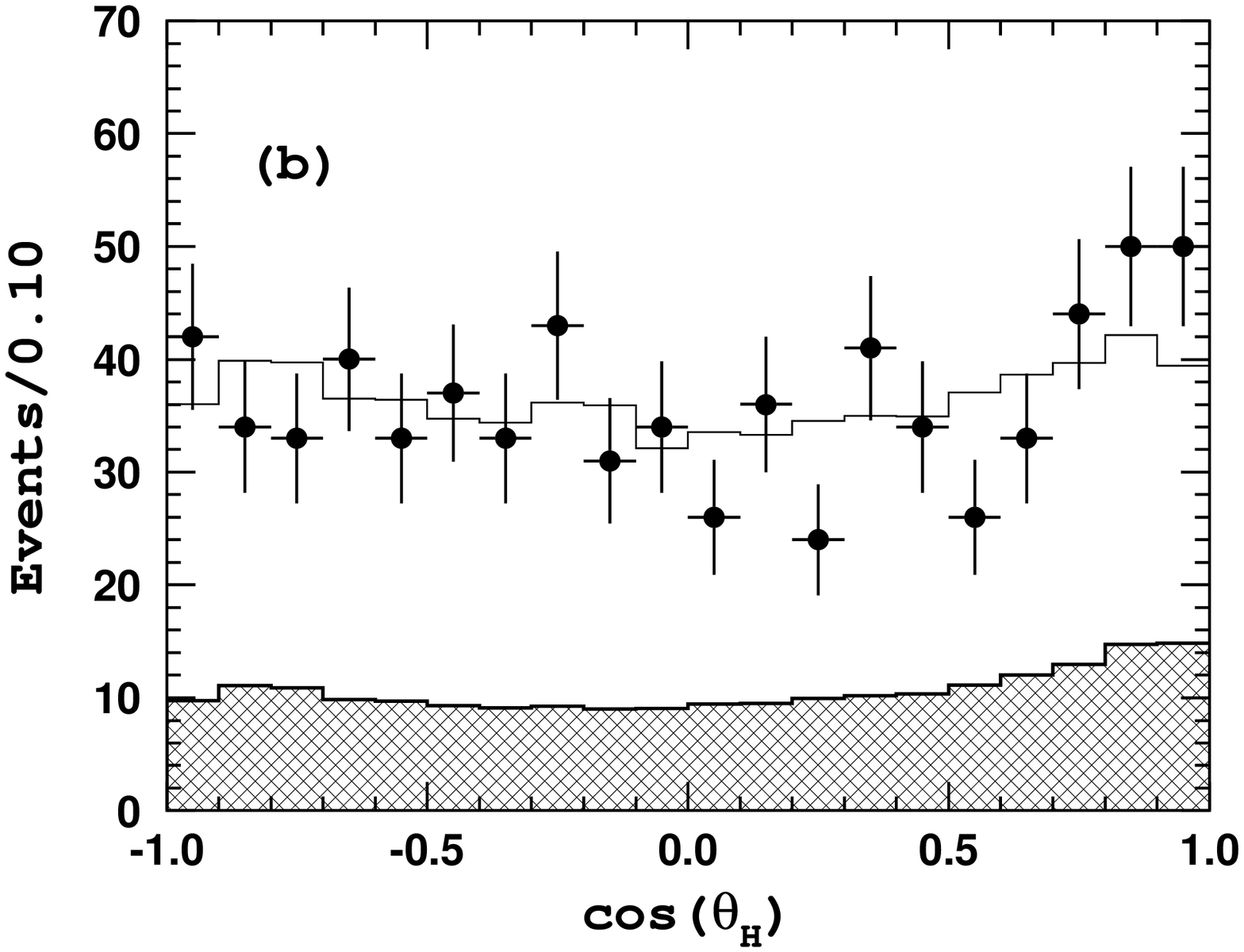}
  \caption{Helicity angle distributions for events in the (a)~$\phi$ mass
           region ($\mkkmin<$~1.05~\Mass) and in the (b)~$f_X(1500)$ region
           (1.05~\Mass~$<\mkkmin<$~3.0~\Mass). Points with error bars
           are data, the open histogram is the fit result with model
           $\KKK-$B$_0$ and the hatched histogram is the background component.}
\label{fig:kkk-heli}
\end{figure}

   The Dalitz plot for $\kkk$ events in the signal region is shown in 
Fig.~\ref{fig:khh-dp-sig}(b). There are 1400 events in the signal region. In
the analysis of the $\kkk$ final state we follow the same strategy as in the
case of the $\kpp$ state.
In an attempt to describe all the features in the $\kpkm$ mass spectrum
mentioned in Section~\ref{sec:khh}, we start with the following minimal
matrix element of the $\bckkk$ decay (referred to as model $\KKK-$A$_J$):
\begin{eqnarray}
S_A(\kkk) &=& a_{\phi}e^{i\delta_{\phi}}\left(\Am_1(K^+_1K^+_2K^-|\phi) 
                         ~+~ \Am_1(K^+_2K^+_1K^-|\phi)\right) \nonumber \\
          &+&  a_{\chic}e^{i\delta_{\chic}}\left(\Am_0(K^+_1K^+_2K^-|\chic)
                         ~+~ \Am_0(K^+_2K^+_1K^-|\chic)\right) \nonumber \\
          &+&  a_{f_X}e^{i\delta_{f_X}}\left(\Am_J(K^+_1K^+_2K^-|f_X)
                         ~+~ \Am_J(K^+_2K^+_1K^-|f_X)\right),
\label{eq:kkk-modA}
\end{eqnarray}
where the subscript $J$ denotes the unknown spin of the $f_X(1500)$ resonance;
amplitudes $a_i$, relative phases $\delta_i$, mass and width of the $f_X(1500)$
%
%
\begin{table}[t]
\caption{Summary of fit results to $\kkk$ events in the signal region. The two
         values given for model $\KKK-$B$_0$ correspond to the two
         solutions (see text for details).}
\medskip
\label{tab:kkk-fit-res}
\centering
  \begin{tabular}{lr|cc} \hline \hline
  \multicolumn{2}{c}{Parameter} &   \multicolumn{2}{|c}{Model}   \\ 
&
& \hspace*{2mm} $\KKK-$A$_0$ \hspace*{2mm} 
& \hspace*{1mm} $\KKK-$B$_0$ \hspace*{1mm} \\
& & & Solution 1/Solution 2
\\
\hline \hline  
 $\phi(1020)K^+$         & 
  fraction, \%~~         &  $14.0\pm1.2$ &   $14.7\pm1.3$/$15.2\pm1.3$   
\\
&  phase,~$^\circ$~~     &   $-17\pm11$  &    $-123\pm10$/$-200\pm10$    
\\ \hline  
 $f_X(1500)K^+$          & 
   fraction, \%~~        &  $83.3\pm2.5$ &   $63.4\pm6.9$/$8.21\pm1.94$  
\\
&  phase,~$^\circ$~~     &     \multicolumn{2}{|c}{ $0~~$ (~f~i~x~e~d~)}
\\
&  Mass, GeV/$c^2$~~     &$1.373\pm0.025$& $1.524\pm0.014$/$1.491\pm0.018$
\\
&  Width, GeV/$c^2$~~    &$0.720\pm0.058$& $0.136\pm0.023$/$0.145\pm0.029$
\\ \hline  
 $\chi_{c0}K^+$          &
   fraction, \%~~        & $4.48\pm1.4$  &$2.67\pm0.82$/$8.01\pm1.35$ 
\\
&  phase,~$^\circ$~~     &  $165\pm15$   &  $-118\pm15$/$127\pm10$    
\\ \hline  
 Non-Resonant            &
   fraction, \%~~        &       $-$     & $74.8\pm3.6$/$65.1\pm5.1$  
\\
&  phase,~$^\circ$~~     &       $-$     &    $-68\pm9$/$61\pm10$     
\\
&  $\alpha$~~            &       $-$     &$0.121\pm0.014$/$0.116\pm0.015$
\\ \hline  
 Charmless Total$^{a}$   &
\\
&  fraction, \%~~        &  $96.0\pm0.7$ & $95.2\pm1.0$/$95.6\pm0.9$ 
\\ \hline \hline  
 \multicolumn{2}{c|}{$-2\ln{\cal{L}}$}       &   $-2140.4$   &   $-2218.2$/$-2177.4$     
\\ \hline  
 \multicolumn{2}{c|}{$\chi^2$}               &    $65.0$     &      $43.3$/$57.1$        
\\
 \multicolumn{2}{c|}{$N_{\rm bins}$}         &     $53$      & $53$  
\\
 \multicolumn{2}{c|}{$N_{\rm fit. var.}$}    &     $6$       &      $9$                  
\\ \hline \hline  
\multicolumn{4}{l}{$^{a}$ Here ``Charmless Total'' refers to the
total three-body  $\bckkk$ } \\
\multicolumn{4}{l}{signal excluding the contribution from $B^+\to\chic K^+$.}
  \end{tabular}
\end{table}
resonance are fit parameters. As there are two identical kaons in the final
state, the amplitude in Eq.~(\ref{eq:kkk-modA}) is symmetrized with respect
to $K_1^+\leftrightarrow K_2^+$ interchange. When fitting the data, we choose
the $f_X(1500)K^+$ signal as our reference by fixing its amplitude and phase
($a_{f_X}\equiv 1$ and $\delta_{f_X}\equiv 0$).
Figures~\ref{fig:kkk-mods}(a,b,c) show the two-kaon invariant mass projections
for model $\KKK-$A$_0$ and the data. The numerical values of the fit parameters
are given in Table~\ref{tab:kkk-fit-res}. Although the data are described
relatively well even with this simple matrix element, there is a region
where the agreement is not satisfactory. The enhancement of signal events in
the higher $\kpkm$ mass range visible in Fig.~\ref{fig:kkk-mods}(a) causes
the width of the $f_X(1500)$ state determined from the fit with model
$\KKK-$A$_0$ to be very large. This results in a poor description of the data
in the $M(\kpkm)\simeq1.5$~\Mass~ region, where the peaking structure is
significantly narrower. On the other hand, as for $\bckpp$, the excess of
signal events at high $M(\kpkm)$ may be evidence for non-resonant $\bckkk$
decay.
%
%
%
%
\begin{figure}[p]
  \centering
 \includegraphics[width=0.48\textwidth,height=45mm]{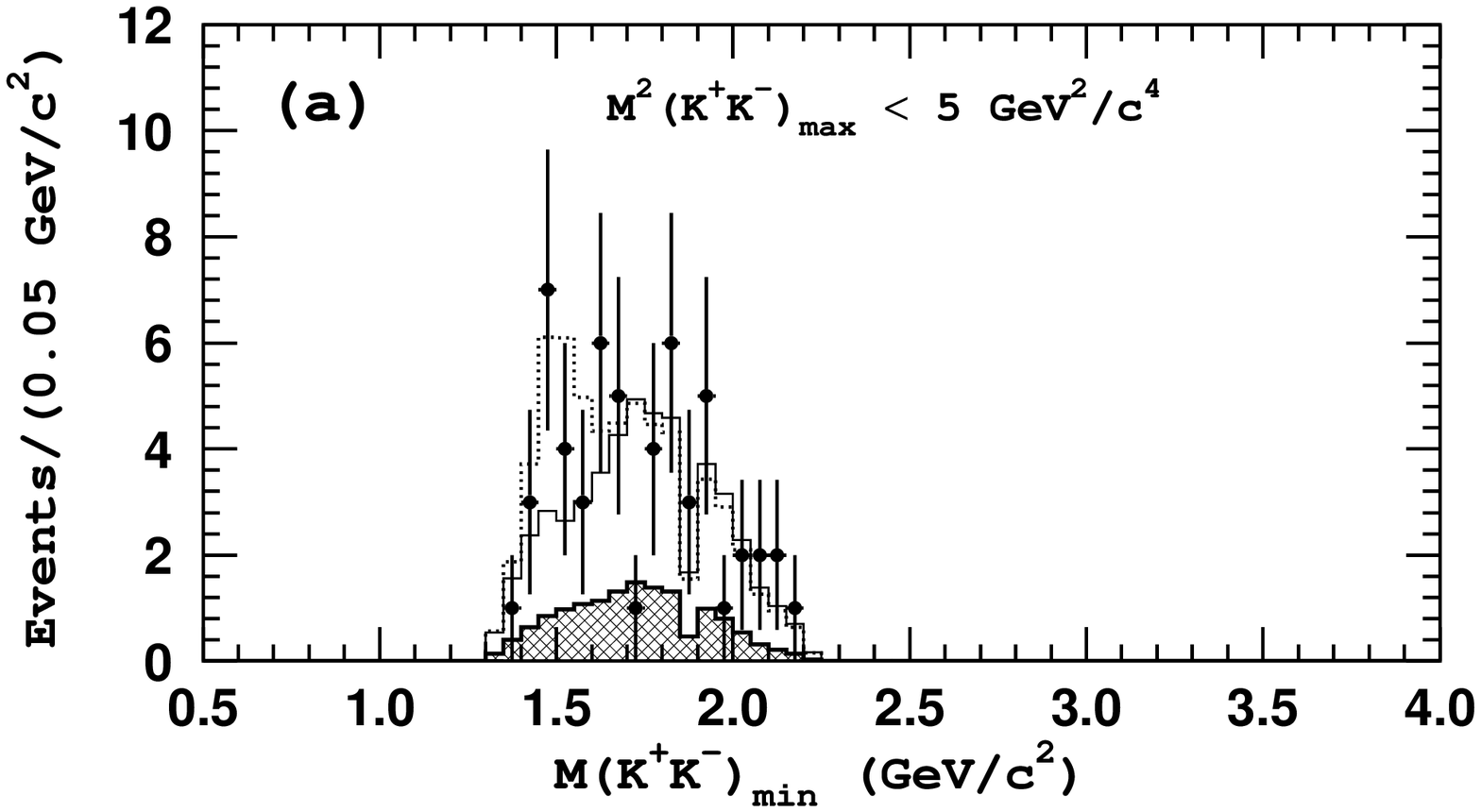} \hfill
 \includegraphics[width=0.48\textwidth,height=45mm]{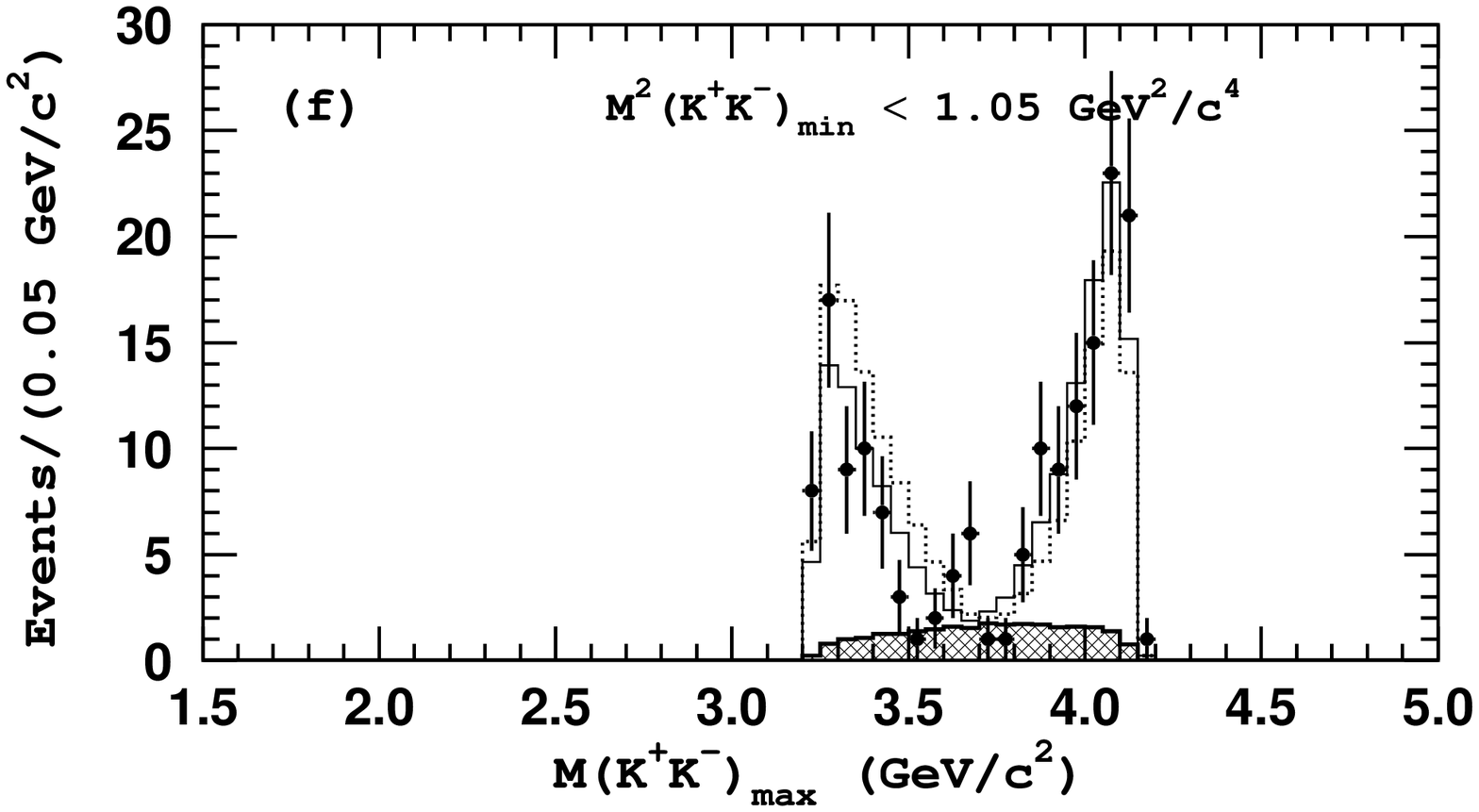} \vspace*{-5mm}\\
 \includegraphics[width=0.48\textwidth,height=45mm]{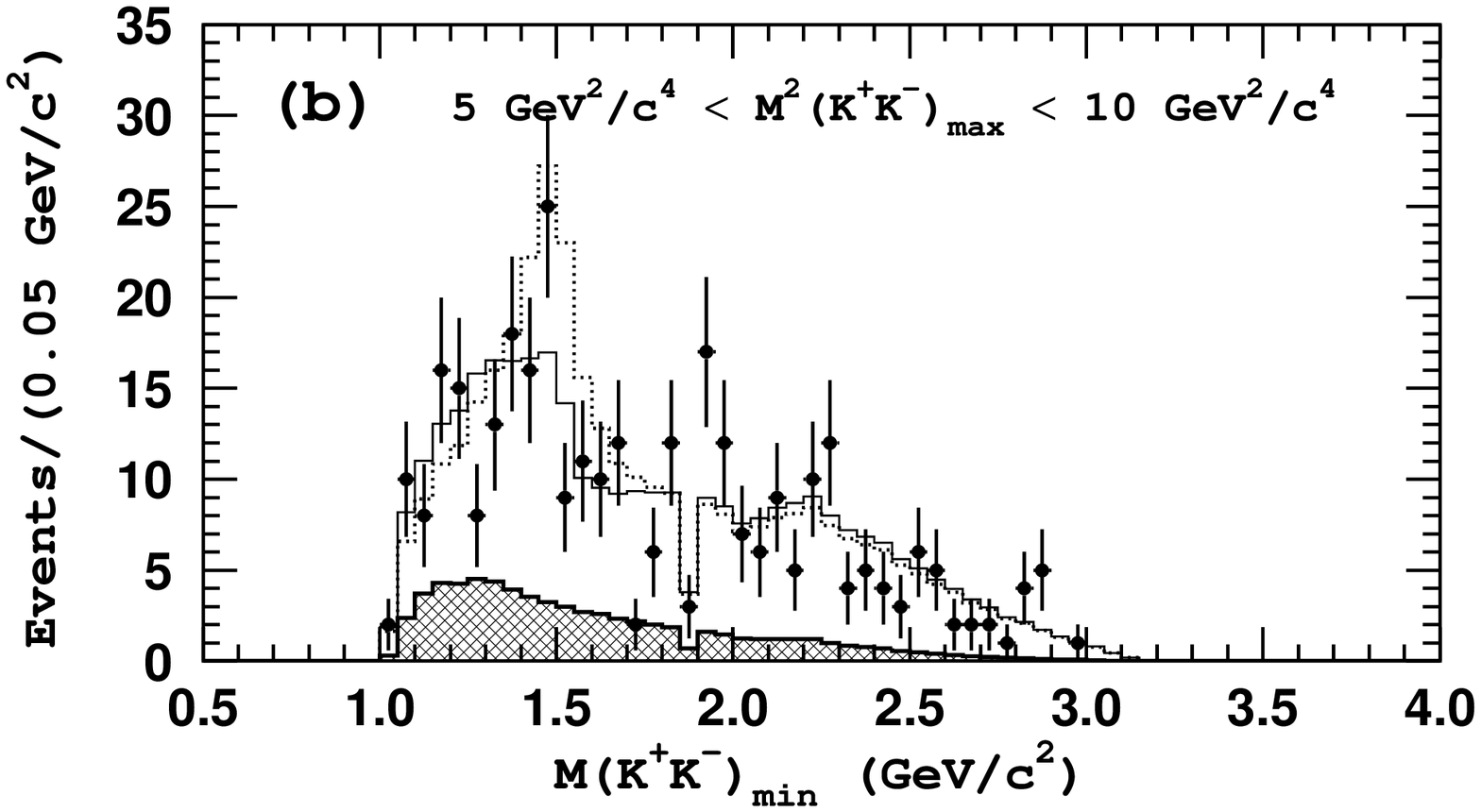} \hfill
 \includegraphics[width=0.48\textwidth,height=45mm]{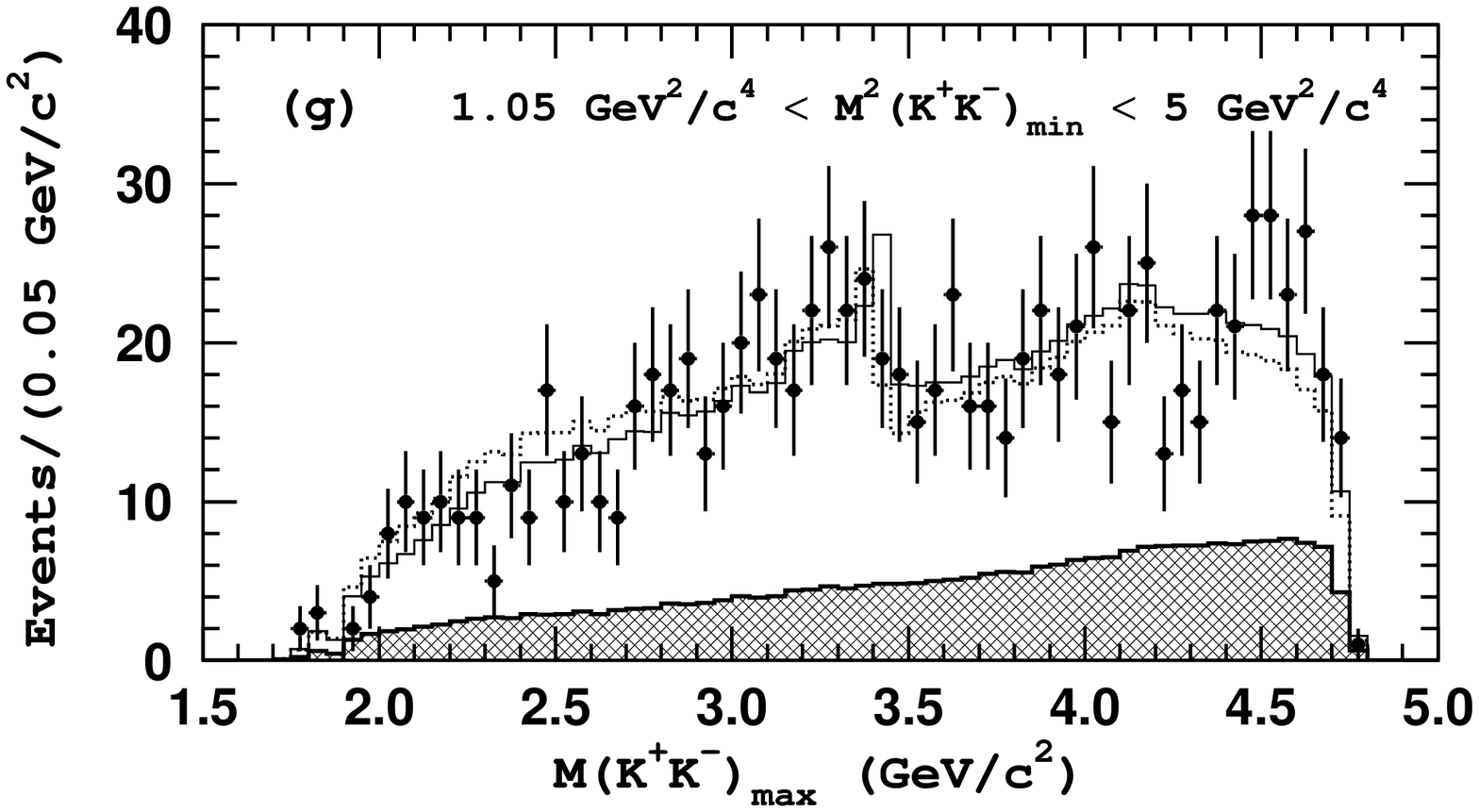} \vspace*{-5mm}\\
 \includegraphics[width=0.48\textwidth,height=45mm]{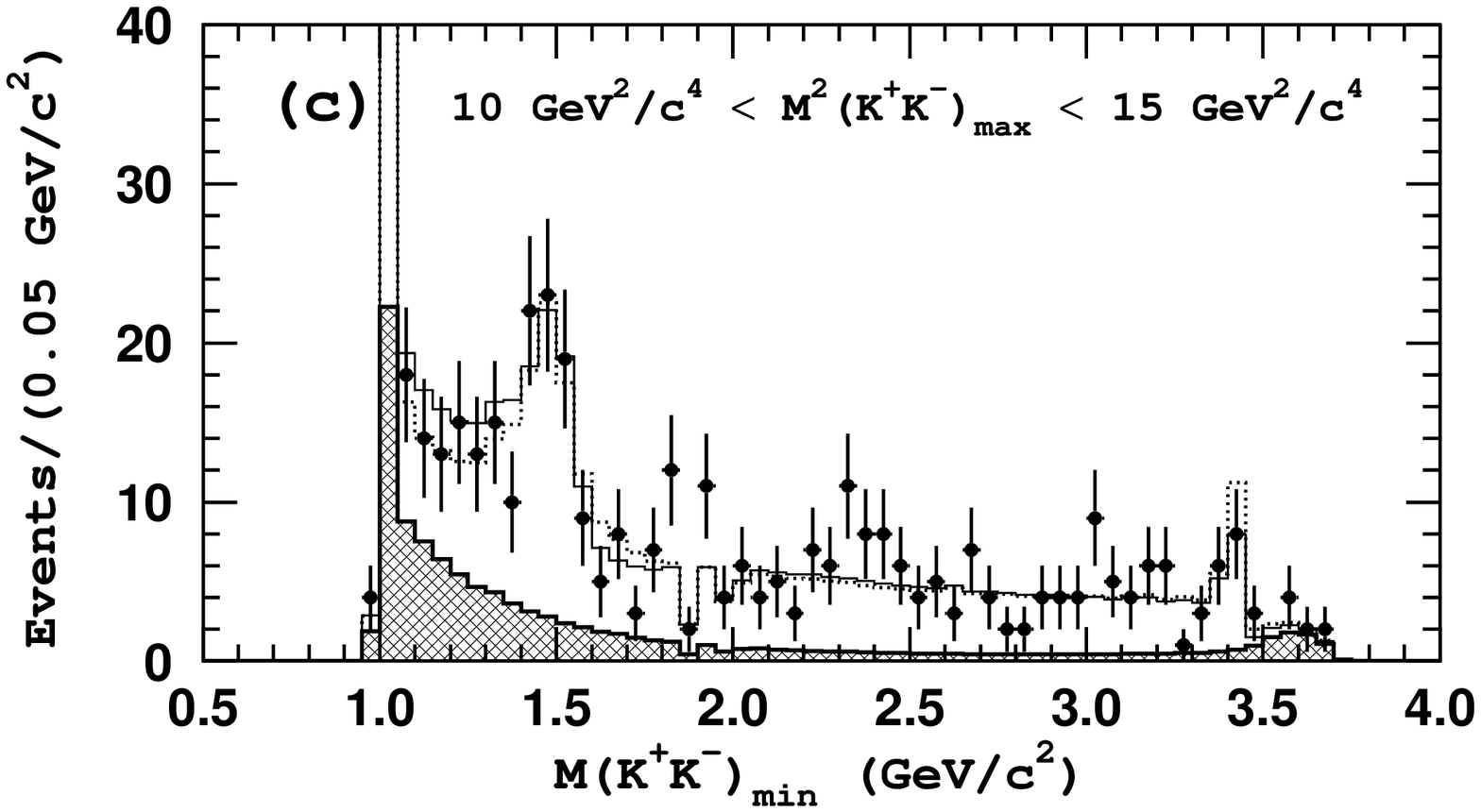} \hfill
 \includegraphics[width=0.48\textwidth,height=45mm]{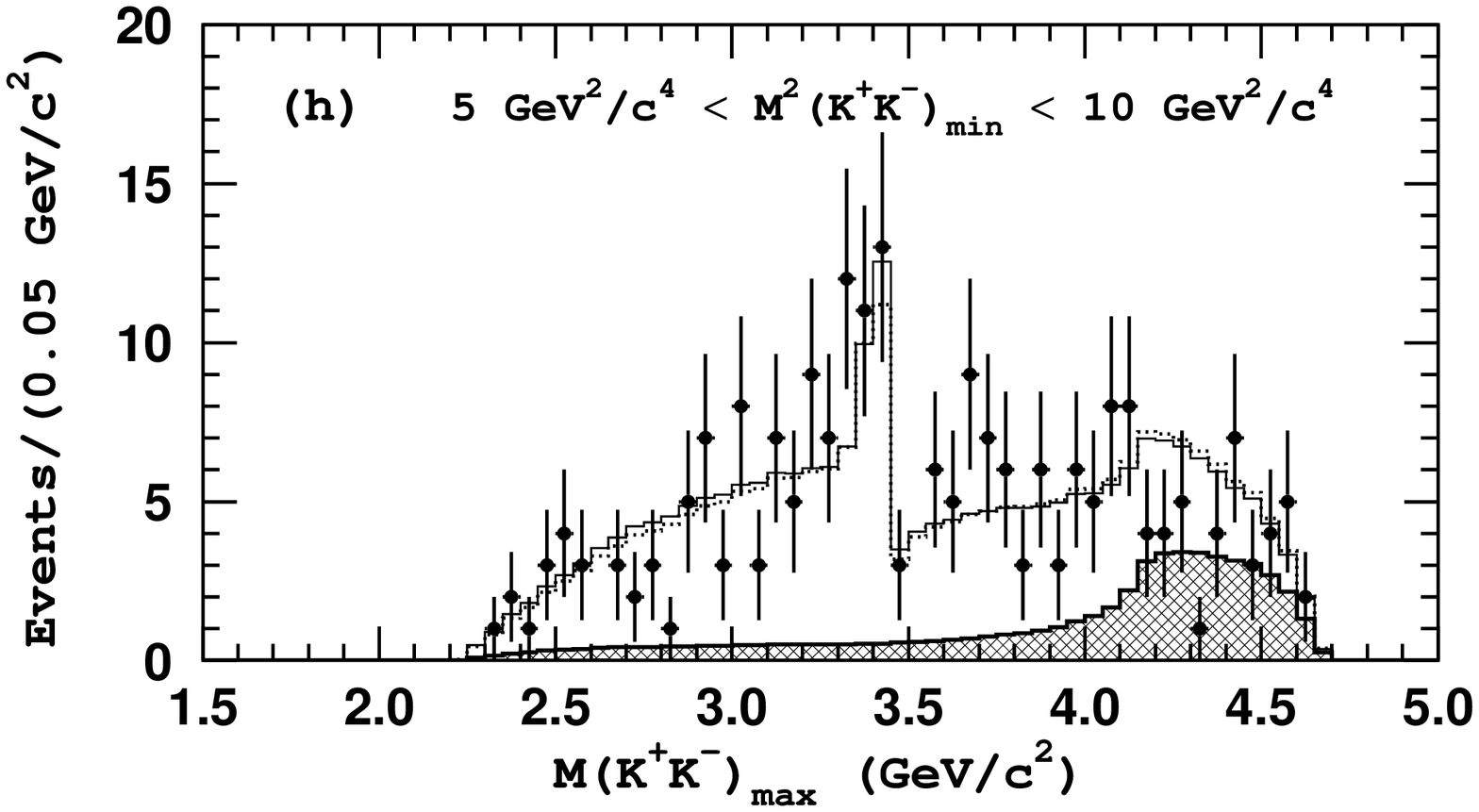} \vspace*{-5mm}\\
 \includegraphics[width=0.48\textwidth,height=45mm]{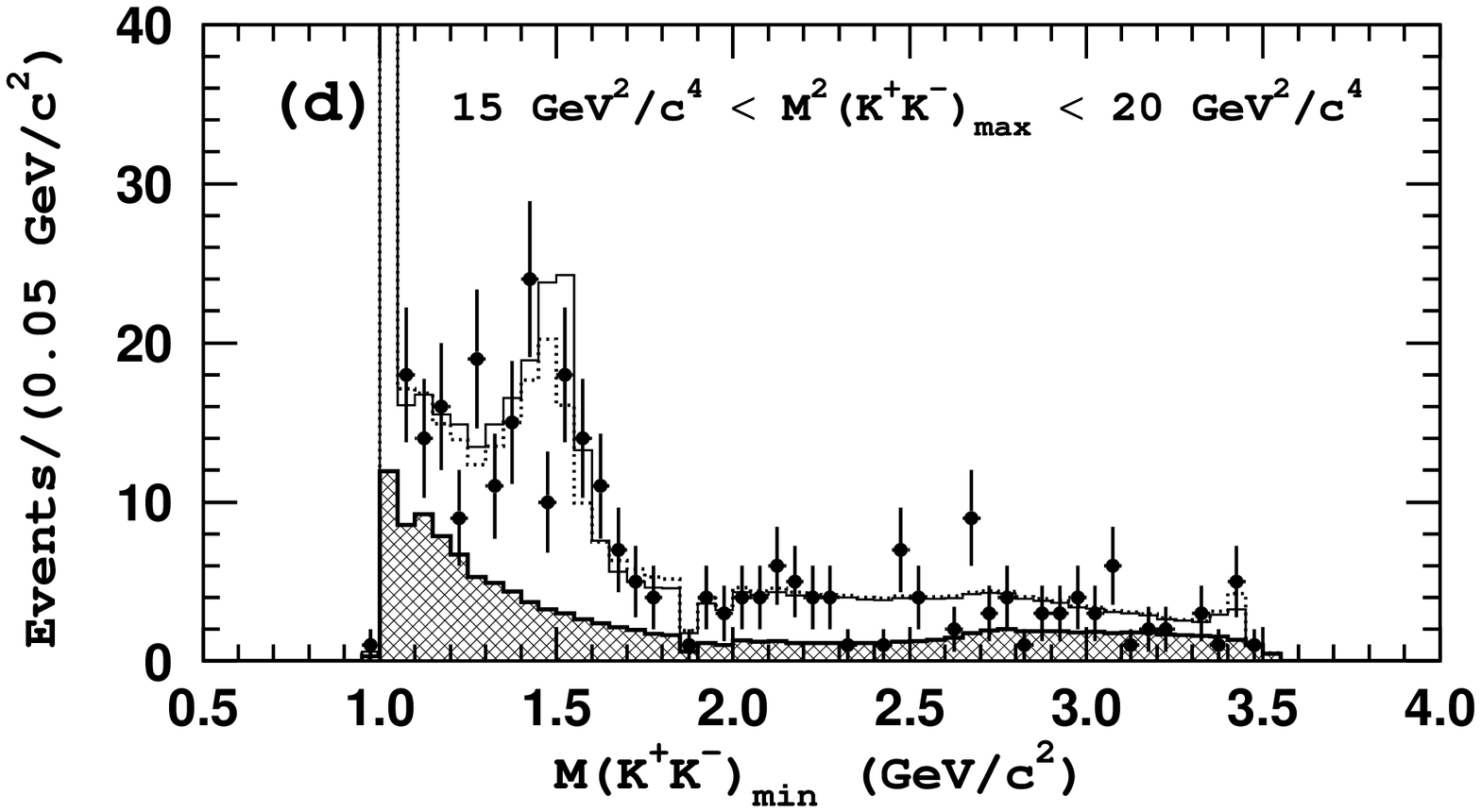} \hfill
 \includegraphics[width=0.48\textwidth,height=45mm]{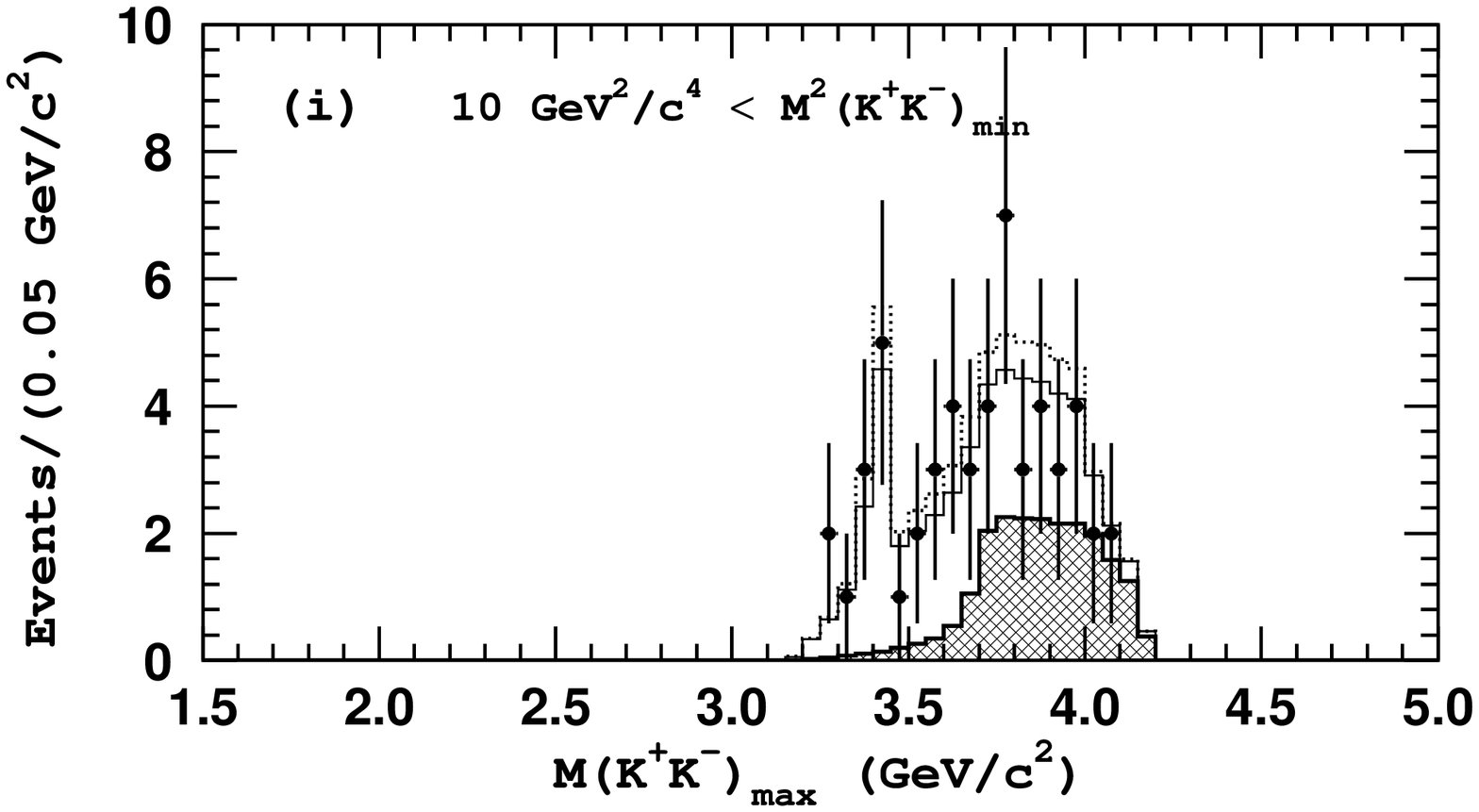} \vspace*{-5mm}\\
 \includegraphics[width=0.48\textwidth,height=45mm]{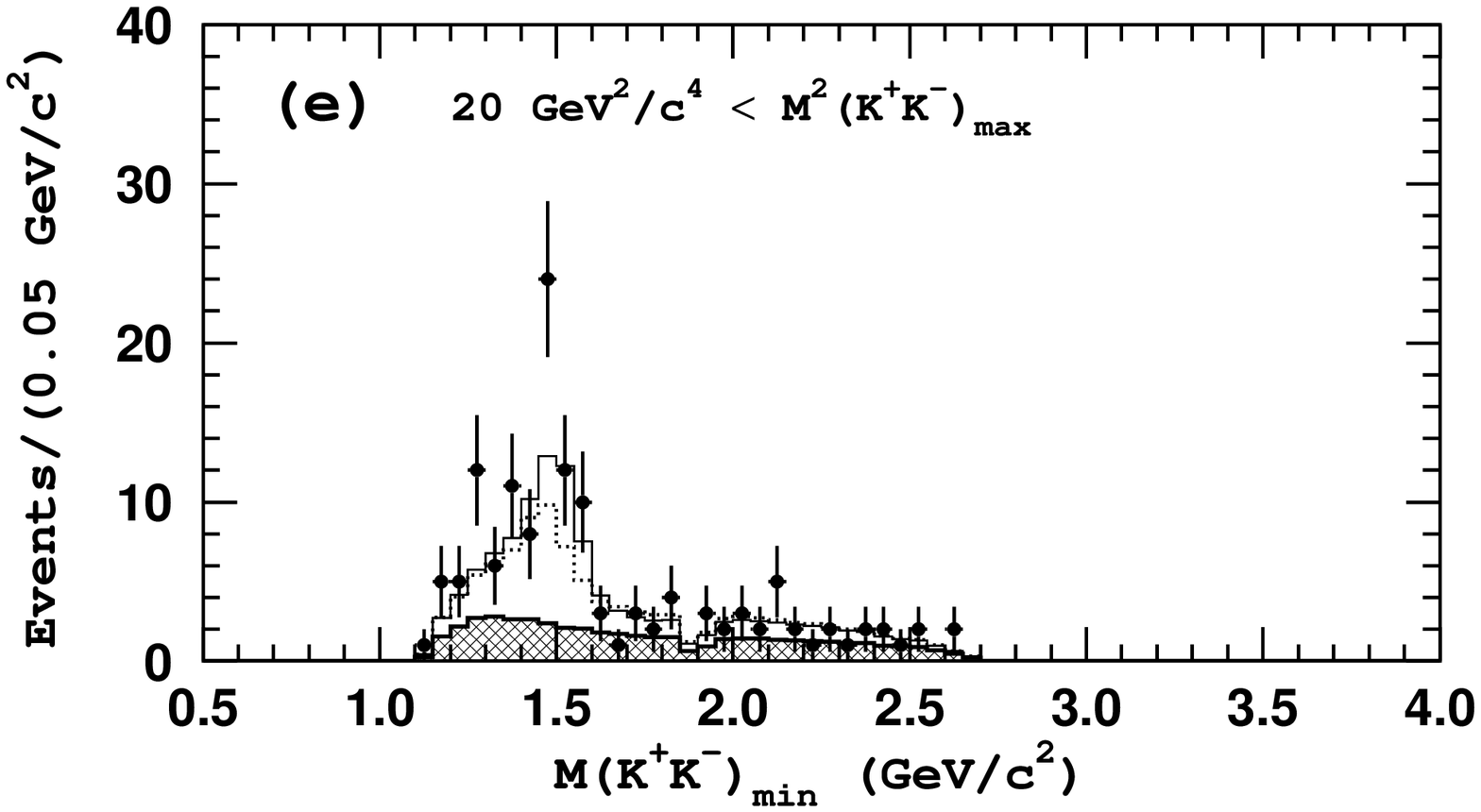} \hfill
 \hspace*{0.48\textwidth}
 \caption{$\mkkmin$ ($\mkkmax$) distributions in slices of $\mkkmax$
          ($\mkkmin$).
           Points with error bars are data, the open histograms are the fit
           results with model $\KKK-$B$_0$ and the hatched histogram is the
           background component. Solid and dotted histograms correspond to
           Solution~1 and Solution~2, respectively 
           (see Table~\ref{tab:kkk-fit-res} and text for details).}
\label{fig:kkk-slices}
\end{figure}
To test this hypothesis, we extend model $\KKK-$A$_J$ to include a
non-resonant amplitude (model $\KKK-$B$_J$) parametrized by
Eq.~(\ref{eq:kkk-non-res}). Results of the fit with model $\KKK-$B$_0$ are
shown in Figs.~\ref{fig:kkk-mods}(d,e,f); numerical values of the fit
parameters are given in Table~\ref{tab:kkk-fit-res}.
The agreement with data is significantly improved compared to model
$\KKK-$A$_0$. In order to check the sensitivity of the
data to the spin of the $f_X(1500)$ state, we replace the scalar amplitude
by a vector (model $\KKK-$B$_1$) or a tensor (model $\KKK-$B$_2$) amplitude
for the $f_X(1500)$ with its mass and width as free parameters. The scalar
hypothesis gives the best fit. Figure~\ref{fig:kkk-slices} shows a detailed
comparison of the fit and the $\mkkmin$ ($\mkkmax$) distributions for
different slices of $\mkkmaxs$ ($\mkkmins$).
Finally, Fig.~\ref{fig:kkk-heli} shows the helicity angle distributions for
the $\phi$ and $f_X(1500)$ regions. Based on these results, we choose model
$\KKK-$B$_0$ as the default. All of the final results for the decay
$\bckkk$ are based on this model.


 To estimate the model dependent uncertainty in the relative fractions of
individual quasi-two-body intermediate states and determine the contribution
of other possible quasi-two-body intermediate states, we modify model
$\KKK-$B$_0$ to include an additional decay channel and repeat the fit to
the data. In particular we test the $\phi(1680)K^+$, $f'_2(1525)K^+$ and
$a_2(1320)K^+$ channels. In all cases the fit finds no statistically
significant signal for the newly added channel. Since we observe a
clear $f_0(980)K^+$ signal in the $\kpp$ final state, we try to include
the $f_0(980)K^+$ amplitude in the $\bckkk$ matrix element as well:
no statistically significant contribution from this channel is found.
As the dominant model
uncertainty is related to the parametrization of the non-resonant amplitude,
we use several alternative, yet also arbitrary, parametrizations to estimate
the relevant uncertainty:
\begin{itemize}
 \item{ $\Am_{\rm nr}(\kkk) =
          a^{\rm nr}_1\left(e^{-\alpha \sft}+
          e^{-\alpha \sst}\right)e^{i\delta^{\rm nr}_1}+
          a^{\rm nr}_2e^{-\alpha \sfs}e^{i\delta^{\rm nr}_2}$;}
 \item{ $\Am_{\rm nr}(\kkk)=
         a_1^{\rm nr}\left(\frac{1}{\sft^\alpha} +
         \frac{1}{\sst^\alpha}\right)e^{i\delta_1^{\rm nr}}$;}
 \item{ $\Am_{\rm nr}(\kkk)=a^{\rm nr}e^{i\delta^{\rm nr}}$.}
\end{itemize}

As in the case of $\bckpp$, we find two solutions in the fit to $\kkk$ events
with model $\KKK-$B$_0$. The comparison between the two solutions and the data
are shown in Fig.~\ref{fig:kkk-slices}. The main difference in these two
solutions is in the fraction of the $B^+\to f_X(1500)K^+$ signal which changes
by about an order of magnitude. Results for both solutions of the $\KKK-$B$_0$
model are given in Table~\ref{tab:kkk-fit-res}.


\subsection{MC pseudo-experiments}

The parameters that are directly determined from the fit to data are the
amplitudes and phases with their statistical errors. However, while the
relative fraction for a particular quasi-two-body channel depends only on
the corresponding amplitude in the matrix element, its statistical error
depends on the statistical errors of all amplitudes and phases. To determine
the statistical errors for quasi-two-body channels, we use a MC
pseudo-experiment technique.

MC pseudo-experiments are MC generated samples which are the proper mixture
of ``signal'' and ``background'' events distributed according to density
functions determined from the fit to experimental events.
For each model we generate 100 statistically independent MC pseudo-experiments
with numbers of signal and background events equal to those found in the
experiment, fit these MC samples, and determine the relative fractions $f_l$
of quasi-two-body channels for each sample. The $f_l$ distributions are then
fit by a Gaussian function; the sigma of the Gaussian determined from the fit
is assigned as the statistical error.


\section{Branching Fraction Results \& Systematic Uncertainties }
\label{sec:results}

In previous sections we determined the relative fractions of various
quasi-two-body intermediate states in three-body $\bckpp$ and $\bckkk$ decays.
To translate those values into absolute branching fractions, we first need to
determine the branching fractions for the three-body decays. To determine the
reconstruction efficiency for the $\bckpp$ and $\bckkk$ decays, we use a MC
simulation where events are distributed over phase space according to the
matrix elements of model $\Kpp-$C$_0$ and model $KKK-$B$_0$, respectively.
The corresponding reconstruction efficiencies are $21.1\pm0.2$\% and
$22.3\pm0.2$\%. Results of the branching fraction calculations for the total
three-body charmless $\bckpp$ and $\bckkk$ decays~\cite{note3} and all the
quasi-two-body intermediate channels are summarized in Table~\ref{tab:branch},
where the first quoted error is statistical, the second is systematic and the
third is the model uncertainty. Branching fractions for $R\to h^+h^-$ decays
are taken from~\cite{PDG}.

For most of the quasi-two-body channels the difference in branching fractions
from the two solutions is relatively small and treated as model error. However
values for the $B^+\to K^*_0(1430)^0\pi^+$ branching fraction are substantially
different for the two solutions and we quote both values in
Table~\ref{tab:branch}. For the $B^+\to\chic K^+$ decay we present both
solutions for the $\chic\to K^+K^-$ channel; for the final result we calculate
the central value combining measurements in $\chic\to\pipi$ channel and the
best fit in $\chic\to K^+K^-$ channel, the second solution in $\chic\to K^+K^-$
channel is used for model error estimation. As the interpretation of the
$f_X(1300)$ and $f_X(1500)$ states is uncertain, we do not quote the
corresponding branching fractions.

For quasi-two-body channels where no significant signal is observed, we
calculate 90\% confidence level upper limits $f_{90}$ for their fractions.
To determine the upper limit we use the following formula
\begin{equation}
0.90=\frac{\int_{0}^{f_{90}}G(a,\sigma;x)dx}{\int_{0}^{\infty}G(a,\sigma;x)dx},
\end{equation}
where $G(a,\sigma;x)$ is a Gaussian function with the measured mean value $a$
for the signal fraction and its statistical error $\sigma$. To account for the
model uncertainty we determine the resonance's contribution with different
parameterizations of the non-resonant amplitude and use the largest value
to evaluate the upper limit. To account for the systematic uncertainty we
decrease the reconstruction efficiency by one standard deviation.


\begin{table}[t]
  \caption{Summary of branching fraction results. The first quoted error is
           statistical, the second is systematic and the third is the model
           error. The branching fraction values in this table are obtained
           from the product of the appropriate fractions in
           Tables~\ref{tab:kpp-fit-res} and~\ref{tab:kkk-fit-res} with the
           branching ratios obtained from the signal yields in 
           Table~\ref{tab:defitall}. Note that the yields in 
           Table~\ref{tab:defitall} include $\chic$ contributions. The
           charmless total fractions in this table exclude the $\chic$
           contribution. The value given in brackets for the
           $K^*_0(1430)^0\pi^+$ and $\chic K^+$
           channels corresponds to the second solution (see text for details).}
  \medskip
  \label{tab:branch}
\centering
  \begin{tabular}{lcr} \hline \hline
\hspace*{20mm}Mode\hspace*{30mm} &
\hspace*{0mm}$\BF(B^+\to Rh^+)\times\BF(R\to h^+h^-)\times10^{6}$ &
\hspace*{10mm}$\BF(B^+\to Rh^+)\times10^{6}$  \\ \hline \hline
 $\kpp$ charmless total & $-$
                        & $46.6\pm2.1\pm4.3$  \\
 $K^*(892)^0\pi^+$, $K^*(892)^0\to K^+\pi^-$
                        & $6.55\pm0.60\pm0.60^{+0.38}_{-0.57}$
                        & $9.83\pm0.90\pm0.90^{+0.57}_{-0.86}$     \\
 $K^*_0(1430)^0\pi^+$, $K^*_0(1430)^0\to K^+\pi^-$
                        & $27.9\pm1.8\pm2.6^{+8.5}_{-5.4}$
                        & $45.0\pm2.9\pm6.2^{+13.7}_{-~8.7}$       \\

                        & ($5.12\pm1.36\pm0.49^{+1.91}_{-0.51}$)
                        & ($8.26\pm2.20\pm1.19^{+3.08}_{-0.82}$)   \\
 $K^*(1410)^0\pi^+$, $K^*(1410)^0\to K^+\pi^-$
                        & $<2.0$ & $-$                             \\
 $K^*(1680)^0\pi^+$, $K^*(1680)^0\to K^+\pi^-$
                        & $<3.1$ & $-$                             \\
 $K^*_2(1430)^0\pi^+$, $K^*_2(1430)^0\to K^+\pi^-$
                        & $<2.3$ & $-$                             \\
 $\rho(770)^0K^+$, $\rho(770)^0\to\pi^+\pi^-$
                        & $4.78\pm0.75\pm0.44^{+0.91}_{-0.87}$
                        & $4.78\pm0.75\pm0.44^{+0.91}_{-0.87}$     \\
 $f_0(980)K^+$, $f_0(980)\to\pi^+\pi^-$
                        & $7.55\pm1.24\pm0.69^{+1.48}_{-0.96}$
                        & $-$                                      \\
 $f_2(1270)K^+$, $f_2(1270)\to\pi^+\pi^-$
                        & $<1.3$ & $-$                             \\
 Non-resonant
                        & $-$
                        & $17.3\pm1.7\pm1.6^{+17.1}_{-7.8}$        \\
\hline
 $\kkk$ charmless total & $-$
                        & $30.6\pm1.2\pm2.3$                       \\
 $\phi K^+$, $\phi\to K^+K^-$
                        & $4.72\pm0.45\pm0.35^{+0.39}_{-0.22}$
                        & $9.60\pm0.92\pm0.71^{+0.78}_{-0.46}$     \\
 $\phi(1680)K^+$, $\phi(1680)\to K^+K^-$
                        & $<0.8$                                   \\
 $f_0(980)K^+$, $f_0(980)\to K^+K^-$                
                        & $<2.9$                                   \\
 $f'_2(1525)K^+$, $f'_2(1525)\to K^+K^-$
                        & $<4.9$                                   \\
 $a_2(1320)K^+$, $a_2(1320)\to K^+K^-$ 
                        & $<1.1$                                   \\
 Non-resonant
                        & $-$
                        & $24.0\pm1.5\pm1.8^{+1.9}_{-5.7}$         \\
\hline
 $\chic K^+$, $\chic\to\pi^+\pi^-$
                        & $1.37\pm0.28\pm0.12^{+0.34}_{-0.35}$
                        & $-$                                      \\
 $\chic K^+$, $\chic\to K^+K^-$
                        & $0.86\pm0.26\pm0.06^{+0.20}_{-0.05}$
                        & $-$                                      \\

                        & ($2.58\pm0.43\pm0.19^{+0.20}_{-0.05}$)
                        & $-$                                      \\
 $\chic K^+$ combined   & $-$
                        & $196\pm35\pm33^{+197}_{-26}$     \\

\hline \hline
  \end{tabular}
\end{table}
 
The dominant sources of systematic error are listed in Table~\ref{khh_syst}.
For the branching fractions of three-body $\bckpp$ and $\bckkk$ decays, we
estimate the systematic uncertainty due to variations of the reconstruction
efficiency over the Dalitz plot by varying the relative phases and amplitudes
of the quasi-two-body states within their errors. The systematic uncertainty
due to requirements on event shape variables is estimated from a comparison
of their distributions for signal MC events and $B^+\to\bar{D}^0\pi^+$ events
in the data. We estimate the uncertainty in the signal yield extraction from
the fit to the $\de$ distribution by varying the parameters of the fitting
function within their errors. The uncertainty due to background
parametrization is estimated by varying the relative fraction of the $\bbbar$
background component and the slope of the $\qqbar$ background function within
their errors. The uncertainty from the particle identification efficiency is
estimated using pure samples of kaons and pions from $D^0\to K^-\pi^+$ decays,
where the $D^0$ flavor is tagged using $D^{*+}\to D^0\pi^+$.
The systematic uncertainty in charged track reconstruction is estimated
using partially reconstructed $D^*\to D\pi$ events and from comparison of
the ratio of $\eta\to\pi^+\pi^-\pi^0$ to $\eta\to\gamma\gamma$ events in data
and MC. The overall systematic uncertainty for the three-body branching
fraction is estimated to be $\pm9.2$\% and $\pm7.4$\% for the $\kpp$ and
$\kkk$ final states, respectively.


\begin{table}[t]
\centering
\caption{Contributions to the systematic uncertainty (in percent) for the
         three-body $\bckpp$ and $\bckkk$ branching fractions.}
\medskip
\label{khh_syst}
  \begin{tabular}{lcc}  \hline \hline
  Source~\hspace*{92mm} &  \multicolumn{2}{c}{Error}          \\
                              &  ~~~$\kpp$~~~ & ~~~$\kkk$~~~  \\
\hline 
 Efficiency nonuniformity     &     $1.2$     &    $0.7$      \\
 Event Shape requirements     &     $2.5$     &    $1.7$      \\
 Signal yield  extraction     &     $5.4$     &    $2.1$      \\
 PID                          &     $6.0$     &    $6.0$      \\
 Charged track reconstruction &     $3.0$     &    $3.0$      \\
 MC statistics                &     $1.0$     &    $1.0$      \\
 $N_{\bbbar}$ estimation      &     $1.0$     &    $1.0$      \\
\hline
 Total                        &     $9.2$     &    $7.4$      \\
\hline \hline
  \end{tabular}
\end{table}


\section{Discussion \& Conclusion}
\label{sec:discussion}

  With a 140~fb$^{-1}$ data sample collected with the Belle detector, we have
performed the first amplitude analysis of $B$ meson decays to three-body
charmless $\kpp$ and $\kkk$ final states. Clear signals are observed in the
$B^+\to K^*(892)^0\pi^+$, $B^+\to\rho(770)^0K^+$, $B^+\to f_0(980)K^+$ and
$B^+\to\phi K^+$ decay channels~\cite{babar-new}. The model uncertainty for
these channels is relatively small due to the narrow width of the intermediate
resonances and (in vector-pseudoscalar decays) due to vector meson
polarization which provides clear signal signatures.

The branching fraction measured for the decay $B^+\to K^*(892)^0\pi^+$
is significantly lower than that reported earlier~\cite{garmash,babar-dalitz}.
The simplified technique used for the analysis of the $\bckpp$ decay
described in~\cite{garmash,babar-dalitz} has no sensitivity to the relative
phases between different resonances, resulting in a large model error. The
full amplitude analysis presented in this paper consistently treats effects of
interference between quasi-two-body amplitudes thus reducing the model error.
The analysis suggests the presence of an additional (presumably
non-resonant) amplitude in the mass region of the $K^*(892)^0$ that absorbs a
significant fraction of the $B$ signal. The $B^+\to K^*(892)^0\pi^+$ branching
fraction measured in our analysis is in better agreement with theoretical
predictions based on the QCD factorization approach~\cite{beneke-neubert}.

The decay mode $B^+\to f_0(980)K^+$ is the first observed example of a $B$
decay to a charmless scalar-pseudoscalar final state. The mass 
$M(f_0(980))=976\pm4^{+2}_{-3}$~\mass ~and width
$\Gamma(f_0(980))=61\pm9^{+14}_{-8}$~\mass ~obtained from the fit are in
agreement with previous measurements~\cite{PDG}. To check the sensitivity of
the results to the parametrization of the $f_0(980)$ lineshape, we repeat the
fit with the Flatt\'e lineshape~\cite{Flatte}. In this case, because of
limited statistics, we fix $g_K$ at the value reported by the E791
Collaboration~\cite{e791}: $g_K = 0.02\pm0.04\pm0.03$. Since the central
value for $g_K$ measured in~\cite{e791} is consistent with zero, we also make
a fit to data with $g_K$ fixed at zero. Finally we repeat the fit with both
$g_\pi$ and $g_K$ floated. In all cases we obtain consistent
results. The sensitivity to the $B^+\to f_0(980)K^+$ decay in the $\kkk$
final state is greatly reduced by the large $B^+\to\phi K^+$ amplitude and by
the scalar non-resonant amplitude. No statistically significant
contribution from this channel to the $\kkk$ three-body final state is
observed, thus only a 90\% confidence level upper limit for the corresponding
branching fractions product is reported.

We report the first observation of the decay $B^+\to\rho(770)^0K^+$. The
statistical significance~\cite{note4} of the signal exceeds $6\sigma$ with all
the models used to fit the $\bckpp$ signal. The measured branching fraction
for this channel agrees well with the theoretical prediction in QCD
factorization~\cite{beneke-neubert}. This is one of the channels where large
direct $CP$ violation is expected~\cite{beneke-neubert}.

Due to the very narrow width of the $\phi$ meson, the branching fraction for
the decay $B^+\to\phi K^+$ is determined with a small model uncertainty.
The obtained value is in good agreement with previous measurements~\cite{phik}.

A clear signal is also observed for the decay $B^+\to \chic K^+$ in both
$\chic\to \pi^+\pi^-$ and $\chic\to K^+K^-$ channels.
Although quite significant statistically, the $B^+\to\chic K^+$ signal
constitutes only a small fraction of the total three-body signal and thus
suffers from a large model error, especially in the $\kkk$ final state,
where the charmless non-resonant amplitude in the $\chic$ mass region is
enhanced compared to the $\kpp$ final state due to the interference caused
by the presence of the two identical kaons.

We also check possible contributions from $B^+\to K^*_2(1430)^0\pi^+$,
$B^+\to K^*(1410)^0\pi^+$, $B^+\to K^*(1680)^0\pi^+$
and $B^+\to f_2(1270)K^+$ decays. In the $\kkk$ final states we check for the
$B^+\to f'_2(1525)K^+$, $B^+\to a_2(1320)K^+$ and $B^+\to\phi(1680)K^+$ 
signals. We find no statistically significant signal in any of these
channels. As a result, we set 90\% confidence level upper limits for their
branching fractions. In the factorization approximation, charmless $B$ decays
to pseudoscalar-tensor final states are expected to occur at the level of
$10^{-6}$ or less~\cite{b2pt}.

For other quasi-two-body channels the interpretation of fit results is less
certain. Although a signal for $B^+\to K^*_0(1430)^0\pi^+$ is observed with a
high statistical significance, its branching fraction is determined with a
large model error. Two solutions with significantly different fractions of the
$B^+\to K^*_0(1430)^0\pi^+$ channel but similar likelihood values are obtained
from the fit to $\kpp$ events. Study with MC simulation confirms the presence
of the second solution. We prepare MC pseudo-experiments where the $\bckpp$
signal is generated with the matrix element of model $\Kpp-$C$_0$ with
parameters corresponding to one of the solutions. In both cases the second
solution is found in the fit to MC samples. It is also worth mentioning that
the two solutions exist with all the parameterizations of the non-resonant
amplitudes we tested. This may indicate that in order to choose a unique
solution additional external information is required. In this sense, the
useful piece of information seems to be the phenomenological estimation of
the $B^+\to K^*_0(1430)^0\pi^+$ branching fraction. The analysis of $B$ meson
decays to scalar-pseudoscalar final states described in Ref.~\cite{b2ps}
suggests that the branching fraction for the $B^+\to K^*_0(1430)^0\pi^+$
decay can be as large as $40\times 10^{-6}$. Unfortunately, the predicted
value suffers from a large uncertainty that is mainly due to uncertainty in
calculation of the $K^*_0(1430)$ decay constant $f_{K^*_0}$. Different methods
used to estimate $f_{K^*_0}$~\cite{b2ps,maltman} give significantly different
results.
We may also try to resolve the ambiguity by employing independent information
from other experiments. For example, analysis of the real and imaginary parts
of the amplitude separately may provide additional useful information.
Following the idea by BaBar Collaboration~\cite{babar-new} (see also discussion
below), we employ LASS results on the partial wave analysis of the elastic
$K$-$\pi$ scattering~\cite{bib:LASS-Kp}. We compare the total scalar $K$-$\pi$
amplitude (which is a sum of the $B^+\to K^*_0(1430)^0\pi^+$ amplitude and the
$K$-$\pi$ component of the non-resonant amplitude Eq.~\ref{eq:kpp-non-res})
with that measured by LASS. From this comparison, we find that results of
the best fit (model $\Kpp-$C$_0$, solution~I) to the $\kpp$ signal events are
in good qualitative agreement with the LASS data.

We cannot identify unambiguously the broad structures observed in the
$M(\pipi)\simeq1.3$~GeV/$c^2$ mass region in the $\bckpp$ decay denoted in our
analysis as $f_X(1300)$ and at $M(\kpkm)\simeq1.5$~GeV/$c^2$ in the $\bckkk$
decay denoted as $f_X(1500)$. If approximated by a single resonant state,
$f_X(1300)$ is equally well described by a scalar or vector amplitude. Analysis
with higher statistics might allow a more definite conclusion. The best
description of the $f_X(1500)$ is achieved with a scalar amplitude with mass
and width determined from the fit consistent with $f_0(1500)$
states~\cite{PDG}.

Amplitude analysis often suffers from uncertainties related to the non-unique
parametrization of the decay amplitude. In our case such an uncertainty
originates mainly from the parametrization of the non-resonant amplitude.
In this analysis, we use a rather simplified empirical parametrization with
a single parameter. In the study of the $\bckpp$ decay by the BaBar
Collaboration~\cite{babar-new} a different approach is used. In their analysis,
an attempt is made to parametrize $K^*_0(1430)^0\pi^+$ and the non-resonant
component by a single amplitude suggested by the LASS collaboration to describe
the scalar amplitude in elastic $K\pi$ scattering~\cite{bib:LASS-Kp}.
Although this approach is experimentally motivated, the use of the LASS
parametrization is limited to the elastic region of
$M(K\pi)\lesssim2.0$~GeV$/c^2$.
Besides, an additional amplitude (a complex constant) is still required for
a satisfactory description of the data~\cite{babar-new}.

It is worth noting that fractions of the non-resonant decay in both $\bckpp$
and $\bckkk$ decays are comparable in size and comprise a significant fraction
of the total three-body signals. Moreover, in the parametrization used in this
analysis the numerical values of the parameter (parameter $\alpha$ in
Eqs.(\ref{eq:kpp-non-res}) and (\ref{eq:kkk-non-res})) for the $\bckpp$ and
$\bckkk$ are very close. This may indicate that the non-resonant amplitudes in
both final states have a common nature, and simultaneous analysis of these two
decay modes may impose additional constraints. An attempt for such an analysis
has been made in~\cite{bib:ochs}. However, the proposed model considers only
the \mbox{$\pi$-$\pi$} component ($a_2^{\rm nr}$ in Eq.~(\ref{eq:kpp-non-res}))
of the non-resonant amplitude and does not account for the \mbox{$K$-$\pi$}
component ($a_1^{\rm nr}$ in Eq.~(\ref{eq:kpp-non-res})), while in our analysis
we find that the \mbox{$K$-$\pi$} component dominates
(see Table~\ref{tab:kpp-fit-res}).

In some cases the uncertainty in the parametrization of the non-resonant
amplitude significantly affects the extraction of relative fractions of other
quasi-two-body channels. Further theoretical progress in this field might
allow reduction of this uncertainty.

Results of the $\bckkk$ Dalitz analysis can be also useful in connection
with the measurement of $CP$ violation in $B^0\to K^0_SK^+K^-$ decay reported
recently by the Belle~\cite{kskk-cp-belle} and BaBar~\cite{kskk-cp-babar}
collaborations. An isospin analysis of $B$ decays to three-kaon final states
by Belle~\cite{garmash2} and independent analysis with moments
technique~\cite{bib:s-plot} by BaBar~\cite{bib:kskk-cp-frac-babar} suggest
the dominance of the $CP$-even component in the $B^0\to K^0_SK^+K^-$ decay
(after the $B^0\to\phi K^0_S$ signal is excluded). This conclusion can be
checked independently by an amplitude analysis of the $K^0_SK^+K^-$ final
state, where the fraction of $CP$-odd states can be obtained as a fraction
of states with odd orbital momenta. Unfortunately, such an analysis is not
feasible with the current experimental data set. Nevertheless, the fact that
we do not observe any significant vector amplitude other than $B^+\to\phi K^+$
in the decay $\bckkk$ supports the conclusion.


\section*{Acknowledgments}

   We thank the KEKB group for the excellent
   operation of the accelerator, the KEK Cryogenics
   group for the efficient operation of the solenoid,
   and the KEK computer group and the National Institute of Informatics
   for valuable computing and Super-SINET network support.
   We acknowledge support from the Ministry of Education,
   Culture, Sports, Science, and Technology of Japan
   and the Japan Society for the Promotion of Science;
   the Australian Research Council
   and the Australian Department of Education, Science and Training;
   the National Science Foundation of China under contract No.~10175071;
   the Department of Science and Technology of India;
   the BK21 program of the Ministry of Education of Korea
   and the CHEP SRC program of the Korea Science and Engineering Foundation;
   the Polish State Committee for Scientific Research under contract No.~2P03B 01324;
   the Ministry of Science and Technology of the Russian Federation;
   the Ministry of Education, Science and Sport of the Republic of Slovenia;
   the National Science Council and the Ministry of Education of Taiwan;
   and the U.S.\ Department of Energy.


\begin{thebibliography}{99}
\bibitem{footnote-2} In Dalitz analysis of $D$ meson three-body decays it is
         common to visualize results of the fit as projections on mass squared
         variables. Because of the large mass of the $B$ meson, we find
         that visualization is more convenient in terms of mass variables.
\bibitem{note3}
Note that $B^+\to \chic K^+$ channel is excluded while calculating branching
fraction for the total $\bckpp$ and $\bckkk$ three-body charmless decays.
\bibitem{note4}
Statistical significance of the signal is
calculated as $\sqrt{-2\ln({\cal{L}}_0/{\cal{L}}_{\rm max})}$, where
${\cal{L}}_{\rm max}$ and ${\cal{L}}_{0}$ denote the maximum likelihood with
the nominal signal yield and with the signal yield fixed at zero, respectively.
\end{thebibliography}
\end{document}